\pdfoutput=1

\documentclass[prb,aps,superscriptaddress,amsmath,amssymb,nofootinbib,twocolumn]{revtex4-2}

\usepackage[T1]{fontenc}
\usepackage[utf8]{inputenc}
\usepackage[english]{babel}

\usepackage{graphicx}
\usepackage[
	colorlinks=true,
	linkcolor=blue,
	urlcolor=blue,
	citecolor=blue,
	bookmarks
	]{hyperref}
\usepackage{stackengine}
\setstackgap{S}{-1.5pt}

\usepackage{amsmath,amssymb,amsfonts}
\usepackage{verbatim} %
\usepackage{mathtools} %
\usepackage{braket} %

\usepackage{bm} %
\usepackage{bbold} %
\usepackage{float}

\usepackage{color}
\usepackage[dvipsnames]{xcolor}
\usepackage{tikz}
\usetikzlibrary{arrows.meta}
\usetikzlibrary{decorations.markings} %

\newcommand{\R}{\mathbb{R}}
\newcommand{\C}{\mathbb{C}}

\newcommand{\Z}{\mathbb{Z}}
\newcommand{\N}{\mathbb{N}}

\newcommand{\Id}{\mathbb{I}}
\newcommand{\Abb}{\mathbb{A}}

\newcommand{\arrow}{{\hspace{3mm} \rightarrow \hspace{3mm}}}

\newcommand{\ccostr}{\hat{c}^\dagger}
\newcommand{\dcostr}{\hat{d}^\dagger}
\newcommand{\etacostr}{\hat{\eta}^\dagger}
\newcommand{\hrho}{\hat{\rho}}
\newcommand{\hA}{\hat{A}}

\newcommand{\hb}{\hat{b}}

\newcommand{\hc}{\hat{c}}

\newcommand{\hd}{\hat{d}}

\newcommand{\heta}{\hat{\eta}}
\newcommand{\hh}{\hat{h}}
\newcommand{\hH}{\hat{H}}

\newcommand{\hn}{\hat{n}}

\newcommand{\hs}{\hat{s}}
\newcommand{\hS}{\hat{S}}
\newcommand{\hT}{\hat{T}}
\newcommand{\hX}{\hat{X}}
\newcommand{\hJ}{\hat{J}}
\newcommand{\hTT}{\hat{\mathbb{T}}}
\newcommand{\hId}{\hat{\mathbb{I}}}
\newcommand{\hLL}{\hat{\mathbb{L}}}
\newcommand{\hmL}{\hat{\mathcal{L}}}
\newcommand{\hM}{\hat{M}}
\newcommand{\hsigma}{\hat{\sigma}}
\newcommand{\imag}{\mbox{Im}}
\newcommand{\mH}{\mathcal{H}}
\newcommand{\mL}{\mathcal{L}}

\newcommand{\mA}{\mathcal{A}}
\newcommand{\real}{\mbox{Re}}
\newcommand{\trace}{\mbox{Tr}}

\def\bbra#1{\mathinner{\langle\langle{#1}|}}
\def\kket#1{\mathinner{|{#1}\rangle\rangle}}
\def\bbraket#1{\mathinner{\langle\langle {#1}\rangle\rangle}}

\definecolor{blueG}{RGB}{51, 102, 204}
\definecolor{magentaG}{RGB}{214.2, 40.8, 132.6}
\definecolor{redG}{RGB}{229.5, 51., 102.}
\definecolor{colA}{RGB}{127, 196, 35}	%
\definecolor{colB}{rgb}{0.71,0.8,0.76}	%
\definecolor{colG}{RGB}{192,192,192}	%
\definecolor{colY}{RGB}{250, 241, 157}	%
\definecolor{colO}{RGB}{204, 209, 161}	%
\definecolor{colP}{RGB}{255, 178, 102}	%
\definecolor{colR}{RGB}{173, 92, 92}	%
\definecolor{colV}{RGB}{177, 153, 255}	%
\definecolor{colLines}{rgb}{0.31,0.31,0.31} %

\usetikzlibrary{decorations.pathmorphing,decorations.markings,patterns,decorations.pathreplacing,shapes.misc}

\newcommand\gridLine[4]{
	\draw [thick,colLines] ({(#1)},{(#2)}) -- ({(#3)},{(#4)});}
\newcommand\gridArrow[4]{
	\draw [thick,colLines,
	postaction={decorate,
		decoration={markings, mark = at position 0.65 with {\arrow[scale=1]{stealth}}}}]
	({(#1)},{(#2)}) -- ({(#3)},{(#4)});}
\newcommand\gridArrowRed[4]{
	\draw [thick,red,
	postaction={decorate,
		decoration={markings, mark = at position 0.65 with {\arrow[scale=1]{stealth}}}}]
	({(#1)},{(#2)}) -- ({(#3)},{(#4)});}
\newcommand\gridArrowBlue[4]{
	\draw [thick,blue,
	postaction={decorate,
		decoration={markings, mark = at position 0.65 with {\arrow[scale=1]{stealth}}}}]
	({(#1)},{(#2)}) -- ({(#3)},{(#4)});}
\newcommand\gridAxis[4]{
	\draw [thick,colLines,
	postaction={decorate,
		decoration={markings, mark = at position 1 with {\arrow[scale=1.5]{stealth}}}}]
	({(#1)},{(#2)}) -- ({(#3)},{(#4)});}

\newcommand\withoutrectangle[5]{
	\draw[#5,fill=#5] ({(#1-#3)},{(#2-#4)})  -- ({(#1+#3)},{(#2-#4)})  -- ({(#1+#3)},{(#2+#4)})  -- ({(#1-#3)},{(#2+#4)})  -- cycle;}

\begin{document}

\title{Going beyond the standard quasiparticle picture: Entanglement spreading after quenching from highly entangled states}

\author{Elisa Sabadini}
\affiliation{\it Dipartimento di Fisica,  Universit\`a di Pisa, Largo Bruno Pontecorvo 3, Pisa, Italy}

\author{Vincenzo Alba}
\affiliation{\it Dipartimento di Fisica, Universit\`a di Pisa, and  INFN Sezione di Pisa, Largo Bruno Pontecorvo 3, Pisa, Italy}

\date{\today}

\begin{abstract}

The quasiparticle picture provides a quantitative description of entanglement spreading after
quantum quenches in integrable systems. In interacting models, however, it relies on the assumption
that the quench creates entangled pairs of quasiparticles, which is justified only for the so-called
integrable initial states. Here we go beyond this setting by studying quenches in the XXZ spin chain
from the $n$-rainbow states, a family of macroscopically homogeneous states formed by increasingly
long-range Bell pairs, which are not integrable for $n>1$. The
local stationary state after the quench approaches the infinite-temperature state as $n\to\infty$, with corrections of
order $1/n$, as we show analytically.
Nevertheless, in the noninteracting XX chain the quench produces entangled \emph{multiplets} of $2n$ species of quasiparticles. Although interspecies correlations
vanish as $1/n$, they strongly suppress the growth of the entanglement entropy, whose rate vanishes as $n\to\infty$.
We capture this behaviour within a generalized quasiparticle picture. In the interacting XXZ
chain, instead, time-dependent density matrix renormalization group (tDMRG) simulations show that the entanglement entropy grows at the rate at which
thermodynamic entropy is exchanged across the cut. Within the time scales accessible to tDMRG, the data show
no trace of the multiplet suppression. This holds both after homogeneous quenches and after joining
the $n$-rainbow state with an infinite-temperature state, for which we conjecture 
a formula for the von Neumann entropy.
\end{abstract}

\maketitle

\section{INTRODUCTION}

The out-of-equilibrium dynamics of one-dimensional quantum many-body systems has attracted considerable interest in recent decades, also thanks to experiments with ultracold atoms~\cite{ColdAtoms1, ColdAtoms2, ColdAtoms3, ColdAtoms4, ColdAtoms5}. A paradigmatic protocol is the {\itshape quantum quench} \cite{quench}, in which an initial state is suddenly evolved under a Hamiltonian of which it is not an eigenstate.
Such dynamics provides a natural setting to investigate how statistical ensembles emerge from the unitary evolution of isolated quantum systems prepared in pure states \cite{RelaxationPedagogical}.
Although these systems cannot relax as a whole,
they can relax locally: local observables can attain stationary values.
More precisely, if in an infinite system the limit
$\hrho_{s,\Abb} \; \coloneq \; \lim_{t \rightarrow +\infty} \hrho_\Abb(t)$ exists for every finite subsystem $\Abb$,
the system is said to relax to a local stationary state (LSS) $\hrho_s$, defined by
\begin{equation}
	\trace_{\, \overline{\Abb}}\,\left[ \hrho_s \right] \; = \; \hrho_{s,\Abb} \; = \; \lim_{t \rightarrow +\infty} \hrho_\Abb(t) \quad \forall \Abb .
	\label{eq:relaxation}
\end{equation}
For generic, nonintegrable systems, where energy is essentially the only conserved quantity, the stationary state is expected to be described by a Gibbs ensemble~\cite{ETH1, ETH2, Thermalization, ETH3}, leading to local thermalization.
In integrable systems, the presence of an extensive number of (quasi)local conserved charges prevents thermalization.
However, if a suitable complete set $\set{\hH_n}$ of (quasi)local conserved charges in involution is known
and an equilibrium state exists,
then (under suitable assumptions \cite{Validity}) the system relaxes to a {\itshape generalized Gibbs ensemble (GGE)} \cite{GGE, Action_GGE_first, Caux1, Caux2, Action_GGE, QuenchActionPedagogical, FirstQuasilocal, StringCharge, GGEComplete, GGE1, GGE2, GGE3},
\begin{equation}
	\hrho_{GGE} = \frac{e^{-\sum_n\beta_n\hH_n}}
	{\trace\left[e^{-\sum_n\beta_n\hH_n}\right]},
\end{equation}
where the Lagrange multipliers ${\beta_n}$ are fixed by the expectation values of the conserved charges in the initial state.
Since the GGE density matrix is impractical to handle directly,
it is convenient to invoke
the generalized eigenstate thermalization hypothesis (ETH)~\cite{GETH1,GETH2,ETH3,alba2015eigenstate,essler2024statistics,rottoli2026eigenstate,senese2026non}, and replace $\hrho_{GGE}$ by a single representative 
eigenstate. Interestingly, the approach to thermalization or equilibration is accompanied by entanglement creation, which acts as the ``driving force'' towards equilibration~\cite{EntThermo4}. This is transparent in the case of integrable systems, for which the interplay between entanglement and thermodynamic entropy is fully characterized via the quasiparticle picture~\cite{QuasiPicture,QuasiProof,AlbaCalabrese2017,Alba_2018,Calabrese2018PhysA,Calabrese2020LectNotes,Alba_2021,klobas2021exact,bertini2022growth}.

\begin{figure}
	\centering
	\begin{tikzpicture}[scale=0.55,baseline={([yshift=-0.6ex]current bounding box.center)}]
		\withoutrectangle{3.25}{1.2}{3.5}{2.4}{cyan, opacity=0.12}
		\gridLine{-0.25}{0}{13.75}{0}
		\foreach \x in {0, 1, 2, 3}{
			\draw[thick,olive] (\x,0.15) arc (180:0:{0.25 + (3-\x)});
		}
		\foreach \x in {0.5, 1.5, 2.5}{
			\draw[thick,olive] (\x,0.15) arc (180:0:{0.25 + (3-\x)});
		}
		\foreach \x in {0,0.5,...,13.5}{
			\draw [colLines,fill=colB] ({\x},{0}) circle (0.15);}
		\node at (-0.6,0) {\scalebox{0.85}{$...$}};
		\node at (14.1,0) {\scalebox{0.85}{$...$}};
		\node at (3.25,-0.7) {\scalebox{1.2}{$ n=7 $}};
		\draw[thick,olive] (8.25,2.15) arc (180:0:0.5);
		\draw [colLines,fill=colB] (8.25,2) circle (0.15);
		\draw [colLines,fill=colB] (9.25,2) circle (0.15);
		\node at (8.25,1.4) {\scalebox{1}{$i$}};
		\node at (9.25,1.4) {\scalebox{1}{$j$}};
		\node at (12,2) {\scalebox{1}{$\displaystyle 
				= \; \frac{\ket{\uparrow_i \downarrow_j} + \ket{\downarrow_i \uparrow_j}}{\sqrt{2}}
				$}};
	\end{tikzpicture}
	\begin{tikzpicture}[scale=0.55,baseline={([yshift=-0.6ex]current bounding box.center)}]
		\withoutrectangle{4.25}{0.2}{4.5}{1.4}{green, opacity=0.12}
		\withoutrectangle{11.25}{0.2}{2.5}{1.4}{orange, opacity=0.12}
		\gridLine{-0.25}{0}{13.75}{0}
		\foreach \x in {0, 1}{
			\draw[thick,olive] (\x,0.15) arc (180:0:{0.25 + (1-\x)});
			\draw[thick,olive] (\x+3,0.15) arc (180:0:{0.25 + (1-\x)});
			\draw[thick,olive] (\x+6,0.15) arc (180:0:{0.25 + (1-\x)});
		}
		\foreach \x in {0.5}{
			\draw[thick,olive] (\x,0.15) arc (180:0:{0.25 + (1-\x)});
			\draw[thick,olive] (\x+3,0.15) arc (180:0:{0.25 + (1-\x)});
			\draw[thick,olive] (\x+6,0.15) arc (180:0:{0.25 + (1-\x)});
		}
		\foreach \x in {0,0.5,...,13.5}{
			\draw [colLines,fill=colB] ({\x},{0}) circle (0.15);}
		\node at (-0.6,0) {\scalebox{0.85}{$...$}};
		\node at (14.1,0) {\scalebox{0.85}{$...$}};
		\node at (4.25,-0.7) {\scalebox{1.2}{$ \hrho_L = \ket{\psi_L}\bra{\psi_L} $}};
		\node at (11.25,-0.7) {\scalebox{1.2}{$ \hrho_R \propto \hId $}};
		\node at (11.25,0.75) {\scalebox{1.5}{\color{RedOrange} $ T \rightarrow +\infty $}};
	\end{tikzpicture}
	\caption{		
		Top: $n$-rainbow state on a spin$-\frac{1}{2}$ chain of length $L$.
		The shaded area shows the unit cell for the $7$-rainbow state ($n=7$).
		The two sites connected by an arc are prepared in the zero-magnetization triplet Bell-pair.
		The unit cell is repeated over the lattice 
		$L/(2n)$ times.
		Bottom: The bipartite quench in which the left and right halves of the chain are 
		prepared in two different states. We consider the situation in which the two states 
		are $n$-rainbow states with different $n$ or one is the $n$-rainbow state and the 
		other one is the infinite-temperature thermal state. 
		The system then evolves under the homogeneous XXZ Hamiltonian.  
}
	\label{fig:cartoon}
\end{figure}
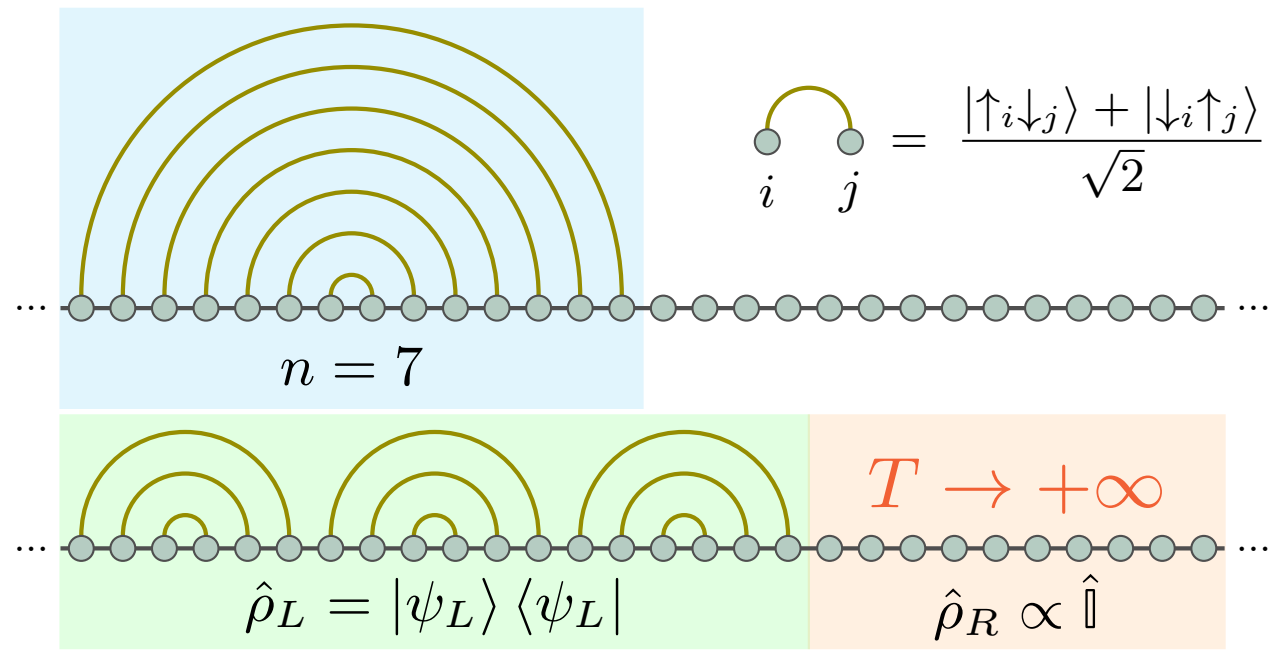

Until recently, most studies focused on quenches from
weakly entangled initial states, whose time evolution can be treated efficiently
both analytically, within the quasiparticle picture~\cite{QuasiPicture} or generalized hydrodynamics (GHD)~\cite{HydroA,Bertini},
and numerically, with matrix product state (MPS) methods such as the density matrix renormalization
group (DMRG)~\cite{DMRG}.

However, quenches from highly entangled initial states are now attracting growing interest, as their experimental realization is becoming feasible, for example through trapped-ion systems \cite{Experimental1, Experimental2}.
Some pioneering results have been obtained recently. 
For instance, the entanglement dynamics from 
crosscap states \cite{Crosscap1, Crosscap2, Crosscap3, Crosscap4} and entangled multipodal states \cite{Multipodal} can be 
described within a generalized quasiparticle picture. The entanglement dynamics from thermofield double states in integrable models was studied in Ref.~\cite{Lagnese2022TFD}, and related questions on entanglement growth from entangled initial states have been addressed recently in Refs.~\cite{zhang2026entanglement, ray2025page}. 

A crucial question is how a large entanglement content of the initial state
affects the relaxation to the steady state.
For instance, does it drive the system towards thermalization?
And how do entanglement spreading
and the transport of local conserved quantities
behave in quenches from highly entangled initial states?

The question of entanglement spreading is particularly relevant for interacting integrable
systems. There, the quasiparticle picture~\cite{QuasiPicture,AlbaCalabrese2017,Alba_2021}
effectively describes the phenomenon in terms of pairs of entangled quasiparticles with opposite
rapidities, which are created by the quench and propagate ballistically. This pair structure is
justified for the so-called integrable initial states~\cite{IntegrableStates}, which have nonzero
overlap only with Bethe eigenstates formed by pairs of opposite rapidities. For more generic initial states the picture remains unclear. In free systems, quenches from generic Gaussian initial states 
produce entangled \emph{multiplets} of quasiparticles~\cite{SpeciesFirst,SpeciesBertini,SpeciesCollura,Caceffo}, 
which can be incorporated in the quasiparticle picture. 
In interacting integrable systems, instead, it is not known how entanglement spreads after quenches
from initial states that are not integrable, and whether multiplets play any role. To the best of our
knowledge, entanglement dynamics in interacting integrable systems has so far been investigated
mostly for integrable initial states. Exact results for the relaxation and the entanglement dynamics in an interacting integrable model, the quantum cellular automaton Rule 54, were obtained in Refs.~\cite{klobas2021exact, Rule54_2, Rule54_3}. 

To address these issues, here we investigate the dynamics starting from 
a special class of initial states, which we call $n$-rainbow states,
inspired by the rainbow state~\cite{Rainbow}. They are constructed 
by repeating $L/(2n)$ times a unit cell containing $n$ concentric long-range Bell pairs 
(see Fig.~\ref{fig:cartoon}). 
This ensures that the system retains some degree of homogeneity, which is essential
for relaxation to a GGE. It also allows us to
efficiently determine the GGE by exploiting integrability~\cite{FirstQuasilocal,Quasilocal,StringCharge}. 
More broadly, the general idea behind this definition is to model initial states that contain a finite density of extended 
highly-entangled ``patches'' (of size $2n$ for the $n$-rainbow states). Similar questions 
were addressed in Ref.~\cite{zhang2015thermalization}. 

In this work, we consider dynamics in the XXZ chain, with anisotropy parameter $\Delta$.
For $\Delta =0$ (XX chain) the Hamiltonian maps onto a free-fermion tight-binding chain~\cite{XXfirst1}.
For generic $\Delta$, the model is exactly solvable by the Bethe ansatz~\cite{Bethe1931,TakahashiBook,Faddeev, ABAfirst, ABAslavnov}.
Here we show that the dynamics after the quench from the $n$-rainbow states leads to 
an infinite-temperature thermal steady state in the limit $n\to\infty$. Corrections to thermal behaviour decay as $1/n$.
This implies that transport of local 
conserved quantities after the quench from the bipartite state $|R_n\rangle\otimes|T\to\infty\rangle$ (see Fig.~\ref{fig:cartoon}) 
is suppressed as $1/n$ in the large $n$ limit. 
In the noninteracting XX limit we prove this analytically, exploiting that the $n$-rainbow states are Gaussian and that Gaussianity is preserved by the time evolution.
Despite local thermalization, the quench from the $n$-rainbow state
produces entangled multiplets of $2n$ different species of quasiparticles.
This is described by a generalization of the quasiparticle
picture~\cite{QuasiPicture, AlbaCalabrese2017} (similar to~\cite{SpeciesFirst, SpeciesBertini}), which
we derive for quenches from both homogeneous and inhomogeneous initial states.
In the limit $n\to\infty$ the number of species diverges, while the
``off-diagonal'' correlations between species vanish as $1/n$.
Remarkably, these correlations cannot be neglected: they strongly suppress the
growth of the entanglement entropy, whose rate vanishes as $n\to\infty$.

Then, we turn to the genuinely interacting XXZ model, whose GGE can be fully characterized by the string-charge 
duality~\cite{StringCharge}. First, we derive a set of recursive equations that determine the GGEs
for the $n$-rainbow states, which we solve numerically up to $n\sim 10^3$.
This allows us to show that, in the limit $n\to\infty$, the GGE describing
the steady state after the quench from the $n$-rainbow state converges to the infinite-temperature 
thermal ensemble. Finite-$n$ corrections decay as $1/n$, as in the XX chain.
A similar result was obtained in Ref.~\cite{random} for quenches from random product states. 
We also investigate quantum transport, considering the setup of Fig.~\ref{fig:cartoon}. Similarly to the XX chain, 
transport of local conserved quantities between the two parts of the chain is suppressed as $1/n$ 
in the large $n$ limit. Using GHD~\cite{Bertini,HydroA}, we derive predictions
for the space-time profiles of local conserved quantities and currents, which are
confirmed by {\itshape ab initio} time-dependent DMRG (tDMRG) simulations. 

Finally, we consider entanglement dynamics in the XXZ chain. Surprisingly,
our numerical results show that the entanglement entropy after the quench from the homogeneous $n$-rainbow state grows linearly in time, even
for moderately large $n=5$, at the rate at which thermodynamic entropy is exchanged between the subsystem and the
rest. This suggests that entanglement spreading is fundamentally different in interacting integrable models, although we cannot rule
out that the linear growth is a long ``prethermal-like'' transient. 
We also consider entanglement dynamics after the bipartitioning protocol in Fig.~\ref{fig:cartoon}.
Since the system is not in a pure state, the von Neumann entropy is not a measure of the
entanglement between the two parts of the chain. We conjecture that the von Neumann entropy  is the sum of the thermal entropy 
produced in the right part and transported to the left and the genuine entanglement entropy produced in 
the left half of the chain. This is supported by tDMRG simulations for the bipartite protocol with the integrable state $|R_1\rangle$ in the
left half. Next, we consider quenches in which the left part is prepared in the $n$-rainbow state with $n>1$. 
Again, the multiplet mechanism, which in the XX chain suppresses the entanglement
growth, appears to be absent in the interacting chain, and the growth rate of the von Neumann entropy coincides with 
the rate at which the two regions exchange thermodynamic entropy. 

This paper is organized as follows. In Section~\ref{sec:setting} we introduce the XXZ chain and the
$n$-rainbow states. In Section~\ref{sec:TBA} we review the thermodynamic Bethe ansatz (TBA) treatment of the
XXZ chain and the string-charge duality, which allows us to construct the GGE.
Section~\ref{sec:GHD} briefly reviews GHD. In Section~\ref{sec:QP} we
discuss the quasiparticle picture for entanglement spreading.
In Section~\ref{sec:exact} we show that quenches from the $n$-rainbow states give rise to thermal
behaviour in the limit $n\to\infty$ in the XX chain. In Section~\ref{sec:Entanglement} we provide
a formula for the rate of entanglement growth after quenches from inhomogeneous initial states
formed by joining $n$-rainbow states. In Section~\ref{sec:homogeneous} we show that quenches from
homogeneous $n$-rainbow states lead to thermalization in the limit $n\to\infty$ in the XXZ chain.
In Section~\ref{sec:homoent} we show that, unlike in the XX chain, the entanglement entropy after
these quenches grows with the rate of exchange of thermodynamic entropy.
In Section~\ref{sec:bipartitioning}
we show that quantum transport after the quench from the state in Fig.~\ref{fig:cartoon} (bottom) is
suppressed as $1/n$ for $n\to\infty$. In Section~\ref{sec:entdynamics} we focus on the
dynamics of the von Neumann entropy after the quench from the bipartitioning protocol with the infinite-temperature state, and
we conclude in Section~\ref{sec:conclusions}. Technical details are reported in the Appendices:
the construction of the quasilocal charges from the transfer matrix (Appendix~\ref{sec:Algebraic}),
the multiplet quasiparticle picture in the XX chain (Appendix~\ref{app:XXdetails}), the charges of the
$n$-rainbow and infinite-temperature states (Appendix~\ref{app:charges}), the quasiparticle picture for
the half-purified protocol (Appendix~\ref{app:halfpurtot}), and the numerical methods
(Appendix~\ref{app:numerics}). Finally, in Appendix~\ref{app:violations} we show that the pair-structure
quasiparticle picture fails for nonintegrable initial states such as the fat N\'eel states.

\section{\label{sec:setting} XXZ MODEL AND $n$-RAINBOW STATES}

\subsection{Definition of the model and notations}

We consider the spin-$1/2$ XXZ chain~\cite{TakahashiBook} on $L$ sites, described by the Hamiltonian
\begin{subequations}	
\begin{equation}
    \hH_{XXZ} \coloneq \sum_{j=1}^{L}\hh_{j,j+1},
    \label{eq:XXZ}
\end{equation}
with periodic boundary conditions (PBC) $\hS^\alpha_{L+1}=\hS^\alpha_1$ and
$\hh_{L,L+1}=\hh_{L,1}$. The local energy density is
\begin{equation}
    \hh_{j,j+1}\coloneq
    \frac{\hS_j^+\hS_{j+1}^-+\hS_j^-\hS_{j+1}^+}{2}
    +\Delta\hS_j^z\hS_{j+1}^z.
    \label{eq:local_energy_density}
\end{equation}
where $\hS^{\set{x, y, z}} = \frac{1}{2}\hsigma^{\set{x, y, z}}$ and $\hS^\pm = \hS^x \pm i \hS^y$.

The XXZ chain displays a global $U(1)$ symmetry corresponding to the conservation of the total magnetization along the $z$ direction. 
It also exhibits $\Z_2$ symmetries associated with global spin flips along the three directions in spin space.
The Jordan-Wigner transformation maps $\hH_{XXZ}$ onto a model of spinless fermions with a density-density interaction, 
\begin{equation}
	\begin{split}
	\hH_{XXZ} \leftrightarrow & \left[\sum_{j=1}^{L-1} \frac{\left(\ccostr_j \hc_{j+1} + \ccostr_{j+1} \hc_j \right)}{2} \right]
	\; +
	\\
	& + \; (-1)^{\hat{N}_{tot}+1} \frac{\left(\ccostr_L \hc_1 + \hc_L \ccostr_1 \right)}{2} \; +
	\\
	& + \;
	\Delta \left[
	\frac{\hId L}{4}
	- \hat{N}_{tot}
	+ \sum_{j=1}^L
	\hn_j \hn_{j+1}
	\right] .
	\label{eq:fermionicXXZ}
	\end{split}
\end{equation}
\end{subequations}
where $\hat N_{tot}$ is the total number of fermions, which is related to the total magnetization,
and $\hId$ is the identity matrix. In Eq.~\eqref{eq:XXZ} we set the exchange coupling to one. In some
formulas below we restore it as an overall energy scale $J$; all numerical results are for $J=1$. 
It is useful to define $\eta$ as 
\begin{equation}
	\left\lbrace
	\begin{array}{ll}
		\eta \coloneq \mbox{arccosh}(\Delta) > 0 & \mbox{if $|\Delta| \geq 1$,} \\
		\eta \coloneq \arccos(\Delta) \in [0,\pi] & \mbox{if $|\Delta| < 1$,} \\
	\end{array}
	\right.
	\label{eq:eta}
\end{equation}
For $|\Delta|>1$ the model is gapped, and for $|\Delta|\rightarrow +\infty$ it reduces to the Ising chain.
Here we focus on the interacting antiferromagnetic regime $\Delta >1$.
For $|\Delta|<1$ the model is in a Luttinger-liquid phase; there we only consider the free-fermion point $\Delta =0$, i.e., the
XX chain, in Sec.~\ref{sec:exact}.

\subsection{$n$-rainbow states}

The rainbow state was first introduced in Ref.~\cite{firstRainbow} as the
concentric singlet phase, and later acquired its current name in Ref.~\cite{rainbow1}.
It has since been studied extensively~\cite{Rainbow, rainbow2, rainbow3, rainbow4, Alba2019Unusual}. 

In terms of fermionic creation and annihilation operators it is defined as
\begin{equation}
\ket{R} \; \coloneq \;  \lim_{L \rightarrow +\infty} \left( \hb^{[(-1)^{L/2 + 1}]}_{-\frac{L-1}{2}, \frac{L-1}{2}}\right)^\dagger
...
\left( \hb^{[-]}_{-\frac{3}{2}, \frac{3}{2}}\right)^\dagger
\left( \hb^{[+]}_{-\frac{1}{2}, \frac{1}{2}}\right)^\dagger \ket{\Omega} ,
\end{equation}
where
\begin{equation}
	\label{eq:Bellpaircreators}
 \left(\hb^{[\pm]}_{ij}\right)^\dagger \; \coloneq \; \frac{\ccostr_i \pm \ccostr_j}{\sqrt{2}}
\end{equation}
and $\ket{\Omega}$ is the vacuum, annihilated by every $\hc_j$.
Sites at the same distance from the centre of the chain form a Bell pair, yielding a rainbow-like structure.
Clearly, the rainbow state exhibits volume-law entanglement for any bipartition of the chain into two contiguous parts.

Here we define the {\itshape $n$-rainbow state} by restricting the rainbow structure to a finite
support of $2n$ sites (i.e., $n$ arcs), and by using it as a unit cell to cover a lattice of $L \rightarrow +\infty$ sites (see Fig.~\ref{fig:cartoon}). 
Thus, unlike the rainbow state, the $n$-rainbow states are macroscopically homogeneous. 

Denoting by $\leftarrow$ ordered products, in which the index runs over the specified set from left to right, we define
\begin{equation}
	\ket{R_n} \; \coloneq \; 
	\prod_{\ell=0}^{\leftarrow} \left[
	\prod_{j=1}^{\leftarrow n}
	\left( \hb^{[(-1)^{j+1}]}_{-\left(j - \frac{1}{2}\right)+2n\ell, \; \left(j - \frac{1}{2}\right)+2n\ell}\right)^\dagger
	\right] \ket{\Omega} ,
\end{equation}
where the outer product runs over the unit cells, labelled by $\ell$. In terms of spins, the
alternating signs of the fermionic pair-creation operators are compensated by the Jordan-Wigner
strings. Hence, every arc corresponds to the zero-magnetization triplet
$(\ket{\uparrow\downarrow}+\ket{\downarrow\uparrow})/\sqrt{2}$ (see Fig.~\ref{fig:cartoon}).
The $n$-rainbow state is a Gaussian state in terms of fermions. 
Indeed, it is straightforward to check that it can be rewritten as 
\begin{equation}
	\ket{R_n} \; = \; \prod_{j=1}^{\leftarrow L/2} \dcostr_j \ket{\Omega} ,
	\label{eq:gaussianstate}
\end{equation}
where $\{\hd_1, ..., \hd_{L}\}$ is a set of fermionic operators
obtained from the original fermions $\hat c_j$ via a unitary transformation.
Finally, the energy density of the $n$-rainbow state only receives
contributions from the $\hh_{j,j+1}$ acting on the two central sites of each unit cell, and it reads 
\begin{equation}
	\frac{E}{L} \; = \;
	\frac{J(2-\Delta)}{8n}
	\; = \; O\left( \frac{1}{n}\right) .
	\label{eq:rainbowenergy}
\end{equation}
In the limit $n\to\infty$ the energy density vanishes, i.e., it approaches that of
the infinite-temperature Gibbs ensemble. 

\section{\label{sec:TBA} THERMODYNAMIC BETHE ANSATZ FOR THE XXZ MODEL}

The XXZ chain is a paradigmatic example of a Bethe-ansatz integrable model \cite{TakahashiBook, Takahashi, Introduction, Bethe1931, Orbach, Walker, Integrabili2}.
In integrable systems, the presence of an extensive number of (quasi)local conserved charges allows one to describe the spectrum
and the eigenstates of the Hamiltonian in terms of stable {\itshape quasiparticles}, similarly to free models.
In its original formulation~\cite{Bethe1931}, the Bethe ansatz writes the wave function of an $N$-body
eigenstate as a sum over $N!$ plane waves with permuted quasimomenta and suitable amplitudes.
Periodic boundary conditions induce a quantization condition for quasimomenta, 
namely the Bethe equations, which are a set of $N$ coupled nonlinear equations.
It is convenient to parametrize the quasimomenta in terms of the so-called rapidities $\lambda$.
In some models, e.g., the Bose gas with repulsive $\delta$-function interactions~\cite{LiebLininger1, LiebLininger2}, all the rapidities are real.
In the XXZ model, complex rapidities are necessary to find a complete basis of eigenstates.

In the thermodynamic limit, the Bethe roots arrange into configurations of
$m$ elements with the same real part, displaced symmetrically along the imaginary direction with
spacing $\eta$ [cf.~\eqref{eq:eta}] (string hypothesis~\cite{TakahashiBook}):
\begin{equation}
	\left\{\lambda_j^{m,\alpha}\right\}\coloneq \left\{\left.\lambda^{m,\alpha}+(m+1-2j)\frac{i \eta}{2}\right|j=1,2,..., m \right\} ,
	\label{eq:strings}
\end{equation}
where $\lambda^{m,\alpha}\in \left[ -\frac{\pi}{2}, \frac{\pi}{2} \right)$ is the central rapidity of the $\alpha-$th string of length $m$ in the eigenstate.
Physically, an $m-$string can be interpreted as a bound state of $m$ quasiparticles~\cite{TakahashiBook, Introduction}.

In the thermodynamic limit, eigenstates can be characterized by their particle/hole  $m$-string 
root densities $\rho_m(\lambda),\rho^h_m(\lambda)$, with total density $\rho^{tot}_m(\lambda) = \rho_m(\lambda)+\rho^h_m(\lambda)$.
For instance,
$L \rho_m(\lambda)d\lambda$ is the number of $m$-strings with central rapidity in $[\lambda, \lambda+d\lambda]$.
Unlike for free particles, the central rapidities of the strings are not quantized independently, which results in a nontrivial $\rho^{tot}_m(\lambda)$.
In the thermodynamic limit, the Bethe equations within the string hypothesis become the so-called Bethe-Gaudin-Takahashi
equations~\cite{TakahashiBook}, which read 
	\begin{equation}
		\left\lbrace
		\begin{array}{l}
			\rho^{tot}_1(\lambda) \; = \;  s(\lambda) + s * \rho^h_2 , \vspace{2mm}\\
			\rho^{tot}_m(\lambda) \; = \;  s * \left(\rho^h_{m-1}+\rho^h_{m+1}\right)
			\quad
			\forall m>1 , \\
		\end{array}
		\right.
		\label{eq:BGT_densities}
	\end{equation}
where
\begin{equation}
	s(\lambda)=\frac{1}{2\pi} \sum_{k=-\infty}^{+\infty} \frac{e^{-i2\lambda k}}{\cosh(k \eta)}.
	\label{eq:a}
\end{equation}
The $*$ symbol denotes the convolution 
\begin{equation}
	(f*g)(\lambda) \coloneq \int_{-\pi/2}^{\pi/2} f(\lambda-\mu) g(\mu) \;\; d\mu .
\end{equation}
For $\Delta>1$ the string length is unbounded.
In the numerical solution of Eq.~\eqref{eq:BGT_densities} one has to
fix a maximum string length $m_{max}$ and truncate the system, e.g.,
by imposing $\rho_{m_{max}}^{tot} = s * \rho^h_{m_{max}-1}$.

The energy of an $m$-string and the energy density of a Bethe eigenstate are, respectively
\begin{equation}
    \begin{split}
        \epsilon_m(\lambda)
        &=\sum_{i=1}^{m}\epsilon_1(\lambda_i)\\
        &=-J\pi\sinh(\eta)\,a_m(\lambda),
    \end{split}
    \label{eq:string_energy}
\end{equation}
and
\begin{equation}
    \frac{E}{L}=\frac{J\Delta}{4}
    +\sum_{m=1}^{\infty}\int_{-\pi/2}^{\pi/2}
    \epsilon_m(\lambda)\rho_m(\lambda)\,d\lambda,
    \label{eq:energy}
\end{equation}
where
\begin{equation}
    a_m(\lambda)\coloneq\frac{1}{\pi}\frac{\sinh(m\eta)}{\cosh(m\eta)-\cos(2\lambda)}.
    \label{eq:bare_string_kernel}
\end{equation}
It is useful to introduce the Yang-Yang entropy~\cite{YangYang} as 
\begin{equation}
		 \frac{S_{YY}[\{\rho_m\}]}{L} \coloneq
		\sum_{m=1}^{+\infty} \int_{-\pi/2}^{\pi/2}
		s^{YY}_m(\lambda)\,d\lambda,
		\label{eq:yang_yang_entropy}
\end{equation}
with 
\begin{multline}
	\label{eq:sYY}
	s^{YY}_m(\lambda) \coloneq \rho_m^{tot}(\lambda) \log \left[ \rho_m^{tot}(\lambda)\right]\\-
		\rho_m(\lambda) \log \left[ \rho_m(\lambda) \right]
		- \rho_m^h(\lambda) \log \left[ \rho_m^h(\lambda) \right] .
\end{multline}
Each set of densities $\rho_m(\lambda),\rho^h_m(\lambda)$ identifies a thermodynamic
macrostate, which corresponds to $e^{S_{YY}}$ microscopic eigenstates.
The thermal root densities at temperature $\beta^{-1}$ are obtained in the
saddle-point approximation by minimizing the free-energy functional $F = E - \beta^{-1}S_{YY}$.
This yields the TBA equations~\cite{XXXthermo, XXZgaudin, Takahashi, TakahashiBook}, which in their partially decoupled form, with $\eta_m\coloneq\rho^h_m/\rho_m$, read
	\begin{equation}
		\left\lbrace
		\begin{array}{l}
			\log \left( \eta_1(\lambda) \right) \; = \; J \beta \pi \sinh(\eta) \; s(\lambda) + \left[ s * \log \left( \eta_2 + 1 \right) \right]_\lambda , \vspace{2mm}\\
			\log \left( \eta_m(\lambda) \right) \; = \;  s * \left(\log \left( \eta_{m-1} + 1 \right)+\log \left( \eta_{m+1} + 1 \right)\right) ,  \vspace{2mm} \\
			\lim_{m \rightarrow +\infty} \log \eta_m/m =0 . \\
		\end{array}
		\right.
		\label{eq:TBA}
	\end{equation}
In this work we only need $\beta=0$. In this case the driving term $J \beta \pi \sinh(\eta) \; s(\lambda)$ vanishes
and the functions $\eta_m(\lambda)$ are constant, which yields the analytical solution~\cite{Takahashi}
\begin{equation}
    \theta_m\coloneq\frac{\rho_m}{\rho_m^{tot}}=\frac{1}{(m+1)^2}.
    \label{eq:TBAanalytic}
\end{equation}

\subsection{\label{sec:SCD} Generalized Gibbs ensemble from string-charge duality}

The root densities describing the XXZ model GGE are fixed by the expectation values of a complete set of
quasilocal conserved charges $\hX_s(\lambda)$, with $s\in\frac{1}{2}\N^{>0}$ and $\lambda$ in the
physical strip $\mathcal{P}_\eta\coloneq\{\lambda\in\C\,|\,\imag(\lambda)\in(-\eta/2,\eta/2)\}$.
Their construction~\cite{FirstQuasilocal,Quasilocal,StringCharge}, and the method to
compute their expectation values on periodic states, are reviewed in
Appendix~\ref{sec:Algebraic}.
The label $s$ refers to the spin-$s$ auxiliary space used in their construction.
Here we only need the final result.
Below we use the notation
$f^{[\pm]}(\lambda)\coloneq f(\lambda\pm i\eta/2)$.

The completeness of the charges $\hX_s(\lambda)$ follows from the
so-called string-charge duality~\cite{StringCharge}: in the XXZ model there
exists a one-to-one mapping between the expectation values of the quasilocal charges
in the initial state and the root densities. 
For an initial state $|\Psi\rangle$, define the charge density
\begin{equation}
    \chi_s(\lambda)\coloneq
    \frac{1}{L}\bra{\Psi}\hX_s(\lambda)\ket{\Psi}.
    \label{eq:charge_expectation}
\end{equation}
The particle density follows from string-charge duality:
\begin{equation}
    \rho_{2s}(\lambda)=\square_s\chi_s(\lambda),
    \label{eq:String_Charge_1}
\end{equation}
while the hole and total densities are
\begin{equation}
    \rho^h_{2s}=a_{2s}-\chi_s^{[+]}-\chi_s^{[-]},
    \label{eq:string_charge_holes}
\end{equation}
and
\begin{equation}
    \rho^{tot}_{2s}=\rho_{2s}+\rho^h_{2s}
    =a_{2s}-\chi_{s-\frac12}-\chi_{s+\frac12},
    \label{eq:string_charge_total}
\end{equation}
where $\square_s f_{s} \coloneq f^{[+]}_{s} + f^{[-]}_{s} - f_{s-\frac{1}{2}} - f_{s+\frac{1}{2}}$ 
is the discrete d'Alembert operator,
which acts on functions that are analytic inside the physical strip $\mathcal{P}_{\eta}$.
We emphasize that, for every $n\in \N^{>0}$, the densities $\{\rho_n, \; \rho_n^h, \; \rho_n^{tot}\}$
are determined exactly by the first $n+1$ charges $\{\chi_s\}$.

As discussed in Appendix~\ref{sec:Algebraic}, all the local and quasilocal charges constructed with the quantum
inverse scattering method (QISM) are invariant under spin flip. Consequently, they cannot generate the
conserved magnetization $\sum_j\hS_j^z$. This is not a contradiction: eigenstates related by a global spin flip
have the same root densities, provided that the reference (vacuum) state is flipped as well.
Hence, $\sum_j\hS_j^z$ is not needed to determine the root densities via the string-charge duality.
To fully characterize the GGE, it is sufficient to supplement the
set $\{\rho_n(\lambda)\}$ with a binary variable $f=\pm$, which encodes
the sign of the magnetization, i.e. $\{\rho^f_n(\lambda)\}$ \cite{Nonballistic}.
In our case, the $n$-rainbow state has $\braket{\hS^z}=0$, so this correction has no effect.

Finally, we recall that there is a special class of initial states, the so-called
{\itshape integrable states}~\cite{IntegrableStates}, for which the GGE can be determined analytically.
All two-site product states, such as the N\'eel state $\ket{\Psi}= \ket{\uparrow \downarrow}^{L/2}$
and the $1$-rainbow state, are integrable~\cite{Pozsgay2018TwoSite}, whereas the $n$-rainbow states with $n > 1$ are not.
More precisely, integrable states are annihilated by all the local conserved quantities that are odd under space 
reflection~\cite{IntegrableStates}. It is straightforward to show that the $n$-rainbow states for $n>1$  
are not annihilated by the third local charge, whose density is $\hat{q}^{(3)}_j\propto i [\hh_{j,j+1}, \hh_{j+1, j+2}]$ \cite{LocalCharges}.
Notice that, since the $n$-rainbow states are invariant under spatial reflection, the expectation value of
this charge vanishes for any $n$. However, $\hat Q^{(3)}|R_n\rangle\neq0$ for $n>1$, while $\hat Q^{(3)}|R_1\rangle=0$. 
We stress that quenches from integrable initial states are expected to produce
entangled pairs of quasiparticles, which is the key assumption of the standard quasiparticle picture (see Sec.~\ref{sec:QP}).

\section{\label{sec:GHD} GENERALIZED HYDRODYNAMICS}

Let us now consider a bipartitioning protocol (see Fig.~\ref{fig:cartoon}), 
where the two halves of the system are prepared at $t=0$ in two different states. Because of the inhomogeneity, the system does not relax to a stationary state.
In lattice models with short-range interactions, local observables display a light-cone dynamics, owing to the existence of a state-dependent maximum velocity $v_{max}$ for the spreading of information~\cite{LiebRobinson, Lightconedynamics}.
In integrable models, this behaviour can be understood in terms of the semiclassical ballistic propagation of stable quasiparticles.
In Bethe-ansatz solvable models these carriers are the $m$-strings,
while in noninteracting systems their velocities are simply obtained from the dispersion relation.

This picture underlies GHD~\cite{HydroA,Bertini,Alba_2021} (see Ref.~\cite{Ruggiero2020QGHD} for its quantum extension), which introduces
a weaker notion of relaxation.
Since the excitations spread ballistically from the junction, the system relaxes to a
{\itshape local quasi-stationary state (LQSS)} that depends on the ray $\zeta=x/t$,
where $x$ is the distance from the junction. GHD holds in the {\itshape space-time} scaling limit  
\begin{equation}
		\lim_{sc}:\quad x,t \rightarrow +\infty
		\quad \text{with} \quad
		\zeta \; = \; x/t\quad\text{fixed.}
		\label{eq:scaling}
\end{equation}
In this limit, the expectation value of a local observable becomes a function of $\zeta$ only:
\begin{equation}
		\lim_{sc}\lim_{th}\langle\psi_0|\hat{O}_x(t)| \psi_0\rangle \; = \; \lim_{th}  \trace \left[\hat{O}_0 \, \hat{\rho}_{LQSS}(\zeta) \right] .
\end{equation}
The state $\hrho_{LQSS}(\zeta)$ is not a proper LSS, since it depends on time through $\zeta$.
However, it can still be characterized by a set of $\zeta$-dependent root densities.
The density profiles and the currents of the conserved (quasi)local charges are
\begin{equation}
    \begin{split}
        \frac{\braket{\hX^{(\alpha)}}(\zeta)}{L}
        ={}&\sum_{m=1}^{\infty}\int_{-\pi/2}^{\pi/2}d\lambda\\
        &\times X_m^{(\alpha)}(\lambda)\rho_{m,\zeta}(\lambda),
    \end{split}
    \label{eq:charges_currents}
\end{equation}
and
\begin{equation}
    \begin{split}
        J_{X^{(\alpha)}}(\zeta)
        \coloneq{}&\sum_{m=1}^{\infty}\int_{-\pi/2}^{\pi/2}d\lambda\\
        &\times X_m^{(\alpha)}(\lambda)v_m(\lambda)
        \rho_{m,\zeta}(\lambda),
    \end{split}
    \label{eq:charge_current}
\end{equation}
where $\set{\rho_{m,\zeta}(\lambda)}$ are the $\zeta$-dependent root densities 
characterizing the LQSS, $X^{(\alpha)}_m(\lambda)$ is the $\hX^{(\alpha)}$-charge 
content of an $m$-string, and $\alpha$ labels the charges.
Assuming that the set $\set{X_m^{(\alpha)}(\lambda)}$ is complete, the scaling limit~\eqref{eq:scaling} and charge conservation imply that the LQSSs satisfy a continuity
equation, with the LSSs of the two homogeneous quenches as boundary conditions for
$\zeta \rightarrow \pm \infty$~\cite{Bertini,HydroA}:
\begin{equation}
    \left[v_{m,\zeta}(\lambda)-\zeta\right]
    \partial_\zeta\theta_{m,\zeta}(\lambda)=0,
    \label{eq:GHD}
\end{equation}
where the filling function is
\begin{equation}
    \theta_{m,\zeta}(\lambda)\coloneq
    \frac{\rho_{m,\zeta}(\lambda)}{\rho^{tot}_{m,\zeta}(\lambda)}.
    \label{eq:ghd_filling}
\end{equation}
Eq.~\eqref{eq:GHD} is the central result of GHD.
Assuming that for all $\lambda, m$ the equation $\zeta = v_{m, \zeta}(\lambda)$ has a unique solution (no exceptions are known), the solution of Eq.~\eqref{eq:GHD} reads 
\begin{multline}
	\theta_{m,\zeta}(\lambda) \; = \;
	\theta^{L}_{m,\zeta}(\lambda) \cdot
	\Theta\left( v_{m,\zeta}(\lambda) - \zeta \right)
	\; + \\ + \;
	\theta^{R}_{m,\zeta}(\lambda) \cdot
	\Theta\left( \zeta - v_{m,\zeta}(\lambda) \right) .
	\label{eq:sol_interacting}
\end{multline}
Here $\Theta$ is the Heaviside step function, while $\theta^{L/R}$ denote the filling functions of the left and right states.
Eq.~\eqref{eq:sol_interacting} is only an implicit solution,
since the dressed velocities $\{v_{m, \zeta}(\lambda)\}$ depend on the LQSS.
In the partially decoupled form, the equations for the dressed velocities read~\cite{Lightconedynamics}
\begin{equation}
	\left\lbrace
	\begin{array}{l}
		v_{1, \zeta} \rho^{tot}_{1, \zeta}
		\; = \;
		- \frac{1}{2} J \sinh(\eta) \; \partial_\lambda s
		\; + \;
		s * \left[v_{2, \zeta} \rho^h_{2, \zeta} \right] , \vspace{2mm} \\
		v_{n, \zeta} \rho^{tot}_{n, \zeta}
		\; = \;
		s * \left[v_{n-1, \zeta} \rho^h_{n-1, \zeta}+v_{n+1, \zeta} \rho^h_{n+1, \zeta} \right] . \\
	\end{array}
	\right.
	\label{eq:dressed}
\end{equation}
The LQSS root densities and the dressed velocities
have to be determined simultaneously, by solving Eqs.~\eqref{eq:BGT_densities}, \eqref{eq:sol_interacting}, and \eqref{eq:dressed} iteratively and self-consistently.

Finally, we recall that, although it is not a quasilocal conserved charge, the Yang-Yang entropy~\eqref{eq:sYY} satisfies the continuity equation
\begin{equation}
    \zeta\partial_\zeta s_{m,\zeta}(\lambda)
    =\partial_\zeta\left[v_{m,\zeta}(\lambda)s_{m,\zeta}(\lambda)\right],
    \label{eq:continuityent}
\end{equation}
with boundary conditions $s_{m,\pm\infty}(\lambda)=s_m^{R/L}(\lambda)$.
Hence, the Yang-Yang entropy is conserved, and it is transported ballistically by the quasiparticles.

\section{\label{sec:QP} STANDARD QUASIPARTICLE PICTURE}

Consider a homogeneous quench from a pure state $\ket{\psi}$, 
with density matrix
$\hrho=\ket{\psi} \bra{\psi}$, and a bipartition $\Abb \; | \; \overline{\Abb}$, where $\Abb$ is a segment of length $\ell$.
The {\itshape entanglement entropy} is the von Neumann entropy of either of the two reduced density matrices:
\begin{equation}
	S_\Abb \; \coloneq \; -\trace\,\left[\hrho_\Abb\log\hrho_\Abb \right]
	\qquad
	\left[ \, S_\Abb=S_{\,\overline{\Abb}} \,\right]
	\label{eq:VonNeumann}
\end{equation}
After a quench, $S_\Abb(t)$ typically grows linearly in time and eventually saturates to a value proportional to $\ell$~\cite{QuasiPicture}.
Because of Eq.~\eqref{eq:relaxation}, this asymptotic value is the von Neumann entropy of the reduced density matrix of $\hrho_s$.
This led to the proposal that the thermodynamic
entropy density of the statistical ensemble (Gibbs or generalized Gibbs) coincides with
the stationary entanglement entropy density of a large subsystem $\Abb$ of an infinite system prepared in a pure state~\cite{AlbaCalabrese2017, EntThermo3, EntThermo4}:
\begin{equation}
	\frac{S_s^{(th)}}{L} \; = \; \frac{S_{\Abb}^\infty}{\ell} .
	\label{eq:thermoentanglement}
\end{equation}
To date, this equivalence has been proven analytically only for noninteracting models \cite{QuasiProof, EntThermo1, EntThermo2}.

The {\itshape quasiparticle picture}~\cite{QuasiPicture, AlbaCalabrese2017, Alba_2018, Calabrese2020LectNotes} provides a semiclassical description of entanglement spreading after a quench in the space-time scaling limit~\cite{SpeciesCollura}.
Its central assumption is that the quench creates pairs of entangled quasiparticles, which propagate ballistically.
In free models, pairs are formed by quasiparticles with opposite momenta and velocities.
In interacting integrable systems, pairs are formed by two $m$-strings with opposite rapidities, which propagate with the dressed velocities $v_m(\pm \lambda)$ determined by the stationary state.
The entanglement between $\Abb$ and $\overline{\Abb}$ is obtained by counting the pairs that are shared between the two subsystems at time $t$. To ensure Eq.~\eqref{eq:thermoentanglement}, each pair is weighted by its Yang-Yang entropy content~\eqref{eq:sYY}:
\begin{subequations}
	\begin{equation}
		S_\Abb (t)\approx
		\sum_{m=1}^{+\infty}
		\int d\lambda
		\; \Omega(\Abb, m, \lambda, t)  s_m(\lambda) ,
		\label{eq:quasiint}
	\end{equation}
	where $\Omega(\Abb, m, \lambda, t)$ is the measure of the set of points $x$ such that a pair of $m$-strings created in $x$ with rapidities $\pm \lambda$ has, at time $t$, the $\lambda$-particle in $\Abb$ and the $(-\lambda)$-particle in $\overline{\Abb}$.
	In the particular case of free models, this expression simplifies to
	\begin{equation}
		S_\Abb (t)\approx
		\int_0^{2\pi} \frac{dk}{2\pi}
		\; \Omega(\Abb, v(k), t)  s(k) ,
		\label{eq:entcontentfree}
	\end{equation}
\end{subequations}
where the entropy content is
\begin{equation}
	\begin{split}
		s(k) ={}& -\braket{n_k} \log \braket{n_k}
		\\
		& -(1-\braket{n_k}) \log (1-\braket{n_k}) ,
	\end{split}
	\label{eq:entropycontentfree}
\end{equation}
and the geometric factor is
\begin{equation}
	\Omega(\Abb, v, t) \coloneq \left[
	x' \in \Abb \; \mbox{s.t.} \;
	(x' - 2vt) \in \overline{\Abb} \,
	\right] .
	\label{eq:kinematicfactorfree}
\end{equation}
In particular, if $\Abb= (-\infty, 0] $ with open boundary conditions, then $\Omega((-\infty, 0], v, t) = t|v|$.

The validity of the pair-structure assumption depends on both the initial state and the model.
In free models, it can be checked directly from the form factor.
In interacting integrable models, the origin of entangled pairs with opposite rapidities is more elusive.
Initial states annihilated by all the parity-odd local charges typically have nonzero overlap only with Bethe eigenstates formed by pairs of opposite rapidities.
This fails, for example, in the XX chain~\cite{NeelOverlaps}, but holds in the XXZ chain at generic $\Delta$.
States with this property are the {\itshape integrable initial states}~\cite{IntegrableStates} introduced in Sec.~\ref{sec:SCD}.
Although this is not a rigorous proof, quenches from integrable initial states are expected to be described by the pair-structure quasiparticle picture.
Conversely, this does not exclude that the pair picture also applies to some states outside this class.
For initial states that produce entangled multiplets, a generalized quasiparticle picture has been
developed for free systems~\cite{SpeciesFirst,SpeciesBertini,SpeciesCollura,Caceffo}, where the
correlations between the members of a multiplet can be obtained from the form factor of the initial
state (see Sec.~\ref{sec:Entanglement}). An analogous description is not available for interacting
integrable systems. The $n$-rainbow states, which produce multiplets in the XX chain and are not integrable
in the XXZ chain, provide a controlled setting to address this problem (see Secs.~\ref{sec:homoent}
and~\ref{sec:entdynamics}).
	
\section{\label{sec:exact} Thermalization and multiplet spreading in the XX chain}

In this section we show that, in the XX chain, quenches from the $n$-rainbow state lead to
thermal behaviour in the limit $n\to\infty$, and that at the same time the dynamics produces entangled multiplets formed by $2n$ quasiparticles. We then
derive a generalized quasiparticle picture for the entanglement dynamics after quenches from both
homogeneous and inhomogeneous initial states. 

\subsection{Entangled multiplets in the XX chain}
	
The XX chain [see Eq.~\eqref{eq:fermionicXXZ}] maps onto a model
of free spinless fermions with particle-number-dependent
boundary conditions (periodic/antiperiodic for odd/even $N_{tot}$).
Here we only consider half-filled states, i.e., $N_{tot}=L/2$.
In Fourier space one has 
\begin{equation}
	 \hc_j \; \coloneq \; \frac{1}{\sqrt{L}} \sum_{k \in BZ_0^L} e^{ijk} \hd_k ,
 \end{equation}
 where the sum over the Brillouin zone is defined as 
 \begin{equation}
	 \sum_{k \in BZ_0^L} \coloneq \sum_{k\in \set{k_0, ..., k_{L-1}}} ,
	\qquad
	k_m \coloneq k_0 + \frac{2 \pi}{L}m ,
	\label{eq:Fourier}
\end{equation}
with $k_0=0$ or $\pi/L$ for periodic or antiperiodic boundary conditions, respectively.
For $L \rightarrow +\infty$, $k_0 \rightarrow 0$
and the set $\{k_m\}$ becomes dense in $[0,2\pi]$ in both cases.
Hence, our results in the thermodynamic limit do not depend on the boundary conditions.
	
The Hamiltonian is diagonal in Fourier space:
	\begin{equation}
    \hH_{XX}=\sum_{k\in BZ_0^L}\hn_k\,\epsilon_k,
    \label{eq:XX_diagonal_hamiltonian}
\end{equation}
where $\hn_k=\dcostr_k\hd_k$ is the mode-occupation operator and
\begin{equation}
    \epsilon_k=\cos(k)
    \label{eq:dispersionXX}
\end{equation}
is the single-particle dispersion.
Although the mode occupations $\{\hn_k\}$ are nonlocal operators, they provide a complete
set of conserved charges to construct the GGE in the XX chain~\cite{modeoccupations1, modeoccupations2, RelaxationPedagogical, Alba_2021}.

\subsection{Correlation matrix, form factor and dynamics}
	
The spatial correlation matrix is defined as
\begin{equation}
	C_{ij}^{(n)} \; = \; \braket{R_n|\ccostr_i \hc_j | R_n} .
\end{equation}
Since the $n$-rainbow states are Gaussian,
$C_{ij}^{(n)}$ determines all the multipoint correlation functions via Wick's theorem.
At \mbox{$t=0$} there are only pairwise correlations between
sites. We denote by $\sigma(i)$ the site paired with $i$, i.e.,
connected to $i$ by an arc (see Fig.~\ref{fig:cartoon}).
Defining $\Delta_i$ through $|\sigma(i)-i|=2 \Delta_i-1$, the initial correlation matrix of the $n$-rainbow state reads
\begin{equation}
	C_{ij}^{(n)} \; = \; \frac{1}{2}\left(\delta_{ji} + (-1)^{\Delta_i+1}\delta_{j \sigma(i)} \right).
	\label{eq:corr}
\end{equation}
To study the dynamics, it is convenient to compute the form factor
\begin{equation}
    \begin{split}
        F_{kk'}^{(n)}
        &\coloneq\braket{R_n|\dcostr_k\hd_{k'}|R_n}\\
        &=\frac{1}{L}\sum_{j,w\in\Lambda}
        e^{ikj-ik'w}C_{jw}^{(n)},
    \end{split}
    \label{eq:formfactor_definition}
\end{equation}
and separate its diagonal and off-diagonal contributions as
\begin{equation}
    F_{kk'}^{(n)}=\frac{\delta_{kk'}}{2}+\widetilde F_{kk'}^{(n)},
    \label{eq:formfactor_decomposition}
\end{equation}
where $\Lambda$ is the set of lattice sites. The calculation of the off-diagonal contribution
$\widetilde F^{(n)}_{kk'}$ is reported in Appendix~\ref{app:XXdetails}. One obtains
\begin{widetext}
	\begin{equation}
		F_{kk'}^{(n)}\; = \;
		\frac{\delta_{kk'}}{2} + 
		\frac{1}{2n}
		\left\lbrace
		\frac{\sin^2\left[\left(\frac{k+k'+\pi}{2}\right)n\right]}
		{\cos\left[\frac{k+k'}{2}\right]}
		\right\rbrace
		\left( \sum_{\alpha=0}^{2n-1} \delta_{|k-k'|,\; \alpha \pi/n}\right)
		\; = \;
		\frac{\delta_{kk'}}{2} + O\left(\frac{1}{n} \right) .
		\label{eq:F-final}
	\end{equation}
\end{widetext}
The off-diagonal term is nonzero only if $k-k'$ is a multiple of $\pi/n$. As a function of $q\coloneq k+k'$,
it is $O(1/n)$, except in a region of width $O(1/n)$ around $q=\pi$ (mod $2\pi$), where the denominator vanishes.
There it exhibits two peaks of opposite sign, whose height remains finite as $n \rightarrow +\infty$, while their width vanishes as $1/n$.
Hence, the off-diagonal term converges to zero both pointwise and in the sense of distributions as $n \rightarrow +\infty$.

The time-evolved form factor and correlation matrix are readily obtained in the Heisenberg picture:
\begin{equation}
    \begin{split}
        F_{kk'}^{(n)}(t)
        &\coloneq\braket{R_n|\dcostr_k(t)\hd_{k'}(t)|R_n}\\
        &=e^{i(\epsilon_k-\epsilon_{k'})t}F_{kk'}^{(n)}.
    \end{split}
    \label{eq:formfactor_dynamics}
\end{equation}
The real-space correlation matrix is then
\begin{equation}
    \begin{split}
        C_{jw}^{(n)}(t)
        &\coloneq\braket{R_n|\ccostr_j(t)\hc_w(t)|R_n}\\
        &=\frac{1}{L}\sum_{k,k'\in BZ_0^L}e^{-i(kj-k'w)}\\
        &\qquad\times e^{i(\epsilon_k-\epsilon_{k'})t}F_{kk'}^{(n)}.
    \end{split}
    \label{eq:corr-matrix}
\end{equation}
Since the off-diagonal contributions dephase, the GGE is determined by the mode occupations
$\braket{\hn_k}=F^{(n)}_{kk}$. For $n\to\infty$ these distributionally approach $1/2$, which is the infinite-temperature
value. Upon increasing $n$, local observables approach their infinite-temperature values, with $O(1/n)$ deviations. 
\begin{figure}
	\centering
	\includegraphics[width=\linewidth]{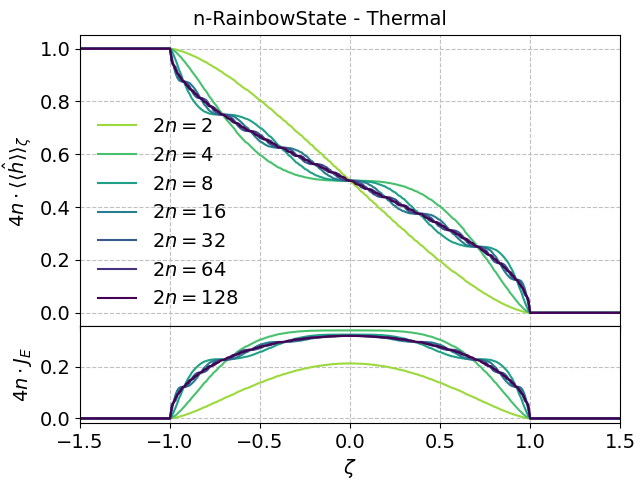}
	\caption{Bipartitioning protocol with the $n$-rainbow state (unit-cell size $2n$ from $2$ to $128$) and the $T \rightarrow +\infty$ thermal state in the XX chain. Top: GHD prediction for the local energy density $\bbraket{\hh}(\zeta)$ in the scaling limit. Bottom: corresponding energy current. Both profiles are divided by $E_n/L = 1/(4n)$, which leads to a collapse of the profiles as $n \rightarrow +\infty$.}
	\label{fig:energyXX}
\end{figure}
In Eq.~\eqref{eq:F-final} we considered a single quasiparticle species with $k \in (0,2 \pi]$.
As $L\rightarrow +\infty$ the form factor becomes an infinite-dimensional matrix. However, because of the $2n$-site unit cell, i.e., \mbox{$\ket{R_n}=\ket{\psi}^{\otimes L/(2n)}$}, only particles with $|k-k'|=\alpha \pi/n$ and $\alpha \in\{0, ..., 2n-1\}$ can be entangled.
Hence, only a vanishing fraction of the entries $F_{kk'}$ is nonzero.
To exploit this property, it is convenient to introduce $2n$
quasiparticle species~\cite{Caceffo, SpeciesBertini, SpeciesCollura, SpeciesFirst} (labelled by capital letters $A \in\{0, ..., 2n-1\}$), each with a reduced Brillouin zone of width $\pi/n$.
Each species is described by a momentum in the reduced Brillouin zone
$BZ_n\coloneq BZ_0^L\cap(0,\pi/n]$. For $k\in BZ_n$, define the corresponding
momentum in the original Brillouin zone as
\begin{equation}
    k_A\coloneq k+\frac{\pi}{n}A.
    \label{eq:species_momentum}
\end{equation}
The annihilation operator of species $A$ is
\begin{equation}
    \heta_{A,k}\coloneq\hd_{k_A},
    \label{eq:species_operator}
\end{equation}
and its dispersion is
\begin{equation}
    \epsilon_A(k)\coloneq\cos\left(k+\frac{\pi}{n}A\right).
    \label{eq:species_dispersion}
\end{equation}
In the thermodynamic limit, the momenta become dense in the Brillouin zone, and
the Hamiltonian can be written as
\begin{equation}
\hH_{XX} \; = \;
L \int_0^{\pi/n} \frac{dk}{2\pi} \; \sum_{A=0}^{2n-1} \epsilon_A(k) \; \etacostr_{A,k} \heta_{A,k}
\end{equation}
Similarly, the form factor becomes a $2n \times 2n$ matrix, 
\begin{equation}
    \begin{split}
        F_{AB}^{(n)}(k)
        &\coloneq\braket{R_n|\etacostr_{A,k}\heta_{B,k}|R_n}\\
        &=\frac{\delta_{AB}}{2}+\widetilde F_{\frac{A+B}{2}}^{(n)}(k),
    \end{split}
    \label{eq:formfactorspecies}
\end{equation}
where
\begin{equation}
    \begin{split}
        \widetilde F_X^{(n)}(k)
        &\coloneq\frac{\sin^2\left[n\left(k+\frac{\pi}{2}\right)+\pi X\right]}
        {2n\cos\left[k+\frac{\pi}{n}X\right]}\\
        &=O\left(\frac{1}{n}\right).
    \end{split}
    \label{eq:species_formfactor_amplitude}
\end{equation}
Again, the number of species grows linearly with $n$, while the correlations between different species vanish as $n \rightarrow +\infty$. 

A striking consequence is that transport is suppressed after
the bipartitioning protocol in Fig.~\ref{fig:cartoon} (bottom), in which
the left part is prepared in the $n$-rainbow state and the right one in the
infinite-temperature thermal state.
Indeed, $\braket{n_k}^L$ is given by Eq.~\eqref{eq:F-final},
while $\braket{n_k}^R = \frac{1}{2}$.
Using the dispersion relation~\eqref{eq:dispersionXX} and the noninteracting
limit of Eq.~\eqref{eq:sol_interacting}, the occupation profile is
\begin{equation}
    \begin{split}
        \braket{n_k}_\zeta
        ={}&\braket{n_k}^L\,\Theta\left(-\sin(k)-\zeta\right)\\
        &+\braket{n_k}^R\,\Theta\left(\zeta+\sin(k)\right).
    \end{split}
    \label{eq:XX_occupation_profile}
\end{equation}
The corresponding energy density is
\begin{equation}
    \bbraket{\hh}(\zeta)=\int_0^{2\pi}\frac{dk}{2\pi}\,
    \epsilon_k\braket{n_k}_\zeta,
    \label{eq:XX_energy_profile}
\end{equation}
and the energy current is
\begin{equation}
    J_E(\zeta)=-\int_0^{2\pi}\frac{dk}{2\pi}\,
    \sin(k)\epsilon_k\braket{n_k}_\zeta.
    \label{eq:XX_energy_current}
\end{equation}
The notation $\bbraket{\hh}(\zeta)$ for the local energy density is defined more formally in Sec.~\ref{sec:bipartitioning}.
	
In Fig.~\ref{fig:energyXX} we display the energy profile
$\bbraket{\hh}(\zeta)$ and the associated energy current $J_E$ versus
$\zeta$. We divide both quantities by the energy density $E_n/L=1/(4n)$ of the $n$-rainbow state,
since both vanish in the limit $n\to\infty$.
Clearly, transport is suppressed in the limit $n\to\infty$.
Interestingly, at intermediate $n$ the energy profile develops a ``tilted washboard'' structure
as a function of $\zeta$, which reflects the periodicity of the $n$-rainbow state. 

\subsection{\label{sec:Entanglement} Entanglement spreading}
	
As discussed in Sec.~\ref{sec:QP}, the standard quasiparticle picture assumes that excitations are produced only in entangled pairs. The quasiparticle picture in the presence of entangled multiplets has been explored in Refs.~\cite{SpeciesFirst, SpeciesBertini, SpeciesCollura, SpeciesBipartitioning, Caceffo} (see also Ref.~\cite{Nested} for nested spin chains). As stressed above, the quench from the $n$-rainbow state produces $2n$ entangled species of quasiparticles. The group velocity of a quasiparticle of species $A$ is
\begin{equation}
	v_A(k) \; \coloneq \; \frac{d \epsilon_A(k)}{dk} \;=\; -\sin\left( k + A \frac{\pi}{n}\right).
	\label{eq:speciesvelocities}
\end{equation}
Clearly, one has 
\begin{equation}
	v_A(k)=-v_{A+n}(k),\quad A\in\{0,\dots,n-1\}, 
\end{equation}
i.e., the multiplet spreads symmetrically around the point where it is created.
The quench produces a uniform density of multiplets of $2n$ entangled quasiparticles
for each value of $k$.
The multiplet quasiparticle description is valid in the scaling regime
$2n \ll \ell \ll L \rightarrow +\infty$ and $t \rightarrow + \infty$, with
$t/\ell$ fixed, where $\ell$ is the size of $\Abb$. 

A key ingredient of the quasiparticle picture is the kinematic
factor $\Omega(\Abb,Q,k,t)$. Let $Q$ denote a configuration of the
multiplet in which a subset $\{j_1,j_2,\dots,j_q\}$, with $q\in \set{0,...,2n}$, of its
quasiparticles is in $\Abb$. Then
$\Omega(\Abb,Q,k,t)$ is the measure of the set of positions $x$ such that at time $t$ a
multiplet with momentum $k$ created in $x$ is in the configuration $Q$.
The quasiparticle prediction for the entanglement entropy is 
\begin{equation}
	S_\Abb (t) = \int_0^{\pi/n} \frac{dk}{2\pi} \sum_Q \; \Omega(\Abb, Q, k, t) \; s(k, Q) ,
	\label{eq:quasimulti}
\end{equation}
where $s(k,Q)$ is the entanglement contribution of the multiplet in configuration $Q$. 
\begin{figure}[t]
	\centering
	\begin{tikzpicture}[scale=0.41,baseline={([yshift=-0.9ex]current bounding box.center)}]
		\withoutrectangle{12}{2}{4}{2}{colO}
		\gridAxis{0}{0}{0}{5}
		\gridAxis{0}{0}{19.6}{0}
		\draw [dashed,colLines] ({0},{3.7}) -- ({19.6},{3.7});
		\foreach \x in {7, 12.8}{
			\gridArrow{\x}{0}{\x-6}{4}
			\gridArrow{\x}{0}{\x+6}{4}
			\gridArrow{\x}{0}{\x-2}{4}
			\gridArrow{\x}{0}{\x+2}{4}
			\node at ({\x-5.8},4.5) {\scalebox{0.85}{$v_1(k)$}};
			\node at ({\x+5.8},4.5) {\scalebox{0.85}{$v_3(k)$}};
			\node at ({\x-2},4.5) {\scalebox{0.85}{$v_0(k)$}};
			\node at ({\x+2},4.5) {\scalebox{0.85}{$v_2(k)$}};
		}
		\node at (-0.6,4.7) {\scalebox{1}{$t$}};
		\node at (-0.5,3.7) {\scalebox{1}{$t^*$}};
		\node at (19.3,-0.6) {\scalebox{1}{$x$}};
		\node at (1.6,0.5) {\scalebox{1}{$\overline{\Abb}$}};
		\node at (10,0.5) {\scalebox{1}{$\Abb$}};
		\node at (17.6,0.5) {\scalebox{1}{$\overline{\Abb}$}};
		\node at (7,-0.6) {\scalebox{0.85}{$Q(t^*)=\{2,3\}$}};
		\node at (13,-0.6) {\scalebox{0.85}{$Q(t^*)=\{0,2\}$}};
	\end{tikzpicture}
	\caption{Entanglement spreading via multiplet propagation after the
		quench from the $n$-rainbow state, for $n=2$. The quench produces multiplets of $2n=4$ entangled quasiparticles.
		We show two possible configurations at time $t^*$, $Q(t^*)=\{2,3\}$ and $Q(t^*)=\{0,2\}$,
		with quasiparticles $2,3$ and $0,2$ in $\Abb$, respectively. 
	\label{fig:multiplet}}
\end{figure}
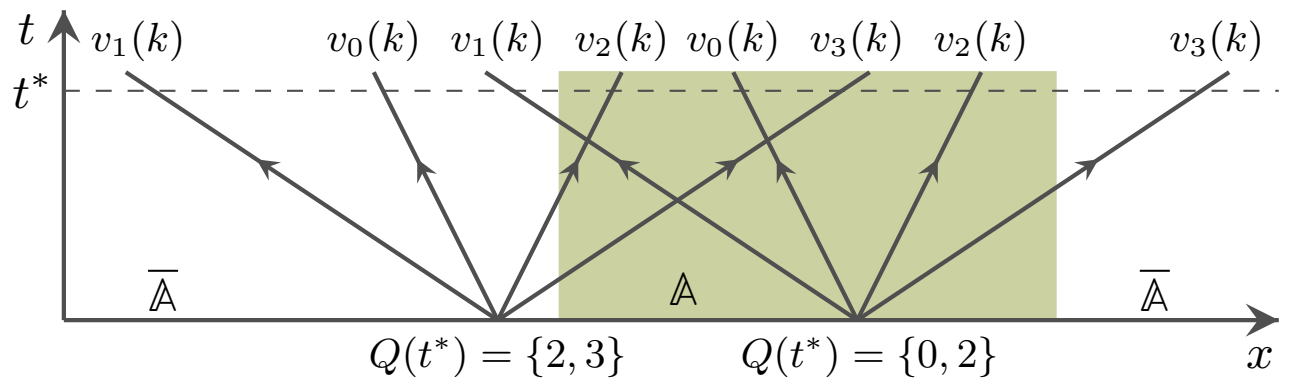
The entanglement content $s(k, Q)$ of a shared multiplet is determined as
follows~\cite{SpeciesBertini, SpeciesFirst}.
We define the vector of creation and annihilation operators of a multiplet with momentum $k$,
\begin{equation}
	\hat{\Gamma}^\dagger (k) \; \coloneq \;
	\left[
	\begin{array}{llllll}
		\heta^\dagger_{0, k} &
		... &
		\heta^\dagger_{2n-1, k} &
		\heta_{0, k} &
		... &
		\heta_{2n-1, k} \\
	\end{array}
	\right] ,
\end{equation}
and the corresponding correlation matrix
	\begin{equation}
		\begin{split}
			\mathcal{C}^{(n)}(k)
			\; \coloneq & \;
			\braket{R_n | \hat{\Gamma}^\dagger(k) \hat{\Gamma}(k) |R_n}
			\; = \\ = & \;
			\left(
			\begin{array}{ll}
				F^{(n)}(k) & 0 \\
				0 &
				\Id_{2n} - F^{(n)}(k)^T \\
			\end{array}
			\right) ,
		\end{split}
	\end{equation}
	whose blocks are $2n \times 2n $ matrices, with $F^{(n)}(k)$
	defined in Eq.~\eqref{eq:formfactorspecies}.
Here $^T$ denotes the transpose, and we used $\{\heta_{A,k}, \heta^\dagger_{B,k}\} = \delta_{A,B}$.
Following Refs.~\cite{entcontent1, entcontent2, entcontent3, SpeciesFirst}, we conjecture that 
\begin{equation}
    \begin{split}
        s(k,Q)={}&-\trace\left[
        \mathcal C_Q^{(n)}\log\mathcal C_Q^{(n)}\right]\\
        ={}&-\trace\Big\{F_Q^{(n)}(k)\log F_Q^{(n)}(k)\\
        &+\left[\Id-F_Q^{(n)}(k)^T\right]
        \log\left[\Id-F_Q^{(n)}(k)^T\right]\Big\},
    \end{split}
    \label{eq:entcontentmulti}
\end{equation}
where $\mathcal{C}^{(n)}_Q$ (and similarly $F^{(n)}_Q$) is the restriction of $\mathcal{C}^{(n)}$
to the species in $\Abb$, i.e., to the rows and columns
corresponding to the species in $Q$.
Notice that $s(k,Q)$ depends on the off-diagonal elements of
$\mathcal{C}$~\cite{SpeciesBertini, Caceffo}.
As $t \rightarrow +\infty$, for any finite $\ell$,
at most one member of each multiplet can be in $\Abb$. This implies that the steady-state
entanglement entropy depends only on the diagonal elements of $\mathcal{C}$, and
it coincides with the thermodynamic entropy of the GGE.
Indeed, if $\Delta v_{min}(k)$ is the minimum difference between the group velocities of
the quasiparticles of the multiplet with momentum $k$, the configurations $Q$ in
which more than one member of the multiplet is in $\Abb$
can have nonzero kinematic factor $\Omega(\Abb, Q, k, t)$
only for $t\lesssim \ell/\Delta v_{min}(k)$, i.e., at finite times.

\begin{figure}[t]
	\includegraphics[width=0.430\textwidth]{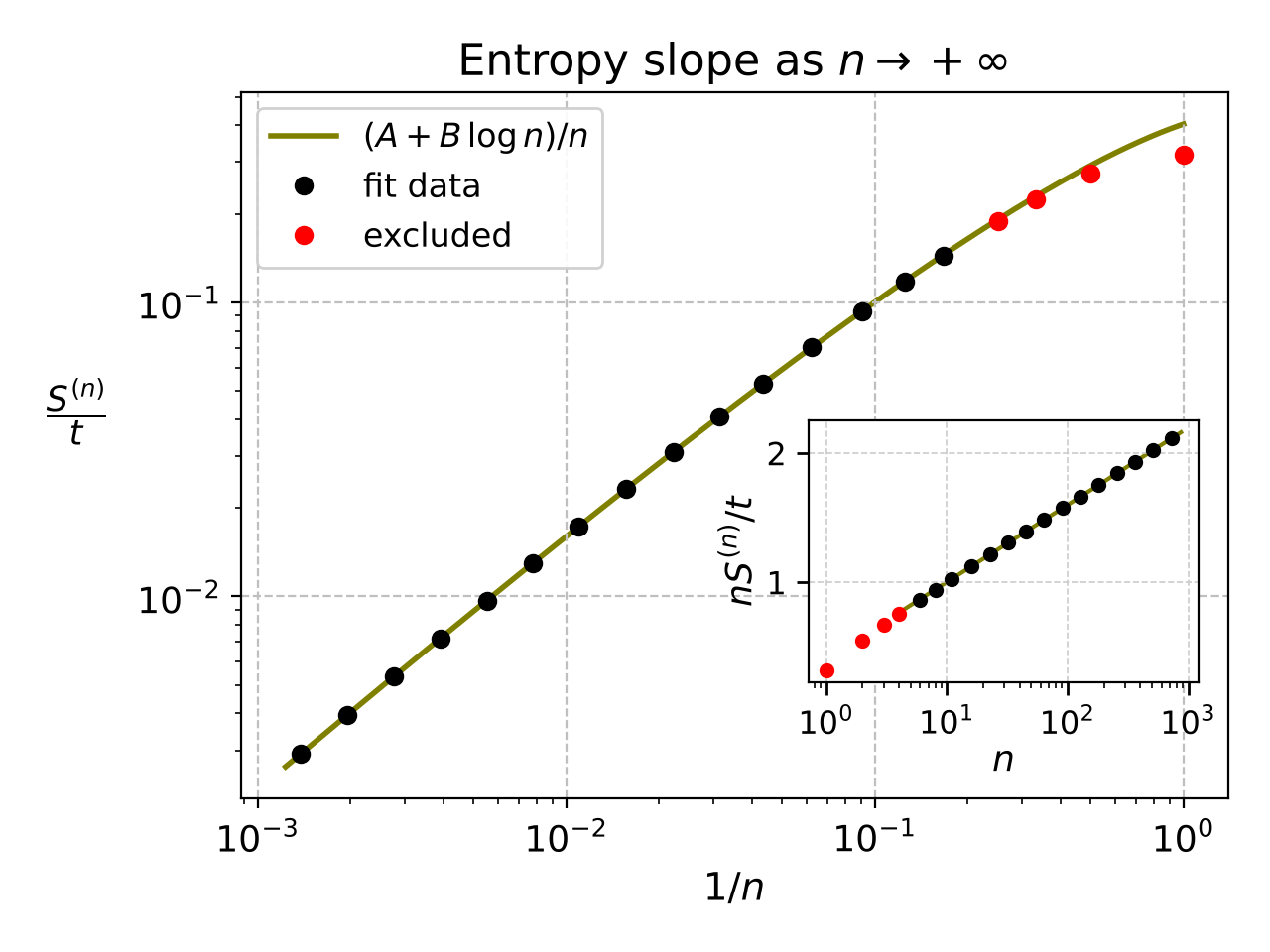}
\caption{ Slope $S^{(n)}(t)/t$ of the half-chain entanglement entropy after the homogeneous
quench from $\ket{R_n}$ in the XX chain (equivalently, twice the slope after the quench from
$\ket{\Omega}\otimes\ket{R_n}$), obtained from the multiplet quasiparticle picture for $n$ up
to $724$ and plotted versus $1/n$. The solid line is the fit to $(A+B\log n)/n$,
Eq.~\eqref{eq:slopefit}, with $A\approx 0.4$ and $B\approx 0.26$. Red symbols denote the
data excluded from the fit. Inset: $nS^{(n)}/t$. 
}	
\label{fig:entropyXXslope}
\end{figure}

We now show that in the space-time scaling limit the slope of $S^{(n)}(t)$ vanishes as
$n\to\infty$, and we determine its scaling with $n$. First, as discussed in Appendix~\ref{app:XXdetails}, we observe that 
we can label the kinematically permitted 
partitions $Q$ as $(q,k)$ with $q+1 \in \set{1, ..., n}$ denoting the number of particles in the left subsystem. 
The crucial observation concerns the structure of the off-diagonal correlations in
Eq.~\eqref{eq:F-final}. As stressed already, in the limit $n\to\infty$ Eq.~\eqref{eq:F-final} 
gives $\mathcal{O}(1)$ correlation only between quasiparticles with $k+k'\simeq \pi$. 
Since $v(\pi-k)=v(k)$, each quasiparticle is mostly
correlated with quasiparticles that move with almost the same velocity. This means that 
in the limit $n\to\infty$ the entangled quasiparticles carrying most of the entanglement 
cross the boundary almost at the same time, and hence do not contribute to the von Neumann entropy. 
This is the opposite of the standard pair structure, which
is recovered only for $n=1$, where the partner of $k$ is $k+\pi$, with velocity $-v(k)$.
Thus, for $n\to\infty$, $F^{(n)}_{Q(q,k)}$ is close to a projector. Only a few
of its eigenvalues are not close to $0$ or $1$.
They originate from the power-law decaying tails of the correlations away from the point $k+k'=\pi$. 
This implies that the contribution of each additional quasiparticle decreases with $q$. We verified that
$s^{(n)}(k,q+1)-s^{(n)}(k,q)\sim 1/q$. By summing over $q\in [0,n]$, we obtain   
\begin{equation}
	\int_{BZ_n}\frac{dk}{2\pi} 
	\sum_q\Omega(q,k,t)\,s^{(n)}(k,q)/t\simeq 
	\frac{A+B\ln n}{n}, 
	\label{eq:slopefit}
\end{equation}
where $A,B$ are constants, and the factor $1/n$ originates from $\Omega/t\sim 1/n$ at large $n$.
Eq.~\eqref{eq:slopefit} is supported by Fig.~\ref{fig:entropyXXslope}, where we show numerical data for
the slope $S^{(n)}/t$ of the entanglement entropy after the quench from the $n$-rainbow state $|R_n\rangle$.
The slope is obtained by fitting the linear growth of the entanglement entropy at asymptotically long times. 
Fig.~\ref{fig:entropyXXslope} shows that the slope vanishes as $\ln n/n$ for large $n$.
The solid line is a fit to Eq.~\eqref{eq:slopefit}, with $A,B$ fitting parameters.
The inset shows $n S^{(n)}/t$ on a logarithmic scale on the $x$-axis. The linear behaviour at large $n$
confirms the presence of the logarithmic term in Eq.~\eqref{eq:slopefit}.
The vanishing as $\ln n/n$ of the slope of the entanglement growth does not contradict the fact
that the entanglement entropy of a finite subsystem of length $\ell$ eventually saturates to
$\ell\, S_{YY}/L$, which tends to $\ell\ln2$ as $n\to\infty$. Saturation requires that at most one
member of each multiplet is in $\Abb$, which happens only for $t\gtrsim\ell/\Delta v_{min}(k)$, with
$\Delta v_{min}(k)$ the smallest velocity difference within the multiplet. Since the velocities
become dense as $n$ grows, and partner velocities are almost degenerate (exactly degenerate for
$k=\pi/(2n)$), this time grows at least linearly with $n$. Notice also that the limits $n\to\infty$
and $t\to\infty$ do not commute. For $t\lesssim n$, before the multiplet structure is resolved, the
entanglement grows at the rate of the pair formula, because on scales smaller than $n$ the
$n$-rainbow state is close to the infinite-temperature state (see Sec.~\ref{sec:homoent} and
Fig.~\ref{fig:homoXX}).

\subsubsection{Quench from the state $|R_{n_L}\rangle\otimes|R_{n_R}\rangle$}

The results of the previous section yield the quasiparticle prediction
for the protocol in which
the left and right halves of the chain are prepared in two $n$-rainbow states with, in general, different $n_L$ and $n_R$.
Eq.~\eqref{eq:quasimulti} becomes
\begin{equation}
	S(t) = S^{(n_L)}(t) + S^{(n_R)} (t) ,
	\label{eq:totalslope}
\end{equation}
where 
\begin{equation}
	S^{(n_R)}(t) = \int_{BZ_{n}} \frac{dk}{2\pi}
	\sum_{Q} \Omega_+(Q, k, t) \; s^{(n_R)}(k, Q) , 
	\label{eq:quasibifree}
\end{equation}
and $\Omega_+(Q,k,t)$ is the measure of the set of positions $x \geq 0$
such that at time $t$ a $k$-momentum multiplet generated in $x$ is in the configuration $Q$. Similarly, one has 
\begin{equation}
	S^{(n_L)}(t) = \int_{BZ_{n}} \frac{dk}{2\pi}
	\sum_{Q} \Omega_-(Q, k, t) \; s^{(n_L)}(k, Q) , 
	\label{eq:quasibifreeL}
\end{equation}
and $\Omega_-(Q,k,t)$ is the measure of the set of positions $x \leq 0$
such that at time $t$ a $k$-momentum multiplet generated in $x$ is in the configuration $Q$. 
The kinematic factors $\Omega_\pm$ follow from simple geometric considerations, detailed in Appendix~\ref{app:XXdetails}.
The contribution $S^{(n_L)}(t)$ describes the entanglement dynamics after the quench from  
$|R_{n_L}\rangle\otimes|\Omega\rangle$, with $|\Omega\rangle$ the fermionic vacuum state. Similarly, $S^{(n_R)}$ 
describes the entanglement dynamics from the state $|\Omega\rangle\otimes |R_{n_R}\rangle$. 
We stress that, if $n_L \neq n_R$, the set of possible partitions $Q$ in the sum $\sum_{Q}$ 
is different for $S^{(n_L)}$ and $S^{(n_R)}$.

\begin{figure}[t]
	\includegraphics[width=0.40\textwidth]{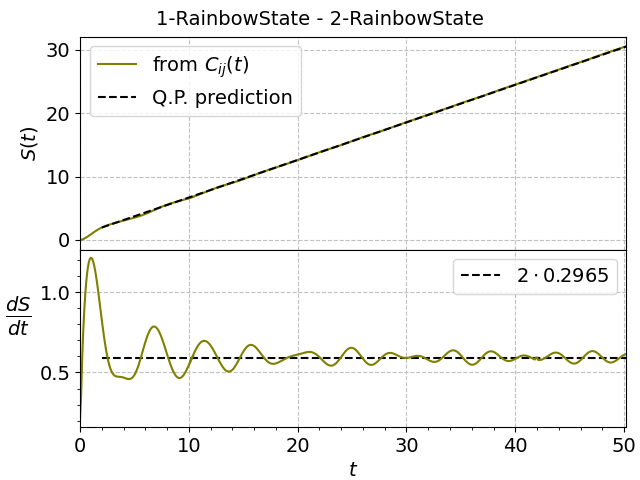}
	\caption{Quench from the state $|R_1\rangle\otimes |R_2\rangle$ in the 
	XX chain. Top: exact numerical data (solid line) for the
von Neumann entropy of the right half of the chain as a function of time, compared with
the quasiparticle prediction (dashed line).
Bottom: time derivative of the same data.
} 	
\label{fig:entropyXXbi}
\end{figure}

To illustrate Eqs.~\eqref{eq:quasibifree} and~\eqref{eq:quasibifreeL}, we first consider the
quench from the state in which the left half of the chain is prepared in $|R_1\rangle$
and the right half in $|R_2\rangle$. In the bulk of the left region the dynamics
creates entangled pairs of quasiparticles.
Indeed, for $n=1$ there are only two species, with $k \in BZ_0^L
\cap \left( 0, \pi \right]$ and velocities $v_0(k)=-\sin(k) = - v_1(k) \;<0$.
Only the configurations in which the two quasiparticles are in different subsystems contribute,
since $s^{(1)}(k, \set{0,1})=0$. 
The entropy content for $Q=\{0\}$ and $\{1\}$ only depends on the diagonal elements of $F^{(1)}(k)$: 
\begin{equation}
	F_{AA}^{(1)}(k) \; = \; \frac{1+(-1)^A\cos(k)}{2}. 
\end{equation}
Notice that $s^{(1)}(k, \{0\}) \; = \; s^{(1)}(k, \{1\})$.
Only the quasiparticle of species $1$, which moves to the right, can reach the right subsystem starting from
the left one.
The corresponding kinematic factor is simply $\Omega_-(\{1\}, k, t)\; = \; t \sin(k)$, and
Eq.~\eqref{eq:quasibifreeL} yields
\begin{equation}
	\frac{S^{(1)}(t)}{t} \; \approx \; 0.159 .
\end{equation}
The contribution of the multiplets produced in $|R_2\rangle$ is slightly more involved, because it requires
the entanglement content of quadruplets that are only partially shared (see
Appendix~\ref{app:XXdetails}). One obtains $S^{(2)}(t)/t\approx0.137$.
Putting everything together, we obtain 
\begin{equation}
	\frac{S(t)}{ t} \; = \; \frac{S^{(1)}(t) + S^{(2)}(t)}{t} \; \approx \; 0.297 .
	\label{eq:entropyfinLR}
\end{equation}
In Fig.~\ref{fig:entropyXXbi} we compare this prediction with exact numerical data. 
The solid line is the half-chain von Neumann entropy obtained from the
exact time evolution of the fermionic correlation matrix (see Appendix~\ref{app:numerics}).
In the numerics $\Abb$ is a finite segment rather than a half-infinite chain, and hence the expected slope is
twice~\eqref{eq:entropyfinLR}, which assumed open boundary conditions. 
The dashed line in Fig.~\ref{fig:entropyXXbi} is the quasiparticle prediction, which is in
excellent agreement with the numerical data.

Using the kinematic factors~\eqref{eq:kinematic} derived in Appendix~\ref{app:XXdetails}, the slope of $S^{(n)}(t)$
can be computed for arbitrary $n$.
The slope of $S(t)$ for an arbitrary protocol $\ket{R_{n_L}} \otimes \ket{R_{n_R}}$ is then obtained by summing the two contributions, as in Eq.~\eqref{eq:totalslope}.

\section{\label{sec:homogeneous} Rainbow quenches in the XXZ chain}

Here we address the effect of interactions. 
We start by determining the GGE describing the steady state after the quench from $|R_n\rangle$ in the
XXZ chain. The $n$-rainbow state is a periodic state with a unit cell of $2n$ sites. Hence, the
expectation values $\chi_s(\lambda)$ of the quasilocal charges (see Sec.~\ref{sec:TBA}) can be obtained from an operator
$\hTT^{(n)}_s(\lambda,\mu)$ acting on two spin-$s$ auxiliary spaces (see Appendix~\ref{sec:Algebraic}).
For the $n$-rainbow state $\hTT^{(n)}_s$ can be constructed recursively, by adding one arc of the
rainbow at a time (Appendix~\ref{app:charges}). Combined with the string-charge
duality~\eqref{eq:String_Charge_1}, this gives the GGE root densities for $n$ up to $n\sim10^3$.
The same approach yields the root densities of the infinite-temperature state
(Appendix~\ref{app:charges}), which provides a check of the $n\to\infty$ limit.

Before proceeding, we observe that, upon mapping $0\equiv \downarrow$ and $1\equiv \uparrow$,
the $n$-rainbow state can be written as 
\begin{equation}
    \ket{R_n}=\ket{\psi}^{\otimes L/(2n)},
    \label{eq:rainbow_product}
\end{equation}
with the unit-cell state
\begin{equation}
    \ket{\psi}=\frac{1}{2^{n/2}}\sum_{k=0}^{2^n-1}
    \ket{r_k}\otimes\ket{\overline{r_k}},
    \label{eq:rainbowrandom}
\end{equation}
where $r_k$ is the $n$-bit binary representation of $k$ and $|\overline{r_k}\rangle$ is obtained
from $r_k$ by exchanging $0\leftrightarrow 1$ and reversing the order of the bits (the arcs are nested). 
For large $n$, the sum in Eq.~\eqref{eq:rainbowrandom} is dominated by configurations $r_k$
with approximately equal numbers of $0$s and $1$s, i.e., by typical random configurations. On the
other hand, Ref.~\cite{random} conjectured that the steady state after quenches from
random product states with a long unit cell in the XXX chain is the infinite-temperature state. This provides a heuristic argument for
thermalization after quenches from the $n$-rainbow states: the random-like
structure of the unit cell for $n \rightarrow +\infty$ suggests that the expectation values of the conserved charges approach their infinite-temperature values.

\subsection{Thermalization after the quench from the $n$-rainbow state}

As discussed in Sec.~\ref{sec:SCD}, by evaluating the first $m+1$
quasilocal charges $\chi_s(\lambda)$ in an initial state,
we can exactly compute the first $m$ GGE densities $\rho_m(\lambda),\theta_m(\lambda)$ 
by string-charge duality \eqref{eq:String_Charge_1}.
For the $n$-rainbow state and the infinite-temperature thermal 
state this can be done employing Eqs.~\eqref{eq:computeXNum} and~\eqref{eq:infinitetempcharges}
(see Appendices~\ref{sec:Algebraic} and~\ref{app:charges}).

\begin{figure*}
	\centering
	\includegraphics[width=\textwidth]{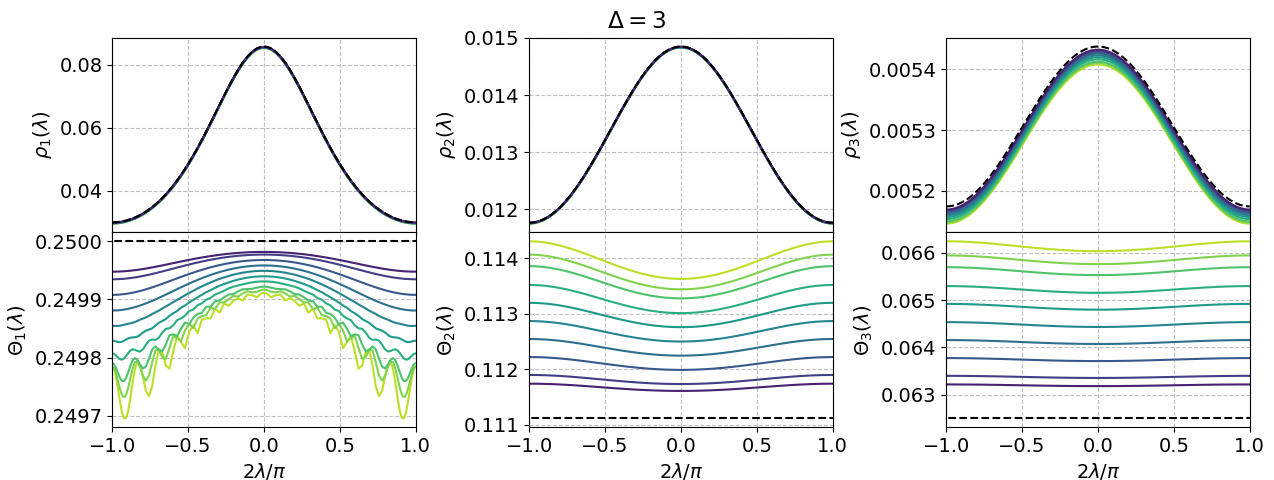}
	\includegraphics[width=\textwidth]{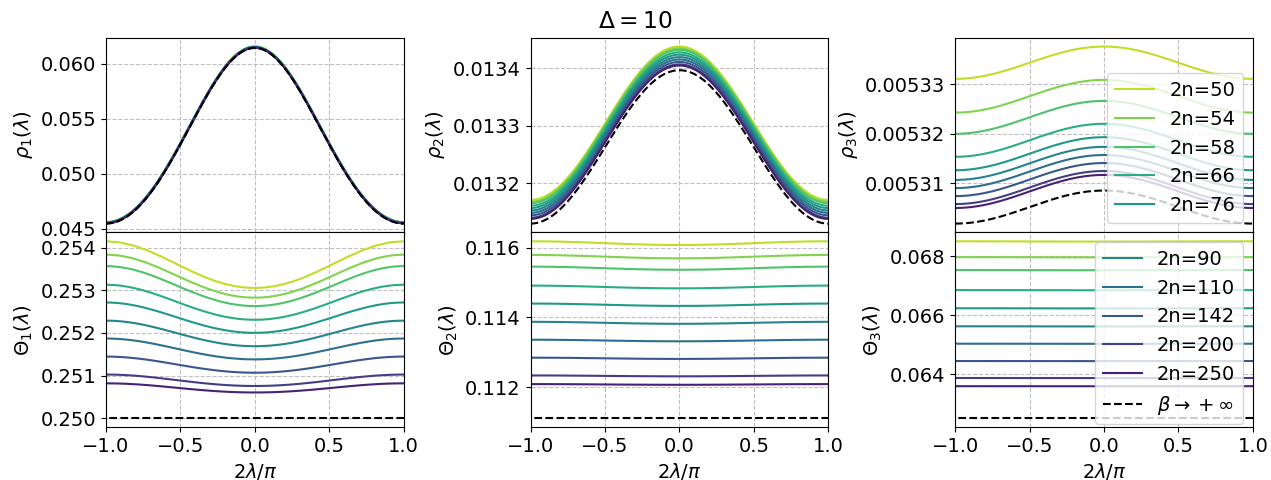}
	\caption{GGE root densities $\rho_m(\lambda)$ and filling functions $\theta_m(\lambda)$ after the quench from $|R_n\rangle$, for $n \in \{25, ..., 125\}$ (coloured solid lines), computed with the string-charge duality~\eqref{eq:String_Charge_1} for $\Delta=3$ (top) and $\Delta=10$ (bottom). Black dashed lines: infinite-temperature root densities.}
	\label{fig:comparison}
\end{figure*}

Fig.~\ref{fig:comparison} shows the densities $\{\rho_m(\lambda), \theta_m(\lambda)\}$
describing the GGE after the quench from the $n$-rainbow state for $\Delta=3$ and $\Delta=10$,
with $n$ up to $125$.
The dashed lines are the infinite-temperature root densities. 
Upon increasing $n$, both $\rho_m$ and $\theta_m$ converge to the
thermal profiles. To characterize the convergence quantitatively, we fit the values at
$\lambda=0$, $\rho^V_m\coloneq\rho_m(0)$ and $\theta^V_m\coloneq\theta_m(0)$, for $n$ up to $2500$.
Eq.~\eqref{eq:rainbowenergy} shows that for the $n$-rainbow state 
$E_n/L \propto n^{-1}$.
Moreover, Eq.~\eqref{eq:string_energy} implies that $\epsilon_m \approx -J \sinh(\eta)$ independently of $m$, if $\eta m \gg 1$. This suggests that  
\begin{equation}
	\rho_m^V(n) \; = \; \rho_m^{V, \infty} + \frac{A_\rho}{n}, 
	\label{eq:RShy}
\end{equation}
and similarly for $\theta^V_m$, with a different amplitude $A_\theta$.
Since the asymptotic value of $\theta_m^V$ is given by
Eq.~\eqref{eq:TBAanalytic}, $A_\theta$ is the only fitting parameter in this case.

\begin{figure*}
	\centering
	\includegraphics[width=\textwidth]{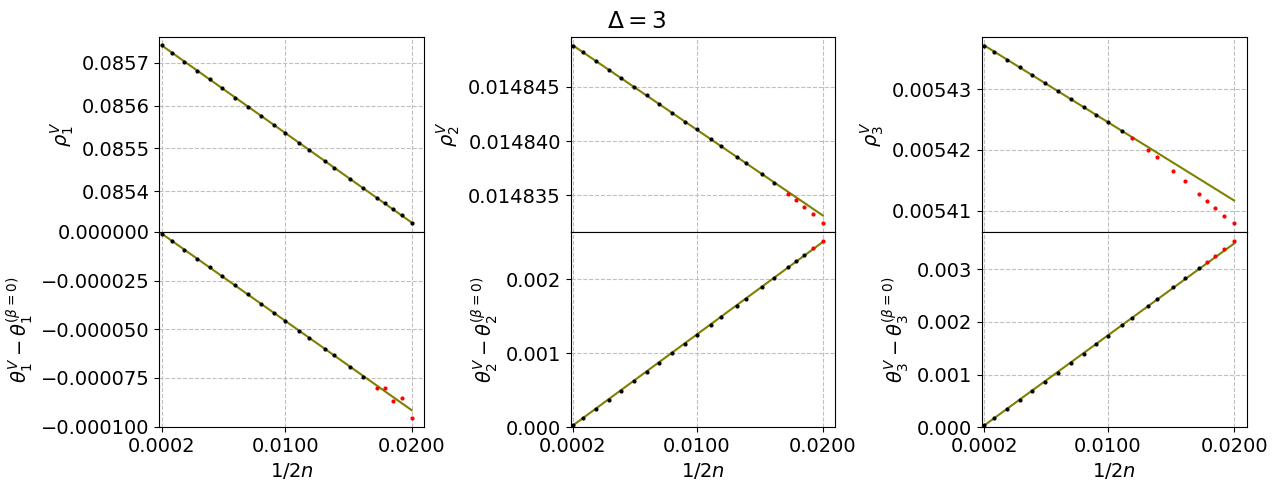}
	\includegraphics[width=\textwidth]{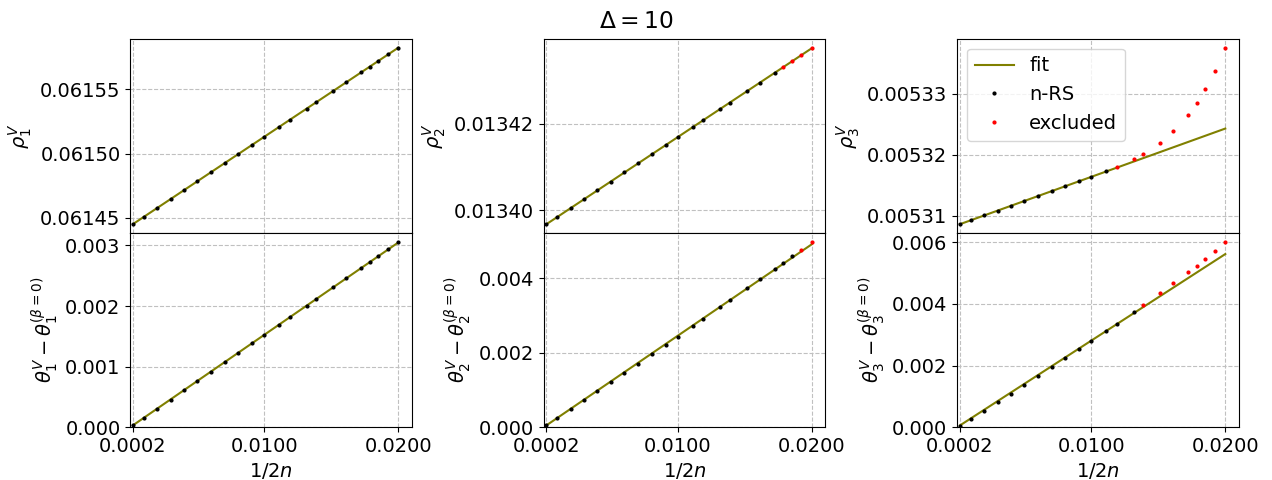}
	\caption{Values at $\lambda = 0$ of $\rho_m(\lambda)$ and $\theta_m(\lambda)$ (black dots) for $m \in \{1,2,3\}$ and unit-cell sizes $2n$ from $50$ to $5000$, for $\Delta=3$ (top) and $\Delta=10$ (bottom). Green solid lines: fits to Eq.~\eqref{eq:RShy}. Red dots: data excluded from the fits.}
	\label{fig:conv_rainbow}
\end{figure*}

For both $\rho_m^V$ and $\theta_m^V$, the data and the corresponding fit curves are shown in Fig.~\ref{fig:conv_rainbow}. 
In the fits we discarded the data at the smallest $n$, since Eq.~\eqref{eq:RShy} is expected to hold only for
large $n$. 

The extrapolated values $\rho_m^{V, \infty}$ are consistent with the $\lambda=0$ values
of the infinite-temperature state, computed directly from Eq.~\eqref{eq:infinitetempcharges}
and the string-charge duality.
These results confirm that the stationary state after the quench from the
$n$-rainbow state approaches the infinite-temperature thermal state as $n \rightarrow +\infty$. 

\section{\label{sec:homoent} Entanglement dynamics after homogeneous quenches in the XXZ chain}

In Sec.~\ref{sec:homogeneous} we showed that the GGE describing the steady state after the
quench from $|R_n\rangle$ approaches the infinite-temperature state as $n\to\infty$. Here we
investigate how the entanglement accompanying this relaxation is generated. We
consider the homogeneous quench from $|R_n\rangle$ and the von Neumann entropy $S(t)$ of half
of the chain, $\Abb=(-\infty,0]$. The bipartition is placed between two unit cells, so that no
Bell pair is cut and $S(0)=0$. This is the natural setting to test the results of
Sec.~\ref{sec:Entanglement}. In the XX chain the quench produces entangled multiplets of $2n$
quasiparticles. Their off-diagonal correlations are only $O(1/n)$, but they suppress the
growth of the entanglement entropy, whose slope vanishes in the limit $n\to\infty$
(see Fig.~\ref{fig:entropyXXslope}). Here we address the mechanism of
entanglement spreading in the presence of interactions for nonintegrable initial states, and
whether the multiplet mechanism survives.

\subsection{Pair-structure prediction and its infinite-temperature limit}

If the quench from $|R_n\rangle$ produced only entangled pairs of quasiparticles with opposite rapidities, the
quasiparticle picture~\cite{QuasiPicture,AlbaCalabrese2017,Alba_2018} would predict a linear growth
\begin{equation}
	\frac{S(t)}{t}\simeq\kappa_{\rm pair}\coloneq\sum_{m=1}^{\infty}\int_{-\pi/2}^{\pi/2}d\lambda\,
	|v_m(\lambda)|\,s^{YY}_m(\lambda). 
	\label{eq:therm-quasi}
\end{equation}
In~\eqref{eq:therm-quasi}, $s^{YY}_m(\lambda)$ is the Yang-Yang entropy
density~\eqref{eq:sYY} of the GGE that describes the steady state, which we construct from the
root densities obtained via Eqs.~\eqref{eq:recursion} and~\eqref{eq:String_Charge_1}.
The dressed velocities $v_m(\lambda)$ are the solution of Eq.~\eqref{eq:dressed} with the
homogeneous root densities, i.e., without any $\zeta$ dependence. Since $|R_n\rangle$ is
invariant under the reflection about the centre of a unit cell, one has
$\rho_m(\lambda)=\rho_m(-\lambda)$ and $v_m(-\lambda)=-v_m(\lambda)$. Hence the two members of
a pair $(\lambda,-\lambda)$ leave their creation point with opposite velocities, and every pair
shared between $\Abb$ and its complement contributes $s^{YY}_m(\lambda)$, which guarantees that
the stationary entanglement density coincides with the thermodynamic entropy density of the stationary state.
We stress that Eq.~\eqref{eq:therm-quasi} can be interpreted as
the rate at which Yang-Yang entropy is exchanged between $\Abb$ and its complement. 

We also stress that for a finite subsystem of length $\ell$ the stationary value of
the entanglement entropy is $\ell\, S_{YY}/L$, which tends to $\ell\ln 2$ for $n\to\infty$
(see Sec.~\ref{sec:homogeneous}). This value is fixed by the GGE, irrespective of the pair or the
multiplet structure, which only affects how it is approached (see Sec.~\ref{sec:Entanglement}).

Since the GGE converges to the infinite-temperature state, $\kappa_{\rm pair}$ has a finite
limit $\kappa_\infty$ for $n\to\infty$. This is obtained by setting
$\theta_m=1/(m+1)^2$ in Eqs.~\eqref{eq:BGT_densities} and~\eqref{eq:dressed} (cf.~\eqref{eq:TBAanalytic}).
Since the filling functions $\theta_m$ do not depend on $\lambda$, the equations are linear with
constant coefficients, and can be solved in Fourier space. We obtain
\begin{equation}
	\begin{array}{ll}
		\kappa_\infty\approx 0.3851 \; \;  & (\Delta=3), \\
		\kappa_\infty\approx 0.3663 & (\Delta=10). \\
	\end{array}
	\label{eq:kappa-inf}
\end{equation}
The sum in~\eqref{eq:therm-quasi} is dominated by the $1$-strings, which contribute
$0.3546$ ($\Delta=3$) and $0.3577$ ($\Delta=10$), followed by the $2$-strings, which
contribute $0.0274$ and $0.0083$, respectively. The contribution of longer strings is negligible
because their velocities are exponentially small in $m\eta$. The dependence on $\Delta$ is weak:
for $\Delta\to\infty$ we find
$\kappa_\infty\to 0.3580$. For comparison, in the XX chain $\kappa_\infty = (2/\pi)\ln 2\approx0.441$ (see below). 
In the following, we show that Eq.~\eqref{eq:therm-quasi} describes the growth of the
von Neumann entropy after quenches from the $n$-rainbow states in the XXZ chain, at least in the limit
$n\to\infty$. Before proceeding, we discuss how Eq.~\eqref{eq:therm-quasi} fails in the
XX chain.

\subsection{Free-fermion benchmark}

For the XX chain, Eq.~\eqref{eq:therm-quasi} follows from Eq.~\eqref{eq:entcontentfree} with $\Omega(\Abb, v(k), t)=t |\sin k|$:
\begin{equation}
	\kappa^{XX}_{\rm pair}=\int_{0}^{2\pi}\frac{dk}{2\pi}\,|\sin k|\,h(\braket{n_k}) ,
	\label{eq:XXpair}
\end{equation}
where $h(x)\coloneq-x\ln x-(1-x)\ln(1-x)$, and $\braket{n_k} = F_{kk}$ is given by Eq.~\eqref{eq:F-final}.
Since $\braket{n_k}\to1/2$ in the sense of distributions as $n\to\infty$, one has $\kappa^{XX}_{\rm pair}\to(2/\pi)\ln2\approx0.441$.
In Table~\ref{tab:XXslopes} we compare Eq.~\eqref{eq:XXpair} with the exact slope of the
half-chain entanglement entropy, which we obtained by evolving the correlation
matrix (see Appendix~\ref{app:numerics}). The exact slopes agree
with the multiplet quasiparticle picture of Sec.~\ref{sec:Entanglement} (cf.
Fig.~\ref{fig:entropyXXslope}).
The pair formula is exact only for $n=1$, which gives $\kappa=1/\pi$, consistent with
$|R_1\rangle$ producing entangled pairs. For $n>1$ Eq.~\eqref{eq:XXpair} overestimates the slope
by a factor that grows with $n$. This is not accidental. In the multiplet picture the contribution of a multiplet
with the quasiparticles in $Q$ inside $\Abb$ is the entropy of the restricted
correlation matrix (cf.~\eqref{eq:entcontentmulti}). Since each multiplet is in a pure state,
this is also the entropy of the complementary set of quasiparticles, and by subadditivity it is
bounded by the sum of the single-particle entropies $h(F^{(n)}_{AA})$ of the quasiparticles that
have crossed the cut. Summing over the multiplets, this bound gives Eq.~\eqref{eq:XXpair}.
Thus, $\kappa_{\rm pair}$ is an upper bound for the slope of the entanglement growth in
the free-fermion chain, and it is attained only if the off-diagonal correlations
$\widetilde F^{(n)}$ vanish.

\begin{table}[t]
	\caption{\label{tab:XXslopes} Slope of the half-chain entanglement entropy after the homogeneous
	quench from $|R_n\rangle$ in the XX chain. $\kappa_{\rm exact}$: exact free-fermion numerics
	(chains of $L\approx1200$ sites, linear fit of $S(t)$ for $t\in[100,400]$).
	$\kappa^{XX}_{\rm pair}$: pair formula~\eqref{eq:XXpair}.}
	\begin{ruledtabular}
	\begin{tabular}{ccccccccc}
		$n$ & $1$ & $2$ & $3$ & $4$ & $8$ & $12$ & $16$ & $\infty$ \\
		\hline
		$\kappa_{\rm exact}$ & $0.318$ & $0.275$ & $0.224$ & $0.188$ & $0.117$ & $0.087$ & $0.070$ & $0$ \\
		$\kappa^{XX}_{\rm pair}$ & $0.318$ & $0.335$ & $0.366$ & $0.382$ & $0.410$ & $0.420$ & $0.425$ & $0.441$ \\
	\end{tabular}
	\end{ruledtabular}
\end{table}

\begin{figure}[t]
	\centering
	\includegraphics[width=\linewidth]{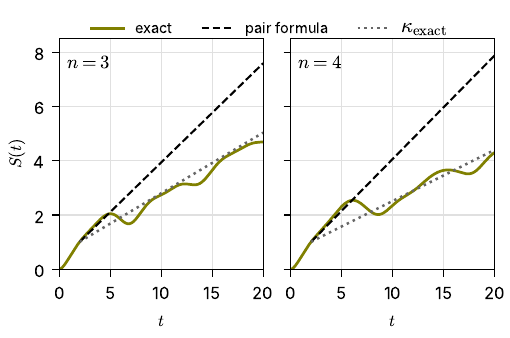}
	\caption{Half-chain entanglement entropy after the homogeneous quench from
	$|R_n\rangle$ in the XX chain, for $n=3$ (left) and $n=4$ (right). Full lines: exact
	free-fermion results. Dashed lines: pair prediction~\eqref{eq:XXpair}. Dotted lines:
	asymptotic slope $\kappa_{\rm exact}$ of Table~\ref{tab:XXslopes}. Both straight lines
	start at $t=2$, as in Fig.~\ref{fig:entropyhomo}. $S(t)$ follows the pair prediction up to
	$t\approx1.3n$, then decreases, and oscillates with period $\approx2n$ around the much
	slower multiplet growth.}
	\label{fig:homoXX}
\end{figure}

The deviation from the pair prediction is not only a long-time effect.
Fig.~\ref{fig:homoXX} shows $S(t)$ for $n \in \set{3,4}$ and $t\le20$. Initially $S(t)$ follows
Eq.~\eqref{eq:XXpair}, up to $t\approx1.3n$. Then, the entanglement entropy decreases, and
it exhibits sawtooth-like oscillations with period $\approx2n$ around a linear growth with
the multiplet slope. We checked that both the position of the first maximum
and the period of the oscillations scale linearly with $n$. This behaviour can be understood as follows. At short times the dynamics
is not sensitive to the structure of the unit cell: on scales smaller than $n$ the $n$-rainbow
state is close to the infinite-temperature state, since all the sites in a block of $r\ll n$
consecutive sites are typically paired with sites outside the block.
After this initial transient, the quasiparticles
that cross the cut at $t\lesssim n$ thus contribute as in the pair picture. At $t\gtrsim n$ the
other members of the multiplets emitted from the same unit cells cross the cut, and the
off-diagonal correlations reduce the entanglement content of these multiplets.

\subsection{Interacting case: tDMRG results}

\begin{figure*}
	\centering
	\includegraphics[width=0.40\textwidth]{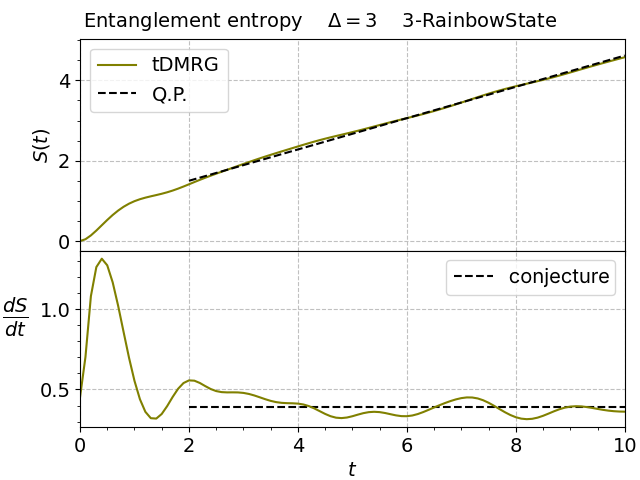}
	\includegraphics[width=0.40\textwidth]{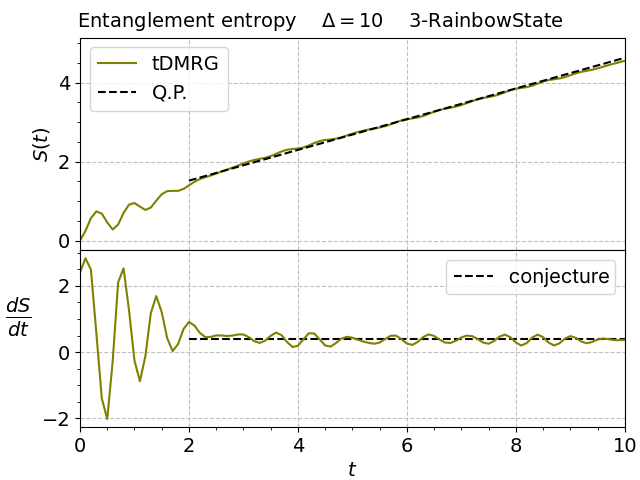}
	\includegraphics[width=0.40\textwidth]{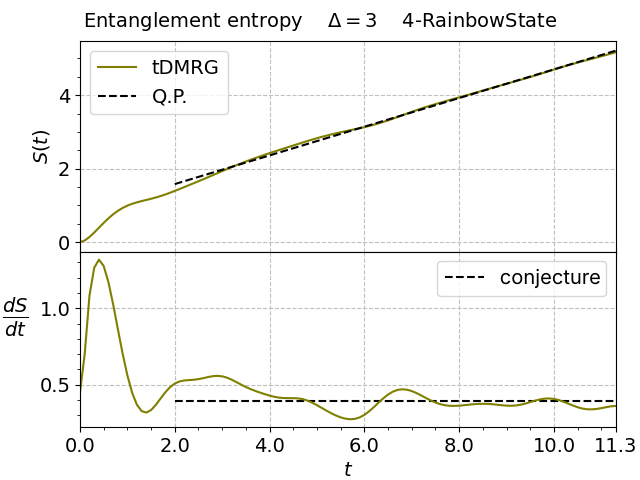}
	\includegraphics[width=0.40\textwidth]{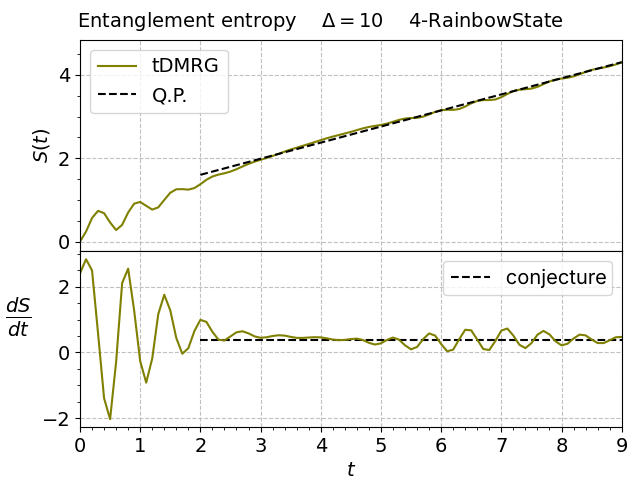}
	\includegraphics[width=0.40\textwidth]{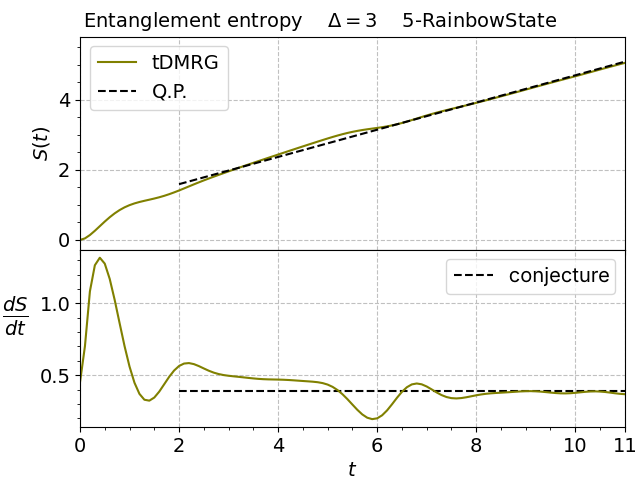}
	\includegraphics[width=0.40\textwidth]{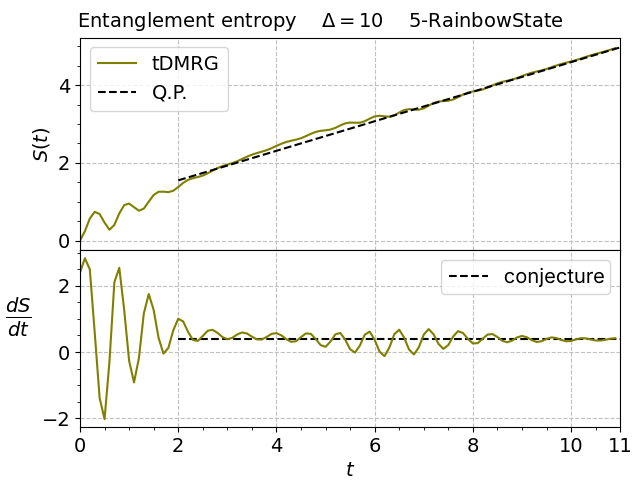}
	\caption{Half-chain entanglement entropy after the homogeneous quench from $|R_n\rangle$ in
		the XXZ chain, for $n \in \set{3,4,5}$ (from top to bottom) and $\Delta \in \set{3,10}$ (left and right). Upper panels: tDMRG data (solid lines) and
		conjectured linear growth with the rate $\kappa_{\rm pair}$ of Eq.~\eqref{eq:therm-quasi} (dashed lines, with fitted intercept).
		Lower panels: numerical derivative $dS/dt$ (solid lines) and $\kappa_{\rm pair}$ (dashed lines).
		For $n=3,4,5$ one has $\kappa_{\rm pair}\approx0.3898, 0.3904, 0.3884$ for $\Delta=3$, and
		$\kappa_{\rm pair}\approx0.3893, 0.3863, 0.3807$ for $\Delta=10$.
		In contrast with the free-fermion case (Fig.~\ref{fig:homoXX}), the entanglement
		entropy grows linearly with the rate~\eqref{eq:therm-quasi}, with no sign of the
		multiplet suppression. Notice the weak dependence of the slope on $\Delta$.}
	\label{fig:entropyhomo}
\end{figure*}

Fig.~\ref{fig:entropyhomo} shows tDMRG results for the half-chain entanglement entropy after
the quench from $|R_n\rangle$ in the XXZ chain, for $n \in \set{3,4,5}$ and $\Delta \in \set{3,10}$.
Details on the simulations are reported in Appendix~\ref{app:numerics}.
The corresponding predictions $\kappa_{\rm pair}$ for the slope are reported in the caption.
These values approach the infinite-temperature slope~\eqref{eq:kappa-inf} upon increasing $n$. 
After an initial transient $t\lesssim2$, during which $S(t)$ quickly reaches $S\approx1.3$--$1.5$,
the entanglement entropy grows linearly. For $\Delta=10$ the transient exhibits strong
oscillations, which could reflect the Ising energy scale $\Delta$, and
which are still visible as small oscillations of $dS/dt$ at later times. The dashed lines are the
pair prediction~\eqref{eq:therm-quasi}, with $\kappa_{\rm pair}$ computed using the GGE of
Sec.~\ref{sec:homogeneous}.
The intercept of the dashed lines is fitted, since the $O(1)$ offset is not universal,
and it is subleading in the space-time scaling limit. The agreement is remarkable.
The derivative $dS/dt$ (lower panels) fluctuates around $\kappa_{\rm pair}$, and it does not
exhibit any systematic drift up to the largest accessible times.

Two features distinguish these results from the free-fermion case of Fig.~\ref{fig:homoXX}.
First, there is no trace of the decrease of $S(t)$ at $t\sim n$, nor of the sawtooth
oscillations. Notice that quasiparticles are faster in the interacting chain:
at infinite temperature the maximal velocity of the $1$-strings is $\approx1.46$ for $\Delta=3$
and $\approx1.50$ for $\Delta=10$, to be compared with $1$ in the XX chain. Thus, if multiplet-like
correlations were present, their effect should be visible even earlier than in the free case,
already at $t\approx3$, well within the time window of the simulations.
Second, at fixed $n$ the entanglement is considerably larger. At the largest times
the tDMRG data give $S\approx4.6$ ($n=3$, $t=10$) and $S\approx4.3$ ($n=4$, $t=9$), whereas in
the XX chain $S\approx2.7$ and $S\approx2.1$, respectively. Finally,
the slope does not decrease from $n=3$ to $n=4$, and $\kappa_{\rm pair}$ for $n=3$ is already
close to its limit~\eqref{eq:kappa-inf}. Taken together, these observations suggest that in the XXZ chain the slope of the
entanglement growth remains finite for $n\to\infty$, and that it approaches $\kappa_\infty$.
This is in sharp contrast with the free-fermion result, for which the slope vanishes.

\subsection{Multiplet effects in the interacting chain}

The results of Fig.~\ref{fig:entropyhomo} are surprising, and they constitute one of the main results of
this work. For $n>1$, $|R_n\rangle$ is not an integrable initial state~\cite{IntegrableStates} (see
Sec.~\ref{sec:SCD}), and in the free-fermion chain it produces entangled
multiplets~\cite{SpeciesFirst,SpeciesBertini,SpeciesCollura}.
Yet, in the interacting chain the entanglement grows with a rate that, within our
numerical accuracy, coincides with the rate of exchange of Yang-Yang entropy across the cut, i.e., it saturates
what in the free-fermion chain is an upper bound. 

In other words, the entanglement dynamics is
indistinguishable from that produced by entangled pairs. We recall that this is not generic: for other nonintegrable initial
states, such as the ``fat'' N\'eel states~\cite{biagetti2024bound}, the pair-structure quasiparticle picture fails (see Appendix~\ref{app:violations}).
Our data thus suggest that the multiplet mechanism of
entanglement spreading, which is exact in the free-fermion chain, may not survive in the interacting
chain, at least within the accessible time window. To the best of our knowledge, this is one of the
first tests of the quasiparticle picture in an interacting integrable system for initial states that are
not integrable. 
The rate~\eqref{eq:therm-quasi} is an upper bound for the growth of the entanglement entropy
within any quasiparticle description, irrespective of the pair or multiplet structure. Since
at late times a finite subsystem contains at most one member of each multiplet, and its
entropy has to be the thermodynamic one, a single quasiparticle with rapidity $\lambda$
carries the Yang-Yang entropy $s^{YY}_m(\lambda)$. At time $t$, the entanglement entropy is
the sum over the shared multiplets of the entropy of the quasiparticles that crossed the cut.
By subadditivity, this cannot exceed the sum of their single-particle entropies, i.e., the
thermodynamic entropy transported across the cut up to time $t$, which gives
$S(t)\le t\,\sum_m\int d\lambda\,|v_m(\lambda)|\,s^{YY}_m(\lambda)$. The bound is saturated if
and only if the quasiparticles that cross the cut together are uncorrelated, which is the
case for entangled pairs because only one member of the pair crosses the cut. Indeed, within any quasiparticle
picture, if \emph{entangled} quasiparticles cross
the cut, their entanglement contribution has to be removed. 
In the XX chain the reduction of the entanglement is due to the coherent sum of $O(n)$ off-diagonal correlations
of order $1/n$ between the members of a multiplet, which are preserved during the dynamics. In the interacting chain,
instead, each quasiparticle scatters with a finite density of other excitations, acquiring
rapidity-dependent displacements, which are at the origin of the dressing of the
velocities~\cite{Bertini,HydroA}. The phase relations between the members of a multiplet are thus
expected to depend on the configuration of the background. This could dephase the collective
correlations that make partially shared multiplets almost pure, leaving only the
diagonal ones, which yield Eq.~\eqref{eq:therm-quasi}.

\section{\label{sec:bipartitioning} Quantum transport of local conserved quantities}

\begin{figure*}
	\centering
	\includegraphics[width=0.40\textwidth]{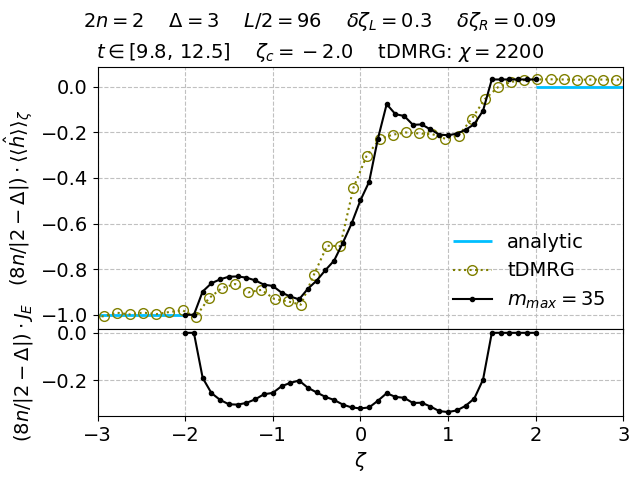}
	\includegraphics[width=0.40\textwidth]{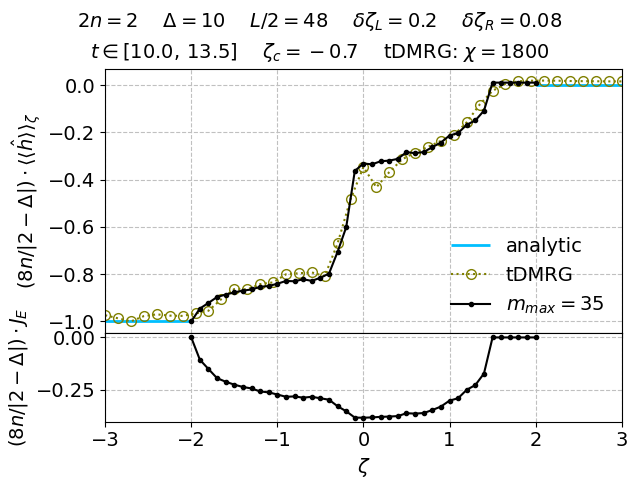}
	\includegraphics[width=0.40\textwidth]{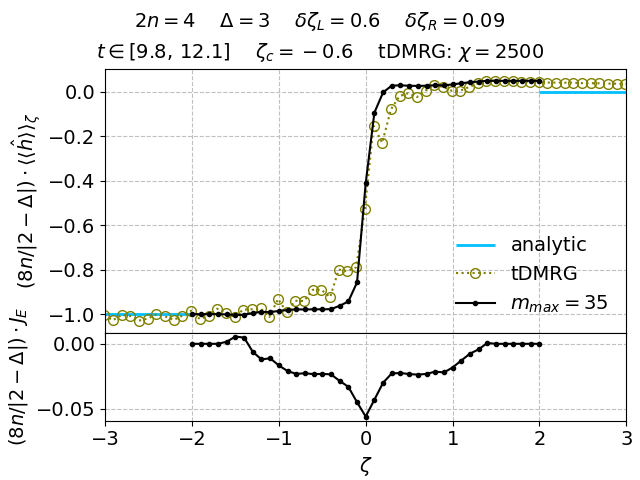}
	\includegraphics[width=0.40\textwidth]{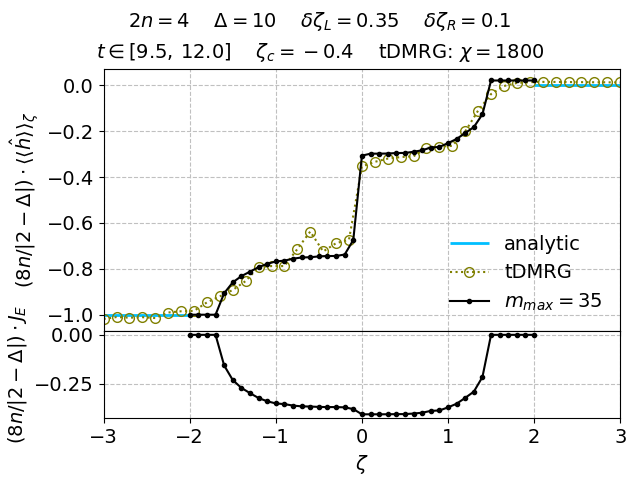}
	\includegraphics[width=0.40\textwidth]{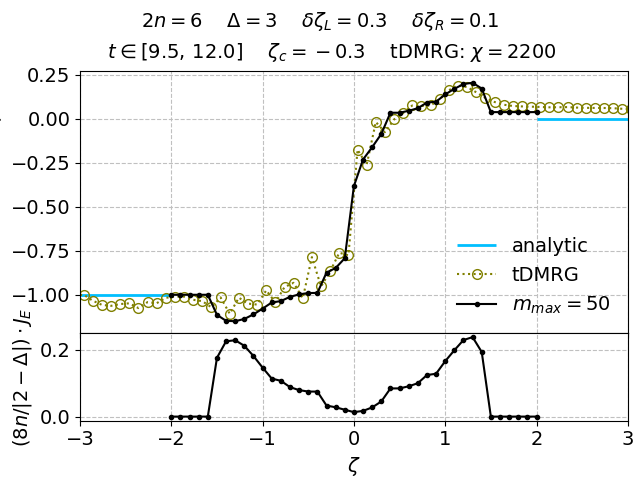}
	\includegraphics[width=0.40\textwidth]{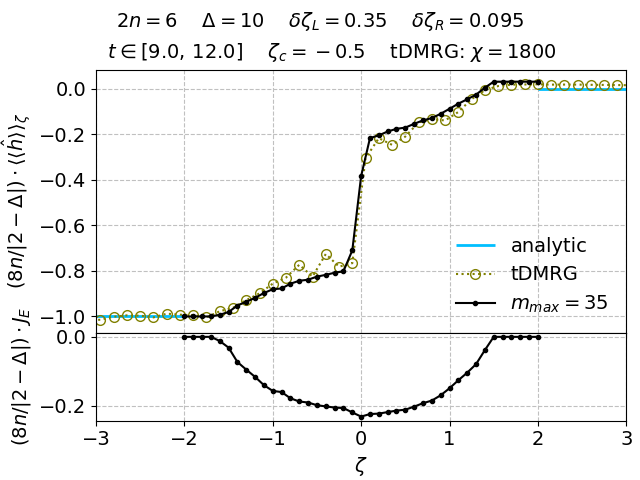}
	\includegraphics[width=0.40\textwidth]{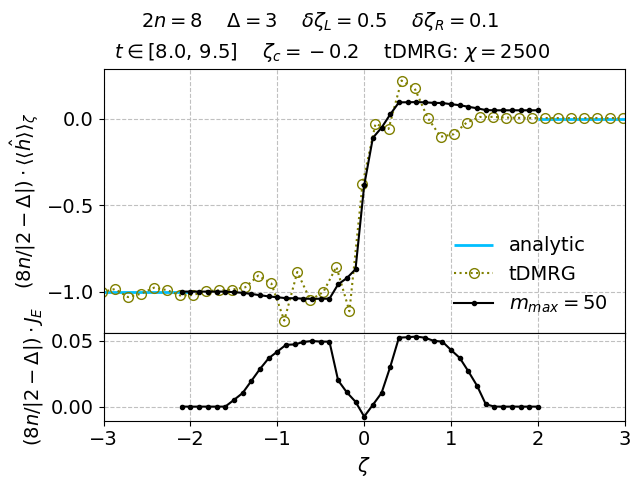}
	\includegraphics[width=0.40\textwidth]{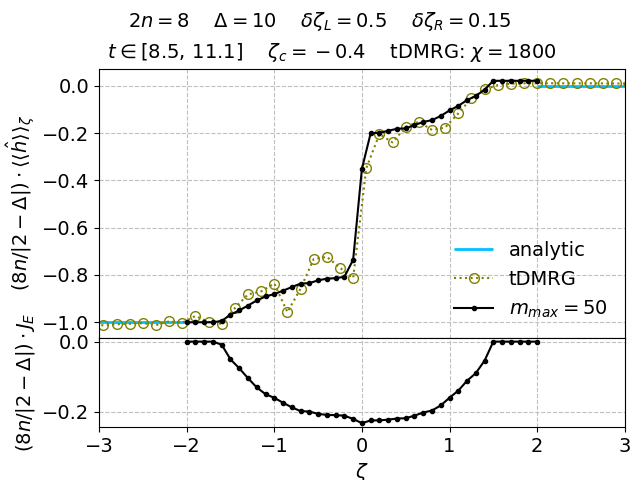}
	\caption{Bipartitioning protocol with the $n$-rainbow state (unit-cell size $2n \in \{2, 4, 6, 8\}$, from top to bottom) and the infinite-temperature thermal state, for $\Delta=3$ (left) and $\Delta=10$ (right). In each panel, the upper subplot shows the $\zeta$-averaged tDMRG energy density $\bbraket{\hh}_\zeta$ (green empty circles) and the GHD prediction (black solid lines). Light-blue horizontal lines: exact bulk energy densities. The lower subplot shows the corresponding GHD energy current. Both the energy density and the current are divided by $|E_n|/L = |2-\Delta|/(8n)$, which removes the trivial $1/n$ suppression and allows for an easier comparison. For $(\Delta, n)=(3, 1)$, resolving the nearly specular profile required $200$--$250$ Gaussian quadrature points.}
	\label{fig:energy1}
\end{figure*}

We now discuss quantum transport, which in the XXZ chain has been extensively studied with GHD~\cite{Bertini, Nonballistic, BertiniPiroli2018}. We focus on the setup of Fig.~\ref{fig:cartoon} (bottom),
in which the left half of the chain is prepared in the $n$-rainbow state and the right half in the infinite-temperature
Gibbs ensemble, and we compare tDMRG results for the local energy density with the GHD predictions in the
scaling limit $x,t\to\infty$ with $\zeta=x/t$ fixed.
Fig.~\ref{fig:energy1} shows results for $n \in \set{1,2,3,4}$ and \mbox{$\Delta\in \{3, 10\}$} (left and right panels, respectively):
the local energy density $\langle\hat h\rangle$ and the energy current $J_E$
as a function of $\zeta$. We rescale both by $8n/|2-\Delta|$, so that the rescaled energy density deep in the left half is $-1$ irrespective of $n$, while it vanishes in the right half.
The finite-time tDMRG data exhibit large oscillations. These are expected, since
the $n$-rainbow state is not homogeneous on the scale of the unit cell. Moreover, in the scaling limit
the profiles develop cusp-like singularities, which typically lead to strong finite-time effects.
For this reason, for each $\zeta$ we average the tDMRG data over a small space-time
window around the ray $x/t=\zeta$ (see Appendix~\ref{app:numerics}).
We denote the averaged observable by double angular brackets, i.e., $\bbraket{\hh}_\zeta$.
For $\zeta$ beyond the maximum quasiparticle velocities $\zeta^\pm_{max}$, the profiles trivially coincide with their bulk values.
The light-blue horizontal lines in Fig.~\ref{fig:energy1} are the exact energy densities
in the left and right bulks, $(8n/|2-\Delta|)\langle\hh\rangle \in \set{-1,0}$,
which provide a benchmark for the GHD and tDMRG results.

Although we are interested in the large-$n$ limit,
here we only show results up to $n=4$. With tDMRG we can access times
$t\approx 10$, which are not sufficient to reach the scaling limit for larger $n$,
since finite-time corrections grow with $n$. Moreover, the computation of the GHD prediction becomes increasingly demanding at larger $n$
(see Appendix~\ref{app:numerics}).
The GHD energy density and current follow from Eqs.~\eqref{eq:charges_currents} and~\eqref{eq:charge_current}, with the charge content in Eq.~\eqref{eq:string_energy} and the vacuum energy in Eq.~\eqref{eq:energy}, evaluated at each ray $\zeta$.

Fig.~\ref{fig:energy1} shows that energy transport between the two halves of the chain
is suppressed as $1/n$ (see Ref.~\cite{jin2021interplay}). This directly follows from the fact that the energy density of the
$n$-rainbow state is $O(1/n)$; the nontrivial information is in the shape of the rescaled profiles. Interestingly, the energy profile as a function of $\zeta$
exhibits nonanalytic points.
This was explained in Ref.~\cite{Nonballistic},
where the nonanalytic points were associated with the presence of multiple light cones, one for each
species of quasiparticles (the $m$-strings, i.e., bound states of $m$ quasiparticles).
More precisely, if $\zeta_m(\lambda)$ is the (unique) solution of $\zeta_m(\lambda)=v_{m,\zeta_m(\lambda)}(\lambda)$,
the light cones of the $m$-strings are $\zeta^\pm_m \coloneq \max/\min_\lambda[\zeta_m(\lambda)]$.
Physically, $\zeta^\pm_m$ are the rays at which quasiparticles of species $m$ start affecting local
observables away from the interface.
While noninteracting models have only two
nonanalytic points (cf. Fig.~\ref{fig:energyXX}), in the XXZ chain the strings give rise to additional light cones.
In practice, only the first few singularities are visible, since the $\zeta^\pm_m$ approach each other upon
increasing $m$~\cite{Nonballistic}. 

Besides the $1/n$ suppression of quantum transport, several remarkable features are 
visible in Fig.~\ref{fig:energy1}. 
First, the rescaled profiles of the local energy develop a very steep, jump-like variation at $\zeta = 0$.
Since the velocities of the $m$-strings are exponentially small in $m\eta$, their light cones accumulate
at $\zeta=0$, and the profile can vary on a very small scale of $\zeta$ around the junction. 
Moreover, the rescaled energy profile is essentially step-like for even $n$, whereas 
for odd $n$ some additional features are visible. Interestingly, for large $\Delta$
the data suggest an $n$-independent behaviour. 

\section{\label{sec:entdynamics} Entanglement dynamics from half-purified states}

Here we consider the entanglement dynamics after the quench from the bipartite state in Fig.~\ref{fig:cartoon} (bottom),
in which the right half of the chain is prepared in the infinite-temperature state. To
represent it as a pure state, we use the standard purification approach (see Ref.~\cite{Lagnese2022TFD} for the entanglement dynamics of purified thermal states in integrable models).
Given a mixed state $\hrho_R$ for the right subsystem, with Hilbert space $\mH_R$,
we introduce an ancillary copy $\tilde{\mH}_R$ of $\mH_R$
and define a pure state $\kket{\psi_R} \in \mH_R \otimes \tilde{\mH}_R$ such that 
$\trace_{\tilde{\mH}_R}\, \left[\kket{\psi_R}\bbra{\psi_R} \right] \; = \; \hrho_R$.
The half-purified initial state is then $\ket{\psi_L} \otimes \kket{\psi_R}$.
For the infinite-temperature state, $\kket{\psi_R}$ is obtained
by forming a Bell pair between each site of $\mH_R$ and a distinct site of $\tilde{\mH}_R$.

We focus on the von Neumann entropy of the left subsystem.
Since the chain is globally in a mixed state, $S(\hrho_R) \neq S(\hrho_L)$, and the von Neumann entropy is not a measure of the
entanglement between the two halves of the chain.
However, $S(\hrho_L)$ coincides with
the entanglement between the left part and the right part together with the ancillas, in the pure state $\ket{\psi_L}\otimes\kket{\psi_R}$.
Since the left part of the system is not purified, $S(\hat\rho_L)$ can be computed
straightforwardly with matrix product states.

\subsection{A conjecture for the entanglement dynamics}

Since the right part of the chain is prepared in a thermal state,
no entangled quasiparticles are created there: genuine entanglement
is produced only in the left part, which is prepared in the $n$-rainbow state.

Let us first consider entanglement dynamics after quenches in the  
XXZ model starting from the globally pure state $|\psi_L\rangle\otimes|\psi_R\rangle$. 
\begin{figure*}
	\centering
	\includegraphics[width=0.40\textwidth]{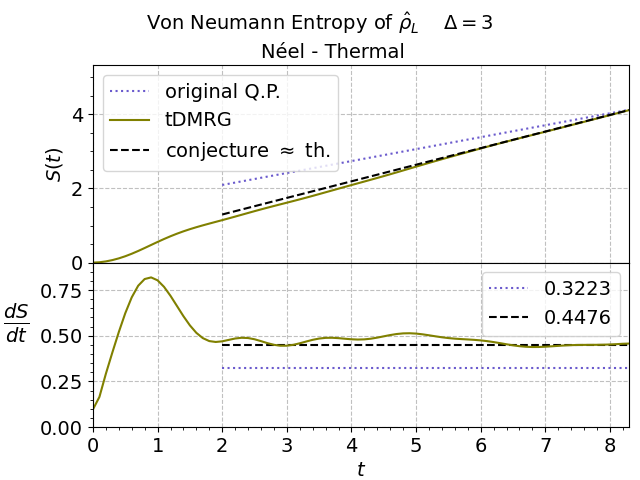}
	\includegraphics[width=0.40\textwidth]{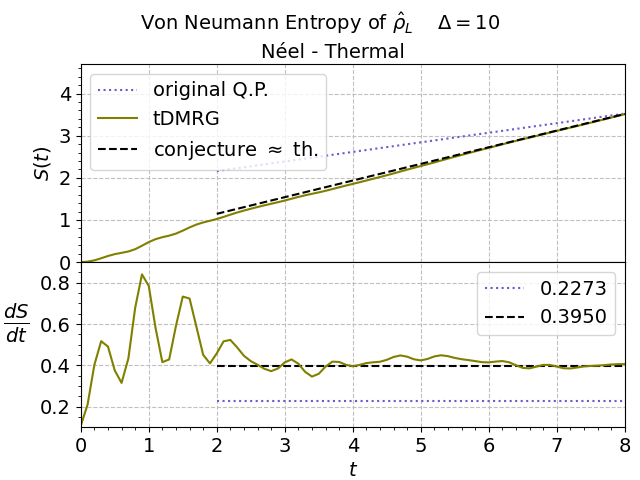}
	\includegraphics[width=0.40\textwidth]{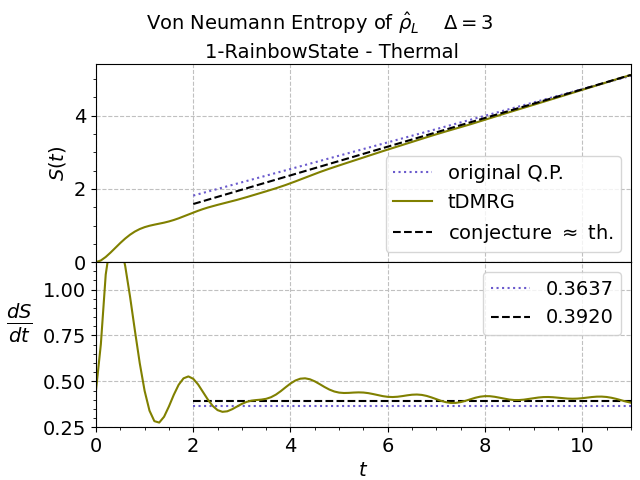}
	\includegraphics[width=0.40\textwidth]{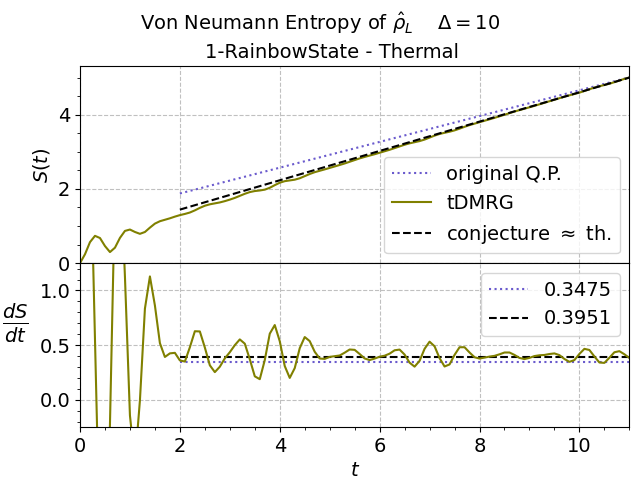}
	\includegraphics[width=0.40\textwidth]{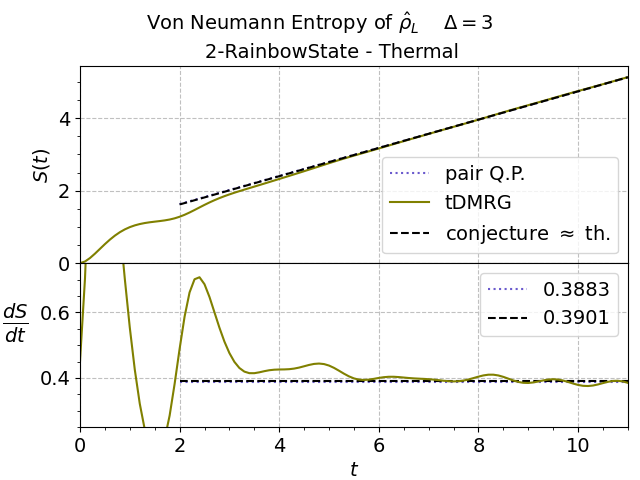}
	\includegraphics[width=0.40\textwidth]{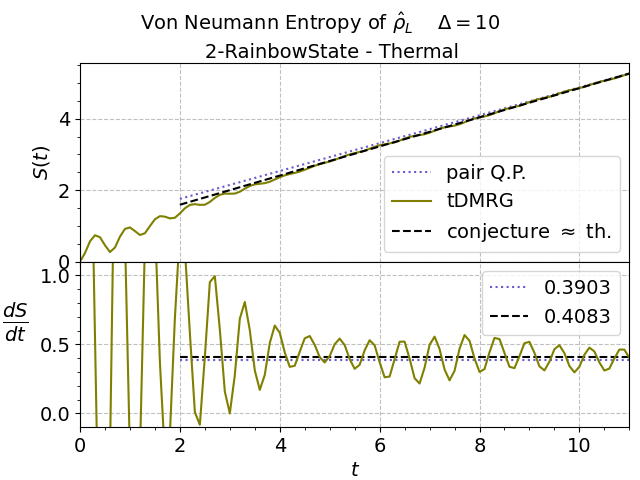}
	\includegraphics[width=0.40\textwidth]{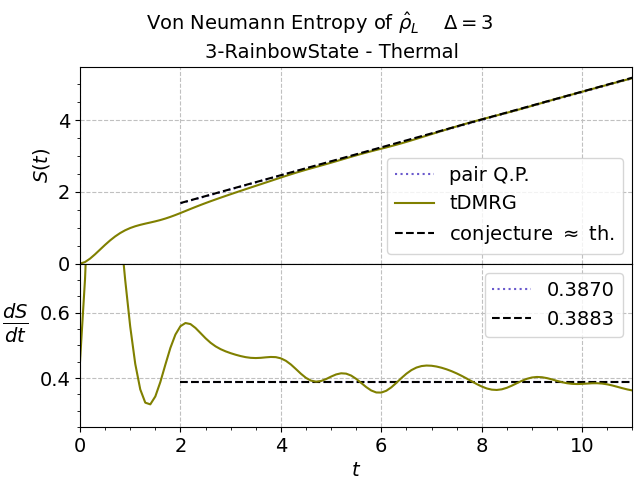}
	\includegraphics[width=0.40\textwidth]{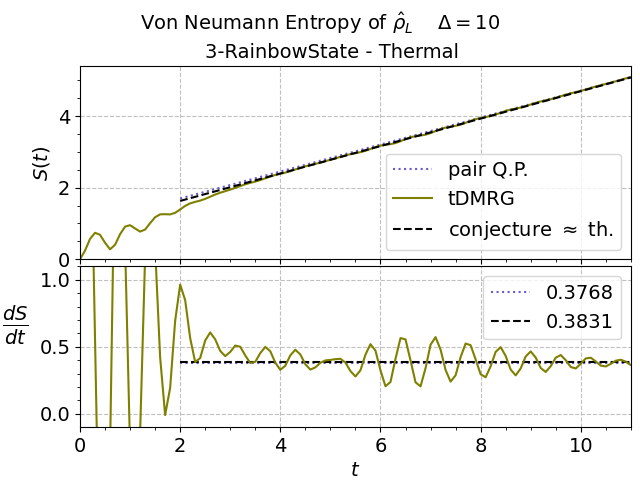}
	\caption{Von Neumann entropy of the left subsystem (green solid lines) after the bipartitioning protocol with $\ket{\psi_L}$ in the left half (from top to bottom: N\'eel state and $n$-rainbow states with $n=1,2,3$) and the infinite-temperature thermal state in the right half, for $\Delta=3$ (left) and $\Delta=10$ (right). Black dashed lines: conjecture~\eqref{eq:finalconjecture}, which is indistinguishable from the rate of exchange of thermodynamic entropy. Violet dotted lines: standard quasiparticle prediction~\eqref{eq:bi_original}.}
	\label{fig:entropyconjecture}
\end{figure*}
Even for integrable initial states $|\psi_L\rangle,|\psi_R\rangle$,
the quasiparticle picture for interacting systems is quite involved, because
the group velocities depend on the initial states and on the ray $\zeta$ (see Sec.~\ref{sec:GHD}).
However, for $\Abb=(-\infty,0]$ one obtains a strikingly simple expression~\cite{Alba_2018_PRB, Alba_2019, Alba_2021}, which only depends on the dressed velocities and root densities at $\zeta=0$: 
\begin{equation}
	\frac{S^{QP}(t)}{t} \; = \sum_{m=1}^{+\infty} \int_{-\pi/2}^{\pi/2} \mbox{sgn}(\lambda) \; v_{m, \zeta=0}(\lambda) \; s_m^{(\zeta=0)}(\lambda) \; d\lambda
	\label{eq:bi_original}
\end{equation}
where we assumed $\mbox{sgn}(\lambda)= \mbox{sgn}[v_{m, \zeta=\pm \infty}(\lambda)]$ and 
$s_m^{(\zeta=0)}(\lambda)$ is the contribution of the $m$-string to the 
Yang-Yang entropy of the thermodynamic state with $\zeta=0$. 
The factor $\mbox{sgn}(\lambda)$ accounts for the possibility that 
interactions are sufficiently strong for the quasiparticles forming an 
entangled pair to end up in the same half of the system as $t\rightarrow +\infty$. 
Indeed, if both quasiparticles of a pair end up on the same side of the interface at $x=0$, they are no longer
shared between the two halves and do not contribute to the entanglement entropy. 

Let us now consider the bipartitioning protocol in which the left chain is prepared in an integrable 
state, and the right one is prepared in the infinite-temperature state. It is natural to expect
that the contribution of the quasiparticles originating in the left part has to be treated
as in Ref.~\cite{Alba_2019}, i.e., only the shared entangled pairs contribute
to the von Neumann entropy. On the other hand, no entangled pairs are produced in the right
part of the chain. Hence, the contribution of the right half of the chain is the 
thermodynamic entropy transported by the quasiparticles that cross the interface and end up in 
the left part. 

\begin{table*}
\caption{\label{tab:entropy} Predicted growth rates of the von Neumann entropy $S(\hrho_L)$ after the
bipartitioning protocol in which the left half of the chain is prepared in the N\'eel state
or in the $n$-rainbow state ($n\in\set{1,2,3}$), and the right half in the infinite-temperature state, in the XXZ chain with $\Delta\in\set{3,10}$.
The rows report $S_\pm^{QP}$ and $S_\pm^{th}$ [cf. Eqs.~\eqref{eq:Spm} and~\eqref{eq:entropyright}], $S^{QP}=S_-^{QP}+S_+^{QP}$, and
$S^{conj}$ [Eq.~\eqref{eq:finalconjecture}].
}
	\begin{ruledtabular}
		\begin{tabular}{l|cc|cc|cc|cc}
			& N\'eel $\Delta = 3$ & N\'eel $\Delta = 10$  &
			$n=1$ $\Delta = 3$ & $n=1$ $\Delta = 10$ &
			$n=2$ $\Delta = 3$ & $n=2$ $\Delta = 10$ &
			$n=3$ $\Delta = 3$ & $n=3$ $\Delta = 10$ \\
			$S^{QP}_-/t \approx S^{th}_-/t$ &
			$0.10610$ &
			$0.01925$ &
			$0.15865$ &
			$0.15328$ &
			$0.19406$ &
			$0.19985$ &
			$0.19313$ &
			$0.19031$ \\
			$S^{QP}_+/t$ &
			$0.21619$ &
			$0.20810$ &
			$0.20502$ &
			$0.19426$ &
			$0.19421$ &
			$0.19043$ &
			$0.19382$ &
			$0.18651$ \\
			$S^{th}_+/t$ &
			$0.34154$ &
			$0.37576$ &
			$0.23336$ &
			$0.24185$ &
			$0.19609$ &
			$0.20841$ &
			$0.19516$ &
			$0.19279$ \\
			$S^{QP}/t$ &
			$0.32229$ &
			$0.22735$ &
			$0.36366$ &
			$0.34754$ &
			$0.38827$ &
			$0.39028$ &
			$0.38695$ &
			$0.37683$ \\
			$S^{conj}/t \approx S^{th}/t$ &
			$0.44764$ &
			$0.39502$ &
			$0.39201$ &
			$0.39513$ &
			$0.39015$ &
			$0.40826$ &
			$0.38830$ &
			$0.38310$ \\
		\end{tabular}
	\end{ruledtabular}
\end{table*}

To derive the resulting quasiparticle picture, we follow the steps of Ref.~\cite{Alba_2019}, but
with the entropy contributions described above (see Appendix~\ref{app:halfpur}).
The contributions of the quasiparticles produced in the left ($-$) and right ($+$) chains read
\begin{multline}
	\label{eq:Spm}
	S_\pm^{QP}= \sum_{m=1}^{+\infty} \int_{-\pi/2}^{\pi/2}d\lambda \Theta \left[
	\mp v_{m, \zeta=0}(\lambda)
	\right] \\\times
	\mbox{sgn}(\lambda) \; v_{m, \zeta=0}(\lambda) \; s_m^{(\zeta=0)}(\lambda) ,
\end{multline}
which coincides with the result of Ref.~\cite{Alba_2019}, up to the factor $\Theta(\mp v_{m,\zeta=0})$, which selects the
quasiparticles coming from the left ($-$) or the right ($+$) part.

The contribution $S_+^{th}$ of the right half of the chain is the thermodynamic entropy transported from the right to
the left part. The rates of exchange of thermodynamic entropy $S_\pm^{th}$ read  
\begin{multline}
	\frac{S^{th}_\pm(t)}{t} \; = 
	\sum_{m=1}^{+\infty} \int_{-\pi/2}^{\pi/2} d\lambda
	\Theta \left[
	\mp v_{m, \zeta=0}(\lambda)
	\right]\\ \times
	\left| v_{m, \zeta=0}(\lambda) \right| \; s_m^{(\zeta=0)}(\lambda).
	\label{eq:entropyright}
\end{multline}
We conjecture that the entanglement dynamics is described by 
\begin{equation}
	\frac{S^{conj}}{t}=\frac{S_-^{QP}+S_+^{th}}{t}. 
	\label{eq:finalconjecture}
\end{equation}
Finally, we observe that
\begin{equation}
	\label{eq:bounds}
	\frac{S^{QP}_+(t)+S^{QP}_-(t)}{t}  \leq \frac{S^{conj}(t)}{t} \leq \frac{S^{th}_-(t)+S^{th}_+(t)}{t}.
\end{equation}
Eq.~\eqref{eq:bounds} follows from the fact that the factor $\mbox{sgn}(\lambda)$ removes the contribution of some of the quasiparticles. 

Table~\ref{tab:entropy} compares the different contributions for the bipartitioning protocol in which the left part is prepared in the
N\'eel state or in the $n$-rainbow states with $n \in \set{1, 2, 3}$, and the right part in the infinite-temperature
state. Details on the numerical solution of the GHD equations are reported in Appendix~\ref{app:numerics}.

In all cases we find, within our numerical precision, $S^{QP}_-/t \approx S^{th}_-/t$.
Hence, the difference between our conjecture and the rate of exchange of thermodynamic entropy cannot be resolved.
This signals that most of the pairs of quasiparticles with opposite rapidities $\pm\lambda$ created in the left part end up on different sides of the
interface. On the other hand, $S^{QP}_+$ differs significantly from $S_+^{th}$, implying that for a significant fraction of the
pairs $(\lambda,-\lambda)$ originating in the right part both members enter the left part. However,
Table~\ref{tab:entropy} suggests that $S_+^{th}\approx S^{QP}_+$ at large $n$. This is expected, because
the GGE describing the $n$-rainbow state approaches the infinite-temperature state for $n\to\infty$, so that the left and
right parts of the chain are described by the same thermodynamic state.
We also observe that $S_-^{QP}$ relies on the assumption that the dynamics in the left part
produces entangled pairs.

We recall that, after the homogeneous quench from the $n$-rainbow state in the XX model,
the slope vanishes in the limit $n\to\infty$ (see Sec.~\ref{sec:homoent}).
This means that in the XX chain the contribution of the left part vanishes, and the slope of the entropy
tends to $\kappa_\infty/2$, the rate at which thermodynamic entropy is transported from the right part.
The main result of this section is that this mechanism fails in the XXZ chain:
our numerical results indicate that, in contrast with the XX chain,
the slope approaches $\kappa_\infty$ in the large-$n$ limit, as predicted by Eq.~\eqref{eq:finalconjecture}.
Fig.~\ref{fig:entropyconjecture} shows tDMRG simulations for the
bipartitioning protocol. The top panels show results for the left part
prepared in the integrable N\'eel state, for $\Delta=3$ and
$\Delta=10$.
The dotted line is the standard
quasiparticle prediction $(S_+^{QP}+S_-^{QP})/t$, with $S_\pm^{QP}$ given in Eq.~\eqref{eq:Spm},
which underestimates the slope of the entropy growth.
Indeed, in the contribution of the right part, the contribution
of the pairs of quasiparticles with opposite rapidities that both end up in the left subsystem has to be restored.
The dashed line is the conjecture~\eqref{eq:finalconjecture},
which is in excellent agreement with the tDMRG data. Fig.~\ref{fig:entropyconjecture} also shows results for the
$1$-rainbow state, which is also integrable, being a two-site product state.
Again, the agreement with the conjecture is excellent. We now discuss the $n$-rainbow states with $n>1$. The tDMRG data for $n\in \set{2,3}$
show that the slope of the entropy does not decrease upon increasing $n$. This is surprising, because
in the multiplet picture developed for the XX chain the entanglement produced by the $n$-rainbow state on the
left vanishes in the large-$n$ limit. One would then expect the growth of the entropy
to be determined only by the transport of thermodynamic entropy from the right part, i.e., by $S^{th}_+/t$.
Instead, the tDMRG data for $n\in \set{2,3}$ are in excellent agreement with the conjecture~\eqref{eq:finalconjecture}.
Hence, the contribution of the left part of the chain, which is prepared in the $n$-rainbow
state, does not vanish, and entanglement spreading in the interacting case differs from
the multiplet mechanism of free models. This is consistent with the results for the
homogeneous quench in Sec.~\ref{sec:homoent}: in both settings, the quasiparticles produced by the
$n$-rainbow state contribute to the entanglement as if they were created in entangled pairs. 
Moreover, already for $n=3$ the standard quasiparticle prediction $S^{QP}$ basically
coincides with the sum $S_+^{th}+S_-^{th}$ of the rates of exchange of thermodynamic entropy between the two
parts of the chain.
Notice, however, that for nonintegrable initial states the standard quasiparticle picture is
expected to fail, as we verify explicitly in Appendix~\ref{app:violations}
for the fat N\'eel states. Hence, $S_-^{QP}$ in
Eq.~\eqref{eq:finalconjecture} cannot be interpreted as the entanglement of shared pairs of quasiparticles. 

\section{\label{sec:conclusions} CONCLUSIONS}

We investigated entanglement spreading, thermalization, and transport after quenches in the XXZ chain
from the $n$-rainbow states, which are macroscopically homogeneous states formed by long-range Bell
pairs, and are not integrable initial states for $n>1$.

Our main result concerns entanglement spreading. In the XX chain the quench produces entangled
multiplets of $2n$ quasiparticles, and the entanglement dynamics is described by a generalized
quasiparticle picture, with correlations given by Eq.~\eqref{eq:formfactorspecies}. Although the
off-diagonal correlations within a multiplet vanish as $1/n$, they suppress the growth of the entanglement
entropy, whose slope vanishes as $n\to\infty$. In the interacting XXZ chain this suppression is absent. After
homogeneous quenches, the entanglement entropy grows at the same rate at which Yang-Yang entropy is exchanged across the cut, and this
rate tends to a finite value as $n\to\infty$. The same holds in the bipartite protocol in which the
$n$-rainbow state is joined with an infinite-temperature state. In this case the system is globally
mixed, and we conjectured that the von Neumann entropy of the initially pure half is the sum of a
pair-structure contribution of the quasiparticles created in the pure half and of the thermodynamic entropy transported from
the thermal half.
This leads to Eq.~\eqref{eq:finalconjecture}, whose predictions are in excellent agreement with tDMRG simulations.
Thus, the multiplet mechanism that governs entanglement spreading in the free-fermion chain does not
seem to survive interactions. To the best of our knowledge, these are among the first results on
entanglement spreading in interacting integrable systems after quenches from nonintegrable initial states.

Several questions remain open. First, it would be important to understand the mechanism that
suppresses the multiplet effects in the interacting chain. A natural scenario is that scattering with
the other excitations dephases the correlations between the members of a multiplet (see
Sec.~\ref{sec:homoent}). This could be tested by studying the crossover from the free-fermion to the
interacting behaviour at weak interactions, and by considering larger values of $n$ and longer times.
Second, quantities that are more sensitive to the structure of the correlations than the von Neumann
entropy, such as the tripartite mutual information~\cite{Caceffo, AlbaCalabrese2019EPL, AlbaCalabrese2019Scrambling} and the R\'enyi
entropies~\cite{AlbaCalabrese2017Renyi, Mestyan2018Renyi, bertini2022growth}, could reveal multiplet effects that are invisible in the von Neumann
entropy. 

\begin{acknowledgments}
VA has been supported by the project ``Artificially devised many-body quantum dynamics in low dimensions - ManyQLowD'' funded by the MIUR Progetti di Ricerca di Rilevante Interesse Nazionale (PRIN) Bando 2022 - grant 2022R35ZBF. This article is based upon work from COST Action ``Many-body Open Quantum Systems'' (QOpen) CA24109 supported by COST (European Cooperation in Science and Technology). VA gratefully acknowledges support from the Simons Center for Geometry and Physics, Stony Brook University, at which some or all of the research for this paper was performed during the Program ``Complexity, information, and tractable simulations of quantum many-body dynamics'' 01/06-02/07 2026.
We acknowledge the use of Claude (Anthropic) and ChatGPT, for help in
editing the manuscript, drafting parts of the text, preparing some of the figures, and
performing independent numerical checks. All results and the text were verified by
the authors, who take full responsibility for the content of this work.
\end{acknowledgments}

\appendix

\section{\label{sec:Algebraic} Construction of the GGE from the transfer matrix}

Here we review the derivation of the GGE from initial-state data. 
We employ the quantum inverse scattering method (QISM), or algebraic Bethe ansatz (ABA),
first introduced in Ref.~\cite{ABAfirst} (see also Refs.~\cite{ABAmath,Faddeev, ABAslavnov,Introduction,Quasilocal, Drinfeld}).
A key insight of the QISM is that the infinitely many (quasi)local conserved
charges of an integrable model can be traced back to an underlying symmetry, encoded in an appropriate algebraic structure $\mA$.
For the XXZ model the latter is the quantum group $U_q(\mathfrak{sl}(2))$ \cite{QuantumGroups}.
The algebra $\mA$ must admit a family of representations $r(s, \lambda)$, labelled by a discrete index $s$ and a continuous {\itshape spectral parameter} $\lambda$.
Here we can restrict to the unitary spin-$s$ representations of $U_q(\mathfrak{sl}(2))$, with $s\in\frac{1}{2}\N^{>0}$.
The Hilbert space $\mH_j$ of the $j$-th site of the lattice carries the fundamental ($s=1/2$) representation of $U_q(\mathfrak{sl}(2))$.
In addition, one introduces ancillary spaces $\nu_s$, which carry spin-$s$ representations of $U_q(\mathfrak{sl}(2))$.

The $U_q(\mathfrak{sl}(2))$ commutation relations are
\begin{equation}
		 [\hJ^z,\hJ^\pm]=\pm \hJ^\pm ,
		\qquad
		[\hJ^+, \hJ^-]=[2\hJ^z]_q ,
\end{equation}		 
where 
\begin{equation}
		[x]_q\coloneq \frac{q^x - q^{-x}}{q - q^{-1}} =
		\frac{\sinh(x \log q)}{\sinh(\log q)} .
\end{equation}
$[x]_q$ is the {\itshape $q$-deformation} of $x$.
As a Hopf algebra, $U_q(\mathfrak{sl}(2))$ has the comultiplication \cite{QuantumGroups, Drinfeld2}
\begin{equation}
	\begin{split}
		& \Delta\left[ \hJ^z \right] \coloneq \hJ^z \otimes \hId + \hId \otimes \hJ^z ,
		\\
		& \Delta\left[ \hJ^\pm \right] \coloneq \hJ^\pm \otimes q^{\hJ^z} + q^{-\hJ^z} \otimes \hJ^\pm .
	\end{split}
\end{equation}
A family of unitary, finite-dimensional representations with $s \in \frac{1}{2}\Z^{>0}$ and $\mbox{dim} \nu_s = 2s+1$ is obtained by denoting by $\ket{k}$, with \mbox{$k \in \lbrace 0, ..., 2s \rbrace$}, the $2s+1$ eigenstates of $\hs^z$ with eigenvalues $s^z\in\{s, ..., -s\}$, respectively~\cite{Symmetry, Quasilocal}:
\begin{equation}
	\begin{split}
		& \hs^z \coloneq r_s\left[ \hJ^z \right] = \sum_{k=0}^{2s} (s-k) \ket{k}\bra{k} ,
		\\
		& \hs^+ \coloneq r_s\left[ \hJ^+ \right] = \sum_{k=0}^{2s-1} \sqrt{[2s-k]_q [k+1]_q} \ket{k}\bra{k+1} ,
		\label{eq:sdeformed}
		\\
		& \hs^- \coloneq r_s\left[ \hJ^- \right] = \sum_{k=0}^{2s-1} \sqrt{[2s-k]_q [k+1]_q} \ket{k+1}\bra{k} .
	\end{split}
\end{equation}
These concrete representations are spin-flip symmetric:
\begin{equation}
	\begin{split}
		\hat{P}_s \hs^\pm \hat{P}_s^{-1} \; & = \; \hs^\mp ,
		\qquad
		\hat{P}_s \hs^z \hat{P}_s^{-1} \; = \; -\hs^z ,
	\end{split}
\end{equation}
with 
\begin{equation}
		 \hat{P}_s \; \coloneq \; \sum_{k=0}^{2s} \ket{k}\bra{2s-k} .
\end{equation}
The elementary building block for the GGE is the family of spin$-s$ 
Lax matrices, each one acting on a spin$-s$ ancillary space $\nu_s$ and the physical single-site space $\mH_j$. 
They are given by 
\begin{subequations}
	\begin{equation}
		\hmL_s^{(j)}(\lambda)=\frac{1}{\sinh(\eta)}
		\left(
		\begin{array}{ll}
			\sin(\lambda + i \eta \hs^z) & i \sinh(\eta) \cdot \hs^- \\
			i \sinh(\eta) \cdot \hs^+ & \sin(\lambda -i\eta \hs^z)
		\end{array}
		\right) ,
		\label{eq:explicit_Lax_matrix}
	\end{equation}
	where $\eta = \log(q)$ and $\Delta = \cosh(\eta) > 1$.
	The $\Delta = 1$ case corresponds to the non-deformed $U(\mathfrak{sl}(2))$ and the Lax matrix becomes simply
	\begin{equation}
		\hmL_s^{(j)}(\lambda) \; = \; \lambda \, \hId \; + \; i\hat{\vec{\sigma}} \otimes 
		\hat{\vec{s}} \; = \; 
		\left(
		\begin{array}{cc}
			\lambda + i \hs^z & i \hs^- \\
			i \hs^+ & \lambda - i \hs^z \\
		\end{array}
		\right) .
		\label{eq:XXXLax_sgen}
	\end{equation}
\end{subequations}
These expressions highlight that $\mL_s(\lambda)$ is a $2\times 2$ 
matrix with respect to its physical indices, but each of its 
four elements is a $(2s+1)\times (2s+1)$ matrix on the auxiliary space $\nu_s$.
$\hmL_s^{(j)}(\lambda)$ preserves the total magnetization.
Moreover, denoting by $\hat{P}_j$ the spin-flip operator on $\mH_j$,
Eq.~\eqref{eq:explicit_Lax_matrix} shows that the Lax matrix is invariant under the
global spin flip $\hat{P}_j \hat{P}_s$.

For $s=0$ (trivial representation) $\hs^\pm=\hs^z=0$, hence $\hmL^{(j)}_0 \propto \hId_j \otimes \hId_{\nu_0}$,
and it is convenient to define it directly as a scalar,
\begin{equation}
    \mL_0(\lambda)\coloneq\frac{\sin(\lambda)}{\sinh(\eta)}.
    \label{eq:Lax_normalization}
\end{equation}
The associated scalar transfer matrix is $T_0(\lambda)\coloneq\mL_0(\lambda)^L$.
The monodromy matrix corresponding to a given $r(s, \lambda)$, 
acting on the ancillary space $\nu_s$ and the total Hilbert space of the lattice, is defined as
\begin{equation}
	\hM_s(\lambda) \; \coloneq \; \hmL_s(\lambda)^{\otimes L}
	\; \coloneq \; \hmL_s^{(L)}(\lambda) \; \otimes ... \otimes \;
	\hmL_s^{(1)}(\lambda) .
	\label{eq:Monodromy}
\end{equation}
Then, the  transfer matrix $\hT_s(\lambda)$ is defined as the trace of $\hM_s(\lambda)$ over the ancillary space
\begin{equation}
	\hT_s(\lambda) \; \coloneq \; \trace_{\nu_s} \hM_s (\lambda) .
	\label{eq:quantumTM}
\end{equation}
The Lax matrix satisfies an intertwining (RLL) relation with an $R$-matrix
that acts on two ancillary spaces and solves the Yang-Baxter equation.
As a consequence, all the transfer matrices commute: 
\begin{equation}
	\left[ \hT_s(\lambda), \hT_{s'}(\mu) \right] \; = \; 0 .
\end{equation}
This property allows one to construct an infinite set of quasilocal charges
in involution, which include the XXZ Hamiltonian for the Lax matrix~\eqref{eq:explicit_Lax_matrix}~\cite{FirstQuasilocal, Quasilocal}:
\begin{equation}
	\hX_s(\lambda)\coloneq  \frac{1}{2\pi i} \frac{\partial}{\partial \lambda}  \log \left( \frac{\hT^{[+]}_s(\lambda)}{T^{[2s+1]}_0(\lambda)}\right) , 
	\quad
	\lambda \in \R ,
	\label{eq:QuasiOriginal}
\end{equation}
This expression can be analytically continued, preserving quasilocality~\cite{Quasilocal}, to the entire ``physical strip'' $\mathcal{P}_\eta \coloneq \left\lbrace
\lambda \in \C \; | \; \imag(\lambda) \in \left( - \frac{\eta}{2} ,\frac{\eta}{2} \right) 
\right\rbrace $.
However, charges are Hermitian only for $\lambda \in \R$.

In Eq.~\eqref{eq:QuasiOriginal} we used the standard ABA notation
\begin{equation}
	\label{eq:traslation}
	f^{[k]}(\lambda)\coloneq f\left( \lambda + k\frac{i\eta}{2}\right) ,
	\quad
	\mbox{where}
	\quad
	[\pm]\equiv [\pm 1] .
\end{equation}
Observe that all transfer matrices are invariant under global spin flip, which is a symmetry 
inherited from the Lax matrices.
This means that all local and quasilocal charges are spin-flip symmetric.
	
The expectation values $\chi_s(\lambda)$ of these charges determine the root densities of the GGE
through the string-charge duality, Eq.~\eqref{eq:String_Charge_1} (see Sec.~\ref{sec:SCD}).

\subsection{Explicit evaluation of charges on periodic product states}

The presence of the logarithm in~\eqref{eq:QuasiOriginal} makes the derivation of the 
GGE root densities challenging. However, one can derive the inversion relation~\cite{StringCharge}  
\begin{equation}
	\lim_{L\rightarrow +\infty}
	\left( \frac{\hT^{[+]}_s(\lambda)}{T^{[2s+1]}_0(\lambda)}\right)^{-1}
	=\lim_{L\rightarrow +\infty} \frac{\hT^{[-]}_s(\lambda)}{T^{[-(2s+1)]}_0(\lambda)} ,
	\label{eq:inversion}
\end{equation}
which allows one to remove the logarithm from Eq.~\eqref{eq:QuasiOriginal}. 
Thus, one obtains 
\begin{equation}
	\hX_s(\lambda)\coloneq  \frac{1}{2\pi i} \left[ \frac{\partial}{\partial \mu}   \left( 
	\frac{\hT^{[-]}_s(\lambda)}{T^{[-(2s+1)]}_0(\lambda)}
	\frac{\hT^{[+]}_s(\mu)}{T^{[2s+1]}_0(\mu)}
	\right) \right]_{\lambda=\mu} .
	\label{eq:Quasilocal}
\end{equation}
Still, for generic initial states $\ket{\Psi}$ the expectation values of the quasilocal charges $\chi_{s}(\lambda) \coloneq
\lim_{L \rightarrow +\infty}\frac{1}{L}\bra{\Psi}\hX_{s}(\lambda)\ket{\Psi}$
are out of reach of both analytical and numerical methods.
An efficient method is available for periodic product states
(which include the $n$-rainbow states) of the form 
\begin{equation}
	\ket{\Psi}=\ket{\psi}^{\otimes L/2n}
	\quad
	\mbox{with $\ket{\psi}$ state in $\mH_{[1,2n]}$} .
	\label{eq:periodic_state}
\end{equation}
We define the composite, normalized Lax operator on site $j$ as
\begin{equation}
	\hLL_{s}^{(j)}(\lambda, \mu) \; \coloneq \;
	\frac{\hmL^{(j)[-]}_s(\lambda)}{\mL^{[-(2s+1)]}_0(\lambda)} \frac{\hmL^{(j)[+]}_s(\mu)}{\mL^{[2s+1]}_0(\mu)}
	\label{eq:composite_Lax}
\end{equation}
Unlike in the monodromy matrix~\eqref{eq:Monodromy}, where the Lax matrices act on different physical sites and on the same ancillary space, here they act on the same physical site $j$ and on two spin-$s$ ancillary spaces $\nu^\pm_s$. This implies 
\begin{equation}
		 \hLL_{s}(\lambda, \mu)^{\otimes L} \; = \;
		\left(\frac{\hmL^{[-]}_s(\lambda)}{\mL^{[-(2s+1)]}_0(\lambda)} \right)^{\otimes L}
		\left(\frac{\hmL^{[+]}_s(\mu)}{\mL^{[2s+1]}_0(\mu)} \right)^{\otimes L} ,
		\label{eq:from_T_to_L}
\end{equation}
and 
\begin{equation}
		 \trace_{\nu_s^- \otimes \nu_s^+}\left[\hLL_{s}(\lambda, \mu)^{\otimes L}\right] =
		\frac{\hT^{[-]}_s(\lambda)}{T^{[-(2s+1)]}_0(\lambda)}
		\frac{\hT^{[+]}_s(\mu)}{T^{[2s+1]}_0(\mu)} .
\end{equation}
Conveniently, the $\hLL_{s}^{(j)}(\lambda, \mu)$ are single-site operators, 
which means that if $|\Psi\rangle$ is of the form~\eqref{eq:periodic_state}, 
the calculation reduces to an operator on the auxiliary space $\mathcal V_s\coloneq\nu_s^-\otimes\nu_s^+$:
\begin{equation}
    \hTT_s(\lambda,\mu)\coloneq
    \braket{\psi|\hLL_s(\lambda,\mu)^{\otimes 2n}|\psi}.
    \label{eq:unitcell_transfer}
\end{equation}
The charge density can then be obtained from
\begin{equation}
    \chi_s(\lambda)=\frac{1}{2\pi iL}
    \left.\partial_\mu\trace_{\mathcal V_s}
    \left[\hTT_s(\lambda,\mu)^{L/(2n)}\right]\right|_{\mu=\lambda}.
    \label{eq:single_sector}
\end{equation}
The Lax matrices preserve the total magnetization on $\nu^\pm_s \otimes \mH_j$.
Hence, $\hTT_{s}(\lambda,\mu)$ preserves the total magnetization on $\nu_s^- \otimes \nu_s^+$,
and it is invariant under spin flip on the two ancillary spaces.
Since Eq.~\eqref{eq:inversion} implies
\begin{equation}
	\lim_{L\rightarrow +\infty}
	\bra{\Psi}\frac{\hT^{[+]}_s(\lambda)}{T^{[2s+1]}_0(\lambda)}
	\frac{\hT^{[-]}_s(\lambda)}{T^{[-(2s+1)]}_0(\lambda)} \ket{\Psi}\; = \;1 ,
\end{equation}
$\hTT_{s}(\lambda,\lambda)$ has a single eigenvalue equal to $1$, while
all the others have magnitude smaller than $1$.
Being nondegenerate, the former corresponds to an eigenvector with zero magnetization, and
\begin{equation}
	\chi_{s}(\lambda) = \frac{1}{2n}\frac{1}{2\pi i} \left[ \partial_\mu
	t^{(max)}_{s}(\lambda,\mu)
	\right]_{\lambda=\mu}
	\label{eq:t_max}
\end{equation}
where $t^{(max)}_{s}(\lambda,\mu)$ is the largest-magnitude eigenvalue of $\hTT_{s}(\lambda,\mu)$, satisfying \mbox{$\lim_{\mu \rightarrow \lambda} t^{(max)}_{s}(\lambda,\mu) = 1$}.
Let $\hTT^{(0)}_{s}$ be $\hTT_{s}$ restricted to its zero-magnetization block.
Since we are assuming that $t^{(max)}_{s}(\lambda,\mu)$ is nondegenerate as well, at least for $\lambda \approx \mu$, its eigenvector has zero total magnetization.
This enables computing $ \partial_\mu t^{(max)}_{s}(\lambda,\mu)_{\lambda=\mu}$ from $\hTT^{(0)}_{s}$.
To avoid the diagonalization and explicit evaluation of $\partial_\mu t^{(max)}_{s}(\lambda,\mu)$, one can exploit Jacobi's formula~\cite{Fagotti_2014}
\begin{equation}
	\partial_\mu \det \hA(\mu) = \trace\left[ \mbox{adj}\left( \hA(\mu)\right) \partial_\mu \hA(\mu) \right]
\end{equation}
where $\mbox{adj}(\hA)$ denotes the {\itshape adjugate}, i.e., the transpose of the cofactor matrix of $\hA$.
Indeed, given that $t^{(max)}_{s}$ for $\lambda \approx \mu$ is an eigenvalue of $\hTT^{(0)}_{s}$, it follows that $\; 0=\det(\hTT^{(0)}_{s}-\hId t^{(max)}_{s})$.
For a compact numerical expression, define the adjugate matrix
\begin{equation}
    \mathsf{A}_s(\lambda)\coloneq
    \operatorname{adj}\left[\hTT_s^{(0)}(\lambda,\lambda)-\hId\right]
    \label{eq:adjugate_matrix}
\end{equation}
and the derivative matrix
\begin{equation}
    \mathsf{B}_s(\lambda)\coloneq
    \left.\partial_\mu\hTT_s^{(0)}(\lambda,\mu)\right|_{\mu=\lambda}.
    \label{eq:charge_derivative_matrix}
\end{equation}
The charge density then reads~\cite{StringCharge}
\begin{equation}
    \chi_s(\lambda)=
    \frac{\trace_{\mathcal V_0}
    \left[\mathsf{A}_s(\lambda)\mathsf{B}_s(\lambda)\right]}
    {4\pi i n\,\trace_{\mathcal V_0}\mathsf{A}_s(\lambda)}.
    \label{eq:computeXNum}
\end{equation}

\section{\label{app:XXdetails} Multiplet quasiparticle picture in the XX chain: technical details}

\subsection{Form factor of the $n$-rainbow state}

In the definition of the form factor $F^{(n)}_{kk'}$ in Sec.~\ref{sec:exact}, the lattice sites
$\Lambda$ are labelled as
\begin{equation}
	\Lambda=[j_{min},j_{min}+(L-1)],
	\qquad
	j_{min}:=-n+1/2, 
\end{equation}
and  we defined the off-diagonal contribution $\widetilde F_{kk'}^{(n)}$ as 
\begin{equation}
	\begin{split}
	& \widetilde F^{(n)}_{kk'} \; = \;
	\frac{1}{2L} \sum_{j \in \Lambda} e^{ikj} e^{-ik'\sigma(j)} (-1)^{\Delta_j+1} =
	\\
	& = \left[\frac{1}{N} \sum_{\ell=0}^{N-1} e^{i\ell 2n(k-k')} \right]
	\real \left[
	\frac{e^{i\frac{1}{2}(k+k')}}{2n}
	\sum_{\Delta=0}^{n-1}
	\left( -e^{i(k+k')} \right)^\Delta
	\right] .
	\end{split}
	\label{eq:ffdef}
\end{equation}
Here we replaced $j \in \Lambda$ by the two indices $\ell$, which labels the $N=L/(2n)$ unit cells along the chain, and $\Delta$, which labels the arcs of a unit cell from the innermost to the outermost. 
From Eq.~\eqref{eq:ffdef} it is straightforward to obtain Eq.~\eqref{eq:F-final}.

\subsection{Ordering of the species and kinematic factors}

To compute Eq.~\eqref{eq:quasibifree} explicitly for generic $n$, i.e., the slope of the entanglement growth after the quench from $\ket{\Omega} \otimes \ket{R_n}$, we focus on the $2n$ quasiparticles of a multiplet produced in the right subsystem.
	
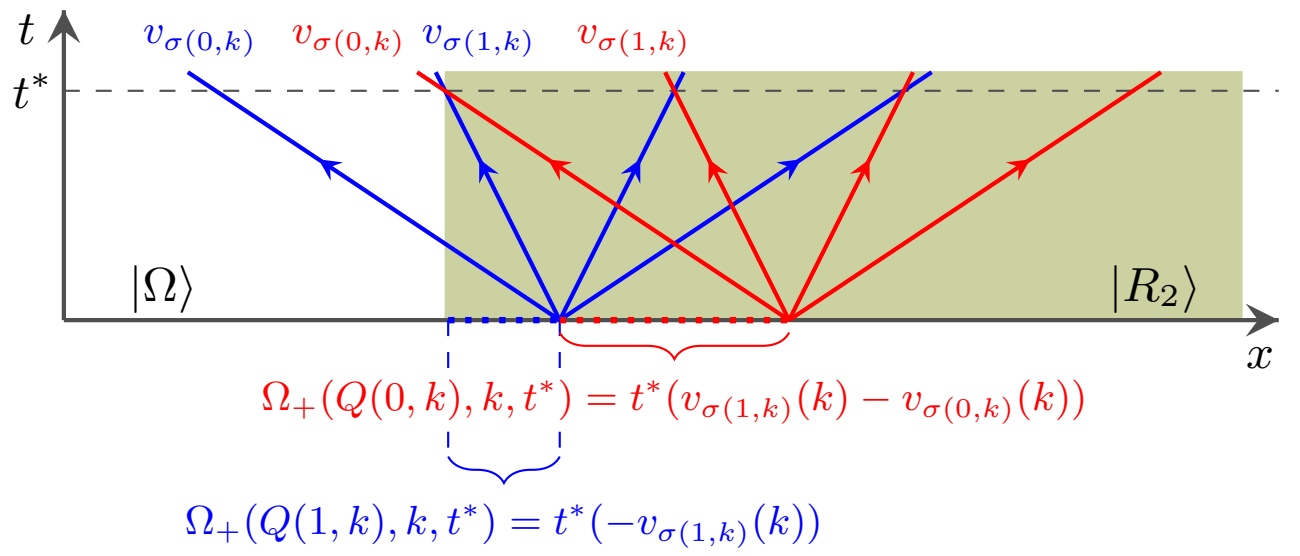
\begin{figure}[t]
	\centering
	\begin{tikzpicture}[scale=0.41,baseline={([yshift=-0.9ex]current bounding box.center)}]
		\withoutrectangle{12.58}{2}{6.42}{2}{colO}
		\gridAxis{0}{0}{0}{5}
		\gridAxis{0}{0}{19.6}{0}
		\draw [dashed,colLines] ({0},{3.7}) -- ({19.6},{3.7});
		\foreach \x in {8}{
			\gridArrowBlue{\x}{0}{\x-6}{4}
			\gridArrowBlue{\x}{0}{\x+6}{4}
			\gridArrowBlue{\x}{0}{\x-2}{4}
			\gridArrowBlue{\x}{0}{\x+2}{4}
			\node at ({\x-5.8},4.5) {\scalebox{0.85}{\color{blue} $v_{\sigma(0, k)}$}};
			\node at ({\x-1.3},4.5) {\scalebox{0.85}{\color{blue} $v_{\sigma(1, k)}$}};
		}
		\draw [very thick,dotted,blue] ({6.2},{0}) -- ({8},{0});
		\draw [dashed,blue] ({6.2},{0}) -- ({6.2},{-0.8});
		\draw [dashed,blue] ({8},{0}) -- ({8},{-0.8});
		\draw [dashed,blue] ({6.2},{-1.8}) -- ({6.2},{-2.2});
		\draw [dashed,blue] ({8},{-1.8}) -- ({8},{-2.2});
		\foreach \x in {11.7}{
			\gridArrowRed{\x}{0}{\x-6}{4}
			\gridArrowRed{\x}{0}{\x+6}{4}
			\gridArrowRed{\x}{0}{\x-2}{4}
			\gridArrowRed{\x}{0}{\x+2}{4}
			\node at ({\x-7.1},4.5) {\scalebox{0.85}{\color{red} $v_{\sigma(0, k)}$}};
			\node at ({\x-2.5},4.5) {\scalebox{0.85}{\color{red} $v_{\sigma(1, k)}$}};
		}
		\draw [very thick,dotted,red] ({8},{0}) -- ({11.7},{0});
		\node at (-0.6,4.7) {\scalebox{1}{$t$}};
		\node at (-0.5,3.7) {\scalebox{1}{$t^*$}};
		\node at (19.3,-0.6) {\scalebox{1}{$x$}};
		\node at (1.6,0.5) {\scalebox{1}{$\ket{\Omega}$}};
		\node at (17.6,0.5) {\scalebox{1}{$\ket{R_2}$}};
		\draw [decorate, red, decoration={brace,amplitude=5pt},xshift=0pt,yshift=0pt]
			(11.7,-0.2) -- (8,-0.2) node [midway,yshift=-13,inner sep=0]{\scalebox{0.85}{$\Omega_+(Q(0,k),k,t^*)=t^*(v_{\sigma(1,k)}(k)-v_{\sigma(0,k)}(k))$}};
		\draw [decorate, blue, decoration={brace,amplitude=5pt},xshift=0pt,yshift=0pt]
			(8,-2.2) -- (6.2,-2.2) node [midway,yshift=-13,inner sep=0]{\scalebox{0.85}{$\Omega_+(Q(1,k),k,t^*)=t^*(-v_{\sigma(1,k)}(k))$}};
	\end{tikzpicture}
	\caption{Quasiparticle-picture entanglement dynamics $S^{(2)}(t)$ after the quench from the 
		state $\ket{\Omega} \otimes \ket{R_2}$, with $\ket{\Omega}$ vacuum state.
		Only the quadruplets produced in the dotted red (or blue) segment,
		of length $\Omega_+(Q(0,k),k,t)$ (or $\Omega_+(Q(1,k),k,t)$),
	    lead to configurations such that at time $t^*$ 
		only the fastest quasiparticle is (or the two fastest quasiparticles are) 
		in the left-hand not-shaded region. 
	\label{fig:kinem}}
\end{figure}

	Quasiparticles $\{0, ..., n-1\}$ are always left-moving and the 
	remaining ones are right-moving with opposite velocities.
	Thus, only quasiparticles with $A \in \set{0, ..., n-1}$ can reach the left subsystem.
	However, the velocities $\{v_A(k)\}$ can cross
	each other upon varying $k \in (0, \pi/ n ]$.
	Thus, it is convenient to reorder them by defining a map $\sigma(A, k)$ such that 
	\begin{equation}
		|v_{\sigma(0, k)}(k)| \; \geq \; ... \; \geq \; |v_{\sigma(n-1, k)}(k)| .
	\end{equation}
	This allows us to label the kinematically allowed configurations by
	$q \in \set{0, ..., n-1}$: 
	\begin{equation}
    Q(q,k)\coloneq\{\sigma(0,k),\ldots,\sigma(q,k)\}.
    \label{eq:multiplet_configuration}
\end{equation}
The corresponding kinematic factor follows from simple geometric considerations (see Fig.~\ref{fig:kinem}):
\begin{equation}
    \begin{split}
        \Omega_+(Q(q,k),k,t)
        ={}&t\Big[\Theta[(n-1)-q]\\
        &\times v_{\sigma(q+1,k)}(k)-v_{\sigma(q,k)}(k)\Big].
    \end{split}
    \label{eq:kinematic}
\end{equation}
	In particular, in the interval $k\in[0,\pi/2n]$ the map reorders velocities as
	$\{...,v_{n+1},v_1,v_{n-1},v_0\}$.
	In the interval $k\in[\pi/(2n),\pi/n]$ pairs are inverted:
	$\{...,v_1,v_{n-2},v_0,v_{n-1}\}$.
	If $n$ is odd, $v_{(n-1)/2}$ is unpaired: it is the fastest particle, and it always occupies the leftmost position in the above sets.
	Explicitly, let us define $p_{max}=\lfloor n/2 \rfloor$. If $n$ is odd, $\sigma(0,k)=p_{max}$.
	Then, for both even and odd $n$ and for all $p \in \set{0, ..., p_{max}-1}$, if $k\in [0,\pi/2n]$
	\begin{equation}
		\left\lbrace
		\begin{array}{l}
			\sigma((n-1)-2p) = p , \\
			\sigma((n-1)-2p-1) = (n-1)-p . \\
		\end{array}
		\right.
		\label{eq:orderingmap}
	\end{equation}
	while the two images are exchanged if $k\in [\pi/2n,\pi/n]$.
	
	After this reordering, the quasiparticle picture can be applied straightforwardly:
	$S^{(n)}(t)$ is given by Eq.~\eqref{eq:quasibifree},
	with $\sum_Q$ replaced by $\sum_{q=0}^{n-1}$.

\subsection{Quench from $|R_1\rangle\otimes|R_2\rangle$: entanglement contribution of $|R_2\rangle$}

The prescription of the previous section allows one to compute $S^{(n)}(t)$ for arbitrary $n$.
As an illustrative example, here we detail the computation of $S^{(2)}(t)$,
for which the diagonalization of $F^{(2)}_Q$ can be performed by hand,
and only two configurations, $q \in \set{0,1}$, contribute.

The only left-moving quasiparticles, which can reach $(-\infty, 0]$,
are those of species $0$ and $1$, with negative group velocities. The reduced 
Brillouin zone is $BZ_2 = (0, \pi/2]$.
Group velocities are 
\begin{equation}
	v_0(k)= - \sin(k) ,
	\quad
	v_1(k) = -\cos(k) .
\end{equation}
Thus, $|v_1(k)| >|v_0(k)| \leftrightarrow k < \frac{\pi}{4}$, in agreement with \eqref{eq:orderingmap}.
The kinematically-permitted partitions are $Q(1, k) = \{0,1\}$ and $Q(0, k)$, which is $Q = \{0\} \leftrightarrow k > \frac{\pi}{4}$, and 
$Q = \{1\} \leftrightarrow k < \frac{\pi}{4}$. Their kinematic factors are
\begin{equation}
	\begin{split}
	&\Omega_+(Q(0,k), k, t) \; = \;
	t \, \,
	\left|\sin(k) - \cos(k) \right| ,
	\\
	&\Omega_+(Q(1,k), k, t) \; = \;
	t \, \, \min\left[\sin(k), \cos(k) \right] .
	\end{split}
\end{equation}
The form factor~\eqref{eq:formfactorspecies} 
restricted to the first two species is 
\begin{equation}
	F^{(2)}(k) \; = \;
	\left[
	\begin{array}{cc}
		F_{00}^{(2)}(k) & F_{01}^{(2)}(k) \\
		F_{01}^{(2)}(k) & F_{11}^{(2)}(k) \\
	\end{array}
	\right] ,
	\label{eq:F-restr}
\end{equation}
with matrix elements given by 
\begin{align}
	& F^{(2)}_{00}(k) \; = \;
	\frac{1}{2} \; + \;
	\sin^2 (k) \cos(k) , \\
	& F^{(2)}_{11}(k) \; = \;
	\frac{1}{2} \; - \;
	\sin (k) \cos^2(k),  \\
	& F^{(2)}_{01}(k) \; = \;
	\frac{1}{4} \frac{\cos^2(2k)}{\cos \left[ k + \frac{\pi}{4}\right]} .
\end{align}
The entropy content of the configurations with exactly one quasiparticle in the left region is 
\begin{multline}
	s(k, \{i\}) \; =  \;
	-  F^{(2)}_{ii}(k) \log F^{(2)}_{ii}(k) \; \\- \;
	\left( 1- F^{(2)}_{ii}(k) \right) \log\left( 1- F^{(2)}_{ii}(k) \right) , 
\end{multline}
with $i \in \set{0, 1}$. The contribution of $Q=\{0,1\}$ is 
\begin{multline}
	s^{(2)}(k, \{0, 1\}) \; =  \;
- \sum_{i \in \{1, 2\}} \Big[\lambda_i (k) \log \lambda_i(k) \; \\+ \;
\left( 1- \lambda_i(k) \right) \log\left( 1- \lambda_i(k) \right) \Big],
\end{multline}
where $\lambda_{i}$ are the eigenvalues of~\eqref{eq:F-restr}, which are easily computed by hand since it is simply a $2\times 2$ matrix. 
Finally, the explicit evaluation of Eq.~\eqref{eq:quasibifree} yields
\begin{equation}
	\frac{S^{(2)}(t)}{t} \; \approx \; 0.137 .
\end{equation}
The contributions of the configurations with one and two particles in
the left subsystem are of the same order, namely $0.0738$ and $0.0635$, respectively. 

\section{\label{app:charges} Quasilocal charges for the $n$-rainbow and the infinite-temperature states}

\subsection{Charges and root densities for the $n$-rainbow state}

We first use Eq.~\eqref{eq:single_sector} to determine the root densities
describing the quench from the $n$-rainbow state in the XXZ chain.
In this case, $\ket{\psi}$ in Eq.~\eqref{eq:periodic_state} is the state of the $2n$-site unit cell.
We can build \mbox{$\hTT^{(n)}_{s}(\lambda,\mu) $} recursively,
by adding at each step two sites forming a triplet Bell pair at
the boundary of the unit cell (i.e., a new outer arc, see Fig.~\ref{fig:cartoon}).
Introducing the notation $\hLL^{(k)}_{s, ij}$ with $(i,j)$ indexing 
the four elements of $\hLL^{(k)}_s$ as a $2 \times 2$ matrix on $\mH_k$, we have 
\begin{multline}
	 \hTT^{(1)}_{s} \; = \; \frac{1}{2}
	\left[
	\hLL_{s, 11}^{\left(-\frac{1}{2}\right)} \hLL_{s, 22}^{\left(\frac{1}{2}\right)} \; + \;
\hLL_{s, 22}^{\left(-\frac{1}{2}\right)} \hLL_{s, 11}^{\left(\frac{1}{2}\right)} \;  \;\right.\\+
\left.
	\hLL_{s, 12}^{\left(-\frac{1}{2}\right)} \hLL_{s, 21}^{\left(\frac{1}{2}\right)} \; + \;
	\hLL_{s, 21}^{\left(-\frac{1}{2}\right)} \hLL_{s, 12}^{\left(\frac{1}{2}\right)}
	\right]
\end{multline}
and the recursive relation is given by 
\begin{multline}
	 \hTT^{(n)}_{s} = \frac{1}{2}
	\left[
	\hLL_{s, 11}^{\left(-n+\frac{1}{2}\right)} \hTT^{(n-1)}_{s} \hLL_{s, 22}^{\left(n-\frac{1}{2}\right)} + 
\right.\\\left. 
	\hLL_{s, 22}^{\left(-n+\frac{1}{2}\right)} \hTT^{(n-1)}_{s} \hLL_{s, 11}^{\left(n-\frac{1}{2}\right)}+
	\hLL_{s, 12}^{\left(-n+\frac{1}{2}\right)} \hTT^{(n-1)}_{s} \hLL_{s, 21}^{\left(n-\frac{1}{2}\right)} 
\right.\\\left. 
	+
	\hLL_{s, 21}^{\left(-n+\frac{1}{2}\right)} \hTT^{(n-1)}_{s} \hLL_{s, 12}^{\left(n-\frac{1}{2}\right)}
\right]. 
	\label{eq:recursion}
\end{multline}
Eq.~\eqref{eq:recursion} combined with~\eqref{eq:computeXNum} and~\eqref{eq:String_Charge_1} allows one to obtain the root densities 
describing the GGE after the quench from the $n$-rainbow state. 

\subsection{Charges and root densities for the infinite-temperature state}

The string-charge duality can also be applied to derive the root densities
of the infinite-temperature ensemble, whose density matrix is $\hrho=\hId/2^L$. To compute the expectation values of the charges~\eqref{eq:QuasiOriginal}, we first define the one-site auxiliary transfer operator
\begin{equation}
    \hat{\bm T}_s(\lambda,\mu)\coloneq
    \frac12\trace_{\mH_j}\hLL_s^{(j)}(\lambda,\mu).
    \label{eq:thermal_transfer}
\end{equation}
Equivalently, $\hat{\bm T}_s=\braket{\hLL_s}_{\beta=0}$.
In terms of this operator, the thermal charge density is
\begin{equation}
    \chi_s^{th}(\lambda)=\frac{1}{2\pi iL}
    \left.\partial_\mu\trace_{\mathcal V_s}
    \left[\hat{\bm T}_s(\lambda,\mu)^L\right]\right|_{\mu=\lambda}.
    \label{eq:XXZthermal}
\end{equation}
As $L \rightarrow +\infty$, $\chi_s^{th}(\lambda)$ only depends on the highest-magnitude eigenvalue $\Lambda^{(max)}_s(\lambda, \mu)$ of $\frac{1}{2} \trace_{\mH_j}\hLL^{(j)}_s(\lambda, \mu)$  and its derivative.
In the XXX chain, Eq.~\eqref{eq:XXXLax_sgen} implies that this operator can be fully expressed in terms of the $U(\mathfrak{sl}(2))$ Casimir $\hat{C} : \nu_s^\pm \rightarrow \nu_s^\pm$ and its comultiplication $\Delta [ \hat{C} ]$, acting on $\nu_s^- \otimes \nu_s^+$.
Since the spectrum of these operators is known, this yields a strikingly simple expression for the infinite-temperature quasilocal charges \cite{random}.
Remarkably, the string-charge duality~\eqref{eq:String_Charge_1} then yields an analytic expression for the infinite-temperature root densities in the XXX chain.

Here we focus on the XXZ chain, for which
$\frac{1}{2} \trace_{\mH_j}\hLL^{(j)}_s(\lambda, \mu)$ does not reduce to the
Casimir operators of $U_q(\mathfrak{sl}(2))$~\cite{Drinfeld2, QuantumGroups, QuantumGroups2} and their comultiplication.
Indeed, one has 
\begin{multline}
	 \frac{1}{2}
	\trace_{\mH_j}
	\hLL^{(j)}_s \; = \;
	\Big\{\sin^{[-]}(\lambda + i \eta \hs^z) \otimes \sin^{[+]}(\mu + i \eta \hs^z)
	\\+ \sin^{[-]}(\lambda - i \eta \hs^z) \otimes \sin^{[+]}(\mu - i \eta \hs^z)\\ 
- (\hs^- \otimes \hs^+ + \hs^+ \otimes \hs^-)\sinh^2(\eta) \Big\}\mathcal{N}^{-1}(\lambda,\mu),
	\label{eq:tracecompositeLaxXXZ}
\end{multline}
where
\begin{equation}
	\mathcal{N}(\lambda, \mu) \; = \;  2 \sin^{[-(2s+1)]}(\lambda) \sin^{[2s+1]}(\mu). 
\end{equation}
Eq.~\eqref{eq:tracecompositeLaxXXZ} is not expressed in terms of the $U(\mathfrak{sl}(2))$ Casimir. 
The operator $\hat{\bm{T}}_{s}$ shares all the properties of the operator $\hTT_{s}$ defined in Appendix~\ref{sec:Algebraic} for periodic product states.
In particular, if \mbox{$\lambda=\mu$} then $\Lambda^{(max)}_s(\lambda, \lambda)=1$ and its eigenvector falls within the zero-magnetization space $\mathcal{V}_0$.
For example, if $s=1/2$ it is $(\ket{\uparrow \downarrow} + \ket{\downarrow \uparrow})/\sqrt{2}$.
Thus, we can restrict to
$\hat{\bm{T}}^{(0)}_{s} \coloneq
\hat{\bm{T}}_{s}|_{\mathcal{V}_0}$. 
The charges can then be computed numerically from the infinite-temperature analogue of Eq.~\eqref{eq:computeXNum},
since Eq.~\eqref{eq:XXZthermal} coincides with Eq.~\eqref{eq:single_sector} upon replacing $\hTT_{s} \rightarrow \hat{\bm{T}}_{s}$ and $2n \rightarrow 1$.
The infinite-temperature density matrix factorizes over sites, so this reduction uses a one-site cell. Define $\mathsf A_s^{th}$ and $\mathsf B_s^{th}$ by replacing $\hTT_s^{(0)}$ with $\hat{\bm T}_s^{(0)}$ in Eqs.~\eqref{eq:adjugate_matrix} and~\eqref{eq:charge_derivative_matrix}. The charge density is then
\begin{equation}
    \chi_s^{th}(\lambda)=
    \frac{\trace_{\mathcal V_0}
    \left[\mathsf{A}^{th}_s(\lambda)\mathsf{B}^{th}_s(\lambda)\right]}
    {2\pi i\,\trace_{\mathcal V_0}\mathsf{A}^{th}_s(\lambda)}.
    \label{eq:infinitetempcharges}
\end{equation}

\section{\label{app:halfpurtot} Quasiparticle picture for the half-purified bipartition}

\subsection{\label{app:halfpur} Sketch of the derivation of Eq.~\eqref{eq:Spm}}

We follow the steps of Ref.~\cite{Alba_2019},
with the entropy contributions described in Sec.~\ref{sec:entdynamics}.
First, we define $X_{m, \lambda}(x_0,t)$ as the position at time $t$ of an $m$-string
with rapidity $\lambda$ and initial position $X_{m, \lambda}(x_0,t_0) =  x_0$.
In the semiclassical picture, $X_{m,\lambda}$ is the solution of
\begin{equation}
	 \frac{\partial X_{m, \lambda}(x_0,t)}{\partial t} = v_{m, \zeta_{m, \lambda}(x_0/t, t)}(\lambda) ,
\end{equation}
where we defined $\zeta_{m,\lambda}(x_0/t,t)$ as 
\begin{equation}
	\zeta_{m, \lambda}(x_0/t, t) \coloneq \frac{X_{m, \lambda}(x_0,t)}{t}.
\end{equation}
Finally, one takes the limit $t_0/t \rightarrow 0$.
As in the derivation of Eq.~\eqref{eq:sol_interacting},
we assume that $\zeta = v_{m, \zeta}(\lambda)$ has a unique solution for all $(\lambda, m)$
(no exceptions are known).

The contribution to the von Neumann entropy of the quasiparticles originating in the left subsystem is 
\begin{multline}
	 \frac{S^{QP}_-(t)}{t} = 
	 \sum_{m=1}^{+\infty} \int_{-\pi/2}^{\pi/2}d\lambda \int_{-\infty}^{+\infty}d\zeta_0 \;\;
	s^{(0)}_m(\lambda, \zeta_0) 
	\\\times\Theta \left[
		\zeta_{m, \lambda} (\zeta_0,t)
	\right]
	\Theta \left[
		-\zeta_{m, -\lambda} (\zeta_0,t)
	\right]
	\Theta \left[
	-\zeta_0
	\right] 
	\label{eq:entropyleft}
\end{multline}
where 
\begin{equation}
	s^{(0)}_{m}(\lambda, \zeta_0) \; \coloneq \; 
	s_m^R(\lambda)\Theta\left[ \zeta_0 \right] \; + \; s_m^L(\lambda) \Theta\left[ - \zeta_0 \right]
\end{equation}
with $s_m^{L/R}$ the thermodynamic entropy densities of the left and right regions. In Eq.~\eqref{eq:entropyleft} 
the integral in $\zeta_0$ is over the points where the quasiparticles are created. The condition $\Theta(-\zeta_0)$ 
ensures that only quasiparticles produced in the left chain are counted. 
The conditions 
$\Theta(\zeta_{m,\lambda}(\zeta_0,t))$ and $\Theta(-\zeta_{m,-\lambda}(\zeta_0,t))$ ensure that the 
quasiparticles forming the entangled pairs at a generic time $t$ are in different subsystems, and hence 
contribute to the entanglement entropy. Starting from~\eqref{eq:entropyleft} one can employ the same 
strategy as in Ref.~\onlinecite{Alba_2019} to obtain the contribution $S_-^{QP}$ to the von Neumann entropy 
due to quasiparticles produced in the left chain, which is reported in Eq.~\eqref{eq:Spm}.

\section{\label{app:numerics} Numerical methods}

\subsection{Free-fermion chains}

The entanglement dynamics is determined from the time-evolution of the 
spatial correlation matrix restricted to subsystem $\Abb$~\cite{entcontent2, entcontent3}.
In the Heisenberg picture, $\hd^\dagger_k(t)=e^{i\epsilon_k t} \hd^\dagger_k(0)$.
In real space, one has 

\begin{equation}
	 C(t) = U(t) C(0) U^\dagger (t),
	\label{eq:dynXX1}
\end{equation}
where, as $L\to\infty$,
\begin{equation}
	U_{jm} = i^{j-m} J_{j-m}(t) ,
\end{equation}
with $J_\alpha(t)$ the Bessel functions of the first kind.
The initial condition $C(0)$ has the simple Bell-pair structure~\eqref{eq:corr}.
Then, given the eigenvalues $\set{\lambda_j}$ of the correlation matrix $C_\Abb(t)$ with indices restricted to $\Abb$, the bipartite entanglement is
\begin{equation}
	S_\Abb(t) = - \sum_j \left[ \lambda_j \log \lambda_j + (1-\lambda_j) \log(1-\lambda_j) \right] .
	\label{eq:dynXX2}
\end{equation}
Since we can only evolve finite chains,
while $U(t)$ is that of an infinite chain,
it is convenient to choose $\Abb$ as a segment initially prepared in $\ket{\psi_R}$, surrounded by two segments initially prepared in $\ket{\psi_L}$. All segments must contain several unit cells and be longer than the distance covered by the fastest quasiparticle in the time window considered.

\subsection{tDMRG simulations}

We implemented tDMRG~\cite{DMRG} using the ITensor library~\cite{ITensor} in Julia.
The time-evolution operator for a time step $\tau=0.1$ was implemented through a fourth-order Trotter decomposition~\cite{trotter4}:
\begin{equation}
	e^{-i\hH'\tau} \; = \; \hat{U}(\tau_1) \hat{U} (\tau_1) \hat{U} (\tau_2) \hat{U}(\tau_1) \hat{U} (\tau_1) \; + \; o(\tau^4)
	\label{eq:evolutor}
\end{equation}
Here the symmetric second-order step is
\begin{equation}
    \hat U(\tau)\coloneq
    e^{-i\hH'_{odd}\tau/2}e^{-i\hH'_{even}\tau}e^{-i\hH'_{odd}\tau/2},
    \label{eq:trotter_second_order}
\end{equation}
and the time increments are
\begin{equation}
    \tau_1=\frac{\tau}{4-4^{1/3}},
    \label{eq:trotter_tau1}
\end{equation}
and
\begin{equation}
    \tau_2=\tau-4\tau_1
    =\tau\left(1-\frac{1}{1-4^{-2/3}}\right).
    \label{eq:trotter_tau2}
\end{equation}
If $\ell$ is the size of the left half of the chain (with open boundary conditions), the size of the purified right subsystem is $2\ell$.
The purification is performed by alternating physical and ancillary sites in the right subsystem,
and by forming triplet Bell pairs between adjacent sites.
In all simulations we used $\ell \sim 50$, chosen to be a multiple of $2n$.
For instance, with this choice $\ket{R_1}\otimes \kket{\psi_R}$ is a homogeneous initial state,
which evolves with an inhomogeneous Hamiltonian: the Hamiltonian acts trivially on the ancillary sites ($\hH_{aux}=0$), and in the purified right subsystem
it has interaction range $2$, since neighbouring physical sites are separated by an ancillary site (for entanglement dynamics with inhomogeneous Hamiltonians, see Ref.~\cite{inHam}).

In Fig.~\ref{fig:entropyhomo} we used a maximum bond dimension $\chi=2500$, except for $(n,\Delta)=(5,3)$, for which
we used $\chi=4000$, since the bond dimension saturates earlier upon increasing $n$.
In Fig.~\ref{fig:energy1}, $\chi$ and the time window are specified in each panel.
In Fig.~\ref{fig:entropyconjecture} we used $\chi=1800$--$2500$, and in Fig.~\ref{fig:entropyparity} $\chi=2500$.

To compare with GHD in Fig.~\ref{fig:energy1}, for every value $\zeta$ we averaged the tDMRG data in a ``small'' space-time window 
with $(x,t)$ such that $t \in \left[t_{min}, t_{max} \right]$ and
$x/t \in \left[ \zeta - \delta \zeta, \zeta + \delta \zeta \right]$.
Here $t_{max}$ should be small enough for the tDMRG data to be reliable (typically $t_{max}\approx 10$),
while both $t_{min}$ and $t_{max}-t_{min}$ should be large enough to average
over the microscopic inhomogeneities.
For the bin width $\delta \zeta$, it is
convenient to use two different values: $ \delta \zeta_{L}$ for $\zeta<\zeta_c$ and
$ \delta \zeta_R$ for $\zeta>\zeta_c$. The values of $\delta\zeta_{L/R}$ and $\zeta_c$ are reported in Fig.~\ref{fig:energy1}. 

\subsection{Numerical solution of the TBA and GHD equations}

Throughout this work, we computed the convolutions in the BGT equations~\eqref{eq:BGT_densities} and in the equations for the dressed velocities~\eqref{eq:dressed}
	with Gaussian quadrature, using $100$ points in $[-\pi/2, \pi/2]$.
	Only for the $(\Delta, n)=(3, 1)$ panel of Fig.~\ref{fig:energy1} we increased the number of points up to $250$, to suppress numerical artefacts.

The computation of the GHD prediction becomes increasingly demanding with $n$.
Indeed, the recursive construction~\eqref{eq:recursion} of the operator $\hTT_s^{(n)}$
requires increasing numerical precision to avoid indeterminate ratios and the growth of numerical errors.
While in Sec.~\ref{sec:homogeneous} we easily reached $n \sim 10^3$ for $s \leq 3/2$, here
$m_{max}= 35$ requires $s_{max}=18$, and $n \sim 5$ is close to the limit of
our computational resources.
The effect of the truncation in $m$ is particularly severe in the right half of the chain,
because the infinite-temperature root densities decay as a power law
with the string length, whereas for generic states, including the $n$-rainbow states,
they usually decay faster.

To compute the predictions in Table~\ref{tab:entropy}, since the infinite-temperature
filling functions $\set{\theta_m}$ decay much more slowly than those of the $n$-rainbow states,
we set to zero the filling functions of the $n$-rainbow states for $m >35$. We then solved Eqs.~\eqref{eq:BGT_densities}, \eqref{eq:sol_interacting}, and \eqref{eq:dressed} with $m_{max}=50$.
Indeed, for $m \geq 35$ the filling functions of the $n$-rainbow states
are generally several orders of magnitude smaller than those of the infinite-temperature state.
Comparing with the results for $m_{max}=49$, we observed deviations of order $10^{-8}$--$10^{-5}$, so that the truncation has negligible effects for our purposes.
Hence, we do not report the uncertainties in Table~\ref{tab:entropy}, and we truncate the values at order $10^{-5}$.

\section{Violations of the quasiparticle picture for nonintegrable initial states}
\label{app:violations}

\begin{figure*}
	\centering
	\includegraphics[width=0.40\textwidth]{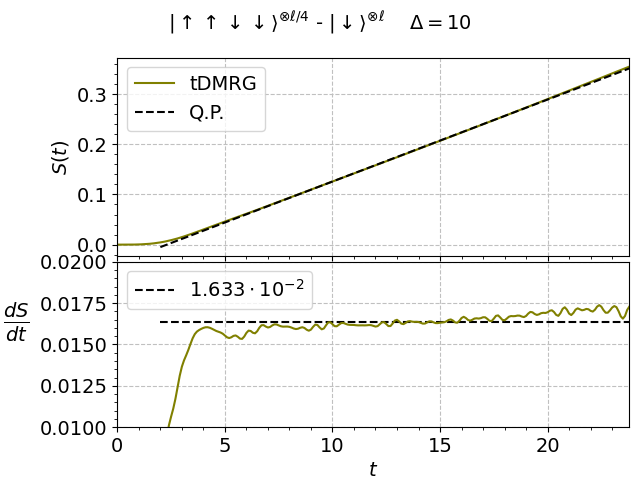}
	\includegraphics[width=0.40\textwidth]{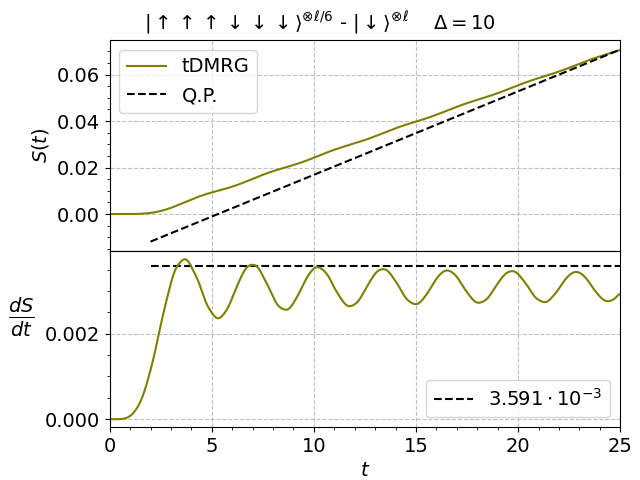}
	\includegraphics[width=0.40\textwidth]{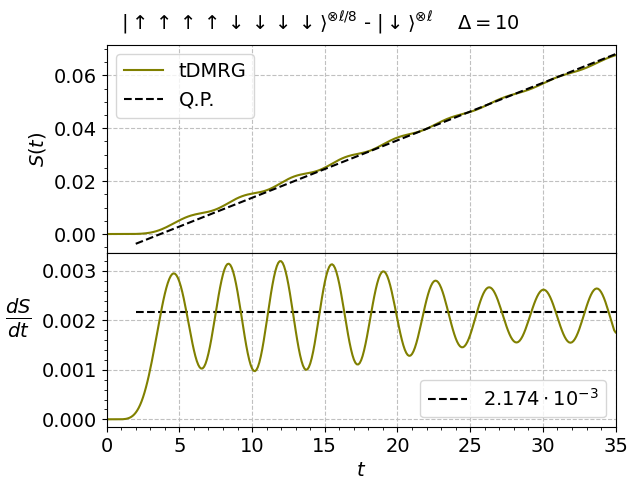}
	\includegraphics[width=0.40\textwidth]{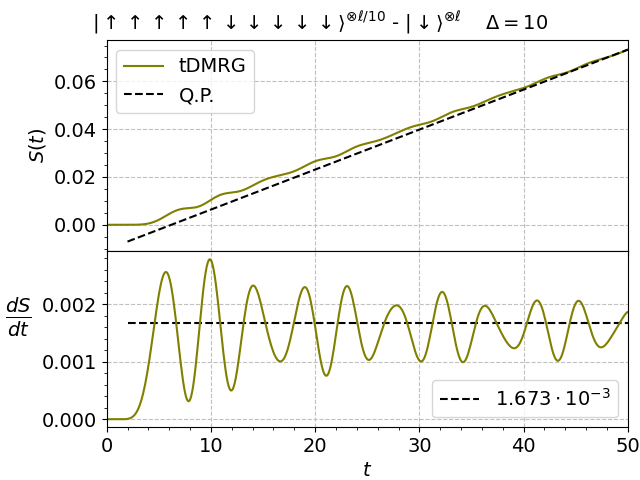}
	\caption{Entanglement entropy (green solid lines) after the bipartitioning protocol with the fat N\'eel state~\eqref{eq:pneel} in the left half and the vacuum in the right half, in the XXZ chain with $\Delta=10$, for $p=2,3$ (top) and $p=4,5$ (bottom). Black dashed lines: pair-structure quasiparticle prediction.}
	\label{fig:entropyparity}
\end{figure*}

Here we provide tDMRG data showing that for quenches from generic nonintegrable initial states 
the standard quasiparticle picture based on entangled pairs fails to describe the entanglement 
dynamics. We consider the dynamics starting from the ``fat'' N\'eel states, which were studied
in Ref.~\cite{biagetti2024bound}. The full state is $|\psi_p\rangle^{\otimes L/(2p)}$, with unit cell
\begin{equation}
    |\psi_p\rangle=|\uparrow^{\otimes p}\downarrow^{\otimes p}\rangle.
    \label{eq:pneel}
\end{equation}
We consider the bipartitioning protocol in which the left part of the chain is prepared in the fat N\'eel state, 
whereas the right one is in the ``vacuum'' state with all the spins down. The chain is long enough that
boundary effects can be neglected. Figure~\ref{fig:entropyparity} shows 
results for 
the XXZ chain with $\Delta=10$ and $p\in\set{2,3,4,5}$. In all the panels the dashed lines are the 
predictions obtained from the quasiparticle picture assuming that the quench produces entangled pairs.
The prediction clearly fails for $p=3$: the derivative $dS/dt$ oscillates around a value that is about
$15\%$ smaller than the prediction. For $p=2$ the agreement is good at intermediate times, but
$dS/dt$ slowly drifts above the prediction at later times. For $p\in\set{4,5}$ the data are compatible with
the prediction within the amplitude of the oscillations, although the accessible times are
limited. Moreover, oscillations seem to be centred slightly below the prediction.
These results show that the agreement with the pair-structure quasiparticle picture is not a
generic feature of nonintegrable initial states, but also that violations can be small and difficult
to detect within the time window accessible to tDMRG.

\bibliographystyle{apsrev4-2}
\bibliography{bibliography02}

\end{document}